\documentclass[a4paper,11pt]{article}
\usepackage{jheppub}
\usepackage{graphicx}
\usepackage{amsmath}
\usepackage{amsfonts}
\usepackage{amssymb}
\usepackage{slashed}
\usepackage[colorlinks=true,linkcolor=blue]{hyperref}
\usepackage{bm}
\usepackage{empheq}
\def\sss{\scriptscriptstyle}
\def\ah{a_{\sss{H}}}
\def\aho{a_{\sss{H}0}}
\def\half{\frac{1}{2}}
\def\OO{\mathcal{O}}
\def\be{\begin{equation}}
\def\ee{\end{equation}}
\def\bea{\begin{eqnarray}}
\def\eea{\end{eqnarray}}
\def\lqcd{\Lambda_\mathrm{QCD}}
\def\Eq#1{Eq.~\eqref{#1}}
\def\Re{\,\mathrm{Re}\,}
\def\Im{\,\mathrm{Im}\,}
\def\nn{\nonumber}
\def\nax{n_{\mathrm{ax}}}

\title{Axion dark matter:  strings and their cores}
\author[1]{Leesa Fleury,}
\author[2]{Guy D. Moore}

\affiliation[1]{McGill University, Department of Physics,\\3600 rue University, Montr\'eal QC H3A 2T8, Canada}
\affiliation[2]{Institut f\"ur Kernphysik, Technische Universit\"at Darmstadt\\
Schlossgartenstra{\ss}e 2, D-64289 Darmstadt, Germany}
\emailAdd{fleuryle@physics.mcgill.ca}
\emailAdd{guy.moore@physik.tu-darmstadt.de}

\abstract{
  Axions constitute a well-motivated dark matter candidate, and if PQ
  symmetry breaking occurred after inflation, it \textsl{should} be
  possible to make a clean prediction for the relation between the
  axion mass and the axion dark matter density.  We show that axion
  (or other global) string networks in 3D have a network density that
  depends logarithmically on the string separation-to-core ratio.
  This logarithm would be about 10 times larger in axion cosmology
  than what we can achieve in numerical simulations.
  We simulate axion production in the early Universe, finding that,
  for the separation-to-core ratios we can achieve, the changing
  density of the network has little impact on the axion production
  efficiency.
}

\begin{document}
\maketitle
\section{Introduction}
\label{sec:intro}

The axion
\cite{Peccei:1977hh,Peccei:1977ur,Weinberg:1977ma,Wilczek:1977pj}
is a proposed particle, the angular excitation of a
new ``Peccei-Quinn'' (PQ) field $\varphi$ that would solve the strong
CP problem \cite{tHooft:1976,Jackiw:1976pf,Callan:1979bg} and which is
also a very interesting dark matter candidate
\cite{Preskill:1982cy,Abbott:1982af,Dine:1982ah}, thereby solving
two puzzles with one mechanism.  In this light, the study of the axion
as a dark matter candidate is, in our view, well motivated.  The axion
model has one undetermined parameter, the vacuum value of $\varphi$,
$f_a$; the axion mass $m_a$ scales as $f_a^{-1}$.  The value of $f_a$
also determines the amount of axion dark matter that would be
produced in the early Universe, which means that (for a well-motivated
initial condition we will describe) it should be possible to
predict the axion mass from the dark matter density.  To do so, we
need to understand the efficiency of axion production in cosmology.

As we will soon explain, the PQ field's evolution in the early
Universe is complicated by the appearance of structures -- cosmic
strings -- that may play a role in determining the production efficiency
of axions around the QCD scale $T \sim 1$ GeV.  Here we will study
these structures and their role in axion production via numerical
simulations.  Ours is not the first study on this problem
\cite{Yamaguchi:1998iv,Yamaguchi:1998gx,Yamaguchi:1999yp,%
Yamaguchi:1999dy,Hiramatsu:2010yn,Hiramatsu:2010yu,Hiramatsu:2012gg,%
Hiramatsu:2012sc,Kawasaki:2014sqa,Hagmann:1990mj,Chang:1998tb,%
Hagmann:2000ja,Davis:1989nj,Dabholkar:1989ju,Battye:1993jv,%
Battye:1995hw,Martins:2003vd}, but we believe it will also not be the last;
we will demonstrate some significant challenges for such studies,
one of which we are not yet able to overcome.  Axion cosmic string
networks are in a family called ``global string networks,'' and
Martins and Shellard have argued \cite{Martins:2000cs,Martins:2003vd} that such
networks are sensitive to the sizes of the string cores, which cannot
be properly simulated numerically.  This
implies that numerical simulations will view string networks with very
different properties -- in particular, a much lower string density --
than the ones that would really occur for physical axion field
parameter values.  We will present strong numerical evidence
supporting this claim.  In particular, we will demonstrate
that the cosmic string density increases as we increase the log of
the ratio of the network age to the string core size.  This ratio
is very large in cosmology, implying that physically relevant
string networks may be an order of magnitude denser than those in
numerical axion-production simulations.
We will also point out another potential pitfall to numerical
simulations, which can be overcome but must be monitored.

We use the remainder of the introduction to review the properties of
axions relevant to cosmology and the overall picture of
axion production in the early Universe.  From the
point of view of cosmology, the relevant features of the axion are
that it is a complex scalar field $\varphi$ with a symmetry-breaking
potential\footnote{Our metric convention is $[{-}{+}{+}{+}]$, and we
  use standard complex-field notation $\varphi=(\varphi_r+i\varphi_i)/\sqrt{2}$,
  explaining some strange-looking factors of 2.},
\be
\label{Lagrangian}
-\mathcal{L}_{\mathrm{axion}} = \partial_\mu \varphi^* \partial^\mu \varphi
+ \frac{\lambda}{8} \left(2 \varphi^* \varphi - f_a^2 \right)^2
\ee
with $f_a$ the vacuum expectation value and $\lambda$ the
self-coupling, which we assume is $\OO(1)$.  The Lagrangian has a U(1)
``PQ'' symmetry $\varphi \to \varphi e^{i\theta}$, spontaneously
broken when $\varphi$ takes on a vacuum value somewhere on the
``vacuum manifold'' $2\varphi^* \varphi = f_a^2$.  The symmetry is
also explicitly broken by an anomalous coupling to QCD and instanton
effects, giving rise to an extra contribution to the potential,
\be
\label{L2}
-\mathcal{L}_{\mathrm{axion,eff.QCD}} =
  \chi(T) \left[ 1 - \cos\left(\mathrm{arg}\:\varphi\right) \right].
\ee
Here $\chi(T)$
is the temperature-dependent topological susceptibility
of QCD, which for our purposes is an input from the theory of QCD.
Its vacuum value%
\footnote{%
  At lowest order in chiral perturbation theory,
%  \be
%  \label{chi0}
  $
  \chi(T=0) \simeq F_\pi^2 m_\pi^2 \frac{m_u m_d}{(m_u+m_d)^2},
  $
  %  \ee
  with $F_\pi$ the pion decay constant, $m_\pi$ the pion mass, and
  $m_u,m_d$ the up and down quark masses \cite{Leutwyler:1992yt}.
  Recent lattice determinations \cite{Basak:2014vca} find
  $m_u/m_d \simeq 0.45$.}
is $\chi(0) \simeq (76\:\mbox{MeV})^4$.

The minimum of the axion potential is at
$\sqrt{2}\: \varphi = \sigma e^{i\theta_a}$ with $\sigma=f_a$ and
$\theta_a=0$.  Radial fluctuations $\sigma=f_a+s$, which we will call
\textsl{saxions}, have a mass $m_s^2 = \lambda f_a^2$;
and angular excitations $\theta_a = a/f_a$, \textsl{axions}, have a
mass $m_a^2(T) = \chi(T)/f_a^2$.  If $f_a \gg \lqcd$, as required by
existing constraints \cite{Turner:1989vc,Raffelt:1990yz,Raffelt:1999tx}, then there is a
large hierarchy between these masses, and only the axion should play a
role cosmologically for $T < f_a$.

$\chi(T)$ is known to be a strong function of $T$
at high temperature $T \gg \lqcd$, varying roughly as
$T^{7+N_f/3}=T^8$ 
\cite{Gross:1980br}.
Therefore as time progresses and temperature falls
cosmologically, the axion goes rapidly
from being effectively massless, $m_a t \ll 1$, to massive,
$m_a t \gg 1$.  The dynamics around $m_a t \sim \pi$ are quite
nontrivial; but once $m_a t \gg 1$, axion fluctuations will be small
and the axion number an adiabatic invariant, so the dynamics around
$m_a t \sim \pi$ determine the efficiency of axion production and
therefore the amount of axion dark matter.

It seems likely that PQ symmetry is restored near the end
of or after inflation.  In particular, PQ symmetry is generically
restored during high-scale inflation (if the Hubble
parameter during inflation is large, $H \geq f_a$) \cite{Visinelli:2014twa}%
\footnote{%
  There may be ways around this argument \cite{Takahashi:2015waa,Choi:2015zra},
  but we believe that it is the \textsl{generic} expectation.}.
Thermal symmetry restoration after inflation would have
occurred if the Universe reheated to a temperature $T \geq f_a$.
If either occurred, then we know the initial conditions for the axion
field: $\theta_a$ would start out uncorrelated
at causally-disconnected points, meaning essentially random initial
conditions should apply.  With the initial conditions known
statistically, the axion cosmology model has only one unknown
parameter, $f_a$.  If we can determine $\chi(T)$ and work out the
axion dynamics near $m_a t \sim \pi$, then it should be possible to
compute the relation
between the value of $f_a$ and the resulting dark matter density.  If
we assume that the axion constitutes the dark matter in the Universe,
the known dark matter density \cite{Ade:2015xua} should give a unique
prediction for $f_a$, and with it the axion mass.  The axion mass is
experimentally measurable, making this scenario testable.
And a narrow search window would also be very valuable for the design
of the most sensitive type of experiment, resonant microwave cavity
detectors \cite{Sikivie:1983ip,Bradley:2003kg,Asztalos:2009yp,Asztalos:2011ei}.

To finish setting the stage, we describe qualitatively how the fields
evolve near $m_at\sim \pi$.  At
early times, $m_a \simeq 0$ and the Lagrangian \Eq{Lagrangian} has a
$U(1)$ symmetry.  This symmetry is locally broken by the phase choice
$\varphi = f_a e^{i\theta_a(x)}$.  Random initial conditions give rise
to a network of topologically stable cosmic strings \cite{Kibble:1976sj},
which then evolve and untangle, entering a scaling solution in which
they maintain a string length per volume of order
\be
\label{length_est}
\frac{\gamma L_{\mathrm{string}}}{V} = \frac{\xi}{t^2} \,,
\ee
where $\gamma$, $L_{\mathrm{string}}$, and $V$ are the typical Lorentz
gamma-factor of a moving string, the physical length of string, and the
physical volume under consideration, respectively, and $t$ is the age of the
Universe. $\xi$ is an order-1 parameter describing the network
evolution. After the axion mass becomes relevant, the potential has a
single global minimum%
\footnote{%
  One can also consider axions with multiple minima,
  so $\cos(\mathrm{arg}\:\varphi)$ in \Eq{L2} becomes
  $\cos(N\:\mathrm{arg}\:\varphi)$.  However, it is difficult to avoid
  problems associated with stable domain walls in this model while
  simultaneously solving the strong CP problem without fine tuning
  \cite{Hiramatsu:2010yn,Hiramatsu:2012sc,Kawasaki:2014sqa}, so we will not consider this possibility here.},
and there are no true topological structures. However, the strings remain
metastable, and there are metastable
domain walls associated with the field's phase varying by $2\pi$ from
one side of the wall to the other; exactly one wall
ends on each string.  The domain wall tension draws the strings
together and accelerates the annihilation of the network \cite{Vilenkin:1982ks}.

We will argue that the ``scaling solution'' for the strings is not
quite as simple as \Eq{length_est} suggests; the string tension, the
parameter $\xi$, and other properties of the string and string-wall
networks should vary logarithmically with the ratio of the inter-string
separation $t/\sqrt\xi\sim t$ to the string core size $\simeq 1/m_s$.
In numerical field-theory simulations this ratio is bounded by the
size of the lattice used to solve the field dynamics: $m_s t$ will not
exceed about $10^3$ in 3D simulations or $10^4$ in 2D simulations.
Physically, we want $m_s t \sim f_a/H$, with $f_a \sim 10^{11}$ GeV and
$H \sim T^2/m_{pl} \sim 10^{-19}$ GeV; so the physically relevant
value for the ratio is more like $m_s t \sim 10^{30}$.  For a fixed
$f_a$, the physical value of $m_s t$ gives strings with 10 times
higher tension than in simulations, leading to
a much denser string network which breaks up more slowly under
the action of domain walls.  The physical
network evolution may be much more efficient at producing axions than
recent simulations (including the ones we present), implying a smaller
value of $f_a$ and heavier value of $m_a$.

The other danger is that even the meta-stability of the domain walls
is dependent on the mass ratio $m_a/m_s$.  We will show below that,
for $m^2_a/m^2_s > 1/39$, the domain walls cease to be even metastable,
and the topological structures abruptly collapse.  This certainly
should not happen cosmologically; but since $m_a$ rises rapidly with
time, it is a challenge to make a numerical simulation with $m_s$
large enough that this condition is maintained until the network
breaks up via the expected physics.  Therefore, there is some danger
that simulations will incorporate unphysical collapse of the
string-wall network or must be stopped before the network has
finished evolving.

The next section reviews the physics of global cosmic strings and
explains the relevance of $\ln(m_s t)$, using analytical arguments.
Next we present numerical evidence that the string network grows
denser with increasing $\ln(m_s t)$.  Then we study the network
evolution and axion production around
$m_a t \sim \pi$ in more detail.  Our study of the string-wall
network's collapse actually does \textsl{not} show evidence that the
final axion density depends strongly on $\ln(m_s t)$; but the dynamic
range we can study is too narrow to make any brave extrapolations
about what happens when the log is increased by another factor of 10.

As an aside, we mention that besides the details of the axion dynamics
around $m_a t \sim \pi$, there are also uncertainties in the axion
production efficiency from our incomplete knowledge of the temperature
dependence of the topological susceptibility $\chi(T)$.
The high-temperature behavior is only known at first order in
perturbation theory \cite{Gross:1980br}, so the
first unknown corrections are suppressed by $\OO(\alpha_s)$.
Therefore the perturbative treatment may not be very
reliable around $T\sim 1$ GeV where the interesting dynamics occurs.
In this temperature range we only have model calculations
\cite{Wantz:2009mi,Wantz:2009it}; lattice calculations
\cite{Berkowitz:2015aua,Borsanyi:2015cka} are currently available only
at lower temperatures and/or in the quenched approximation.  We will
find that the axion production is not too sensitive to the exact value
of $\chi(T)$, but it would nevertheless be valuable to have a reliable
lattice calculation for $T$ in the range from 0.5 to 1.5 GeV.

\section{String networks and string cores}
\label{sec:string}

We begin with a lightning review of why the strings arising from
\Eq{Lagrangian} have logarithmically large string tensions.  We assume
some familiarity with string defects; a reader who needs some
background can look in \cite{vilenkin2000}.  We will make the
classical field approximation throughout, which is an excellent
approximation for the IR axion field dynamics since the mean
occupancy is $\sim f_a^2/H^2 \sim 10^{60}$.

\subsection{Cosmic strings}

\indent
Consider a string lying along the $z$ axis in polar coordinates.  The
field varies as
$\varphi = f_a f(r) e^{i(\phi-\phi_0)}$ with $\phi$ the azimuthal
angle and $f(r)$ a function obeying $f(r) \to_{r\to 0} 0$,
$f(r) \to_{m_sr \gg 1} 1-\OO(1/m^2_s r^2)$.  The associated energy is
\be
E = \int d^3 x |\nabla \varphi|^2 = 
\int dz \int rdr \int d\phi \left(
\half (f_a \partial_r f)^2
+ \half \left| \frac{f f_a}{r} \partial_\phi e^{i\phi-\phi_0}
\right|^2 + \frac{\lambda f_a^4}{8}(1-f^2)^2 \right).
\label{Estring}
\ee
Far from the string's core, $f\to 1$ and all terms become negligible
except for the $\phi$-derivative term, which becomes $f_a^2 / 2r^2$.
Therefore the energy density decays as $1/r^2$ as we move away from
the string core.  This is in contrast to ``local'' strings, where the
U(1) symmetry is gauged (local) and a gauge field compensates for the
$\phi$-derivative except in the string's core.  The local string case
has received much more attention in the literature.

The large-distance part of the string tension (energy per length) in
\Eq{Estring} is approximately
\be
\label{tension}
T = \frac{E}{L} \simeq \pi \int_{\sim 1/m_s}^{\ell} rdr \: \frac{f_a^2}{r^2}
= \pi f_a^2 \ln(m_s \ell) \,,
\ee
where $\ell$ is an IR length scale where this description breaks down,
for instance the distance to the next string, $\ell \sim 1/H$.  This
dependence on the distance to the next string indicates that there is
a long-range force-per-length (gradient of string energy-per-length)
between strings, with a strength of order $d(E/L)/d\ell =\pi f_a^2 / \ell$.

There are two important facts here.  First, there are long-range
interactions between strings, mediated by the (nearly) massless axion
field.  The force between strings of opposite winding sense is
attractive, which helps them to find each other and annihilate.
Though we have not shown it, the presence of a massless mode also
helps accelerating or bent strings to radiate energy more efficiently
than for the local string case.  Both of these effects help
the string network to annihilate more efficiently, leading to a much
lower-density string network.

Second, the attractive forces between strings, and the radiation of
energy into long-wavelength axions, scale with $f_a^2$ in the same way
as the string tension, but they are not enhanced by
$\ln(m_s \ell)$, while the energy-per-length, or string tension, is.
Therefore the ratio of tension to radiation/force effects is
proportional to this log, but not to $f_a$.
To keep track of this difference, we will name this large logarithm
$\kappa \equiv \ln(m_s \ell) \simeq \ln(f_a/H)$.

\subsection{Scaling in 2D}
\label{sec:2Dscale}

Let's first see how $\kappa$ plays a role in string density for 2D
networks, where sensitivity to this logarithm is generally
accepted \cite{Yamaguchi:1998iv}.  The 2D network is described by the 3D
theory but enforcing that the field $\varphi$
does not vary along the $z$ axis, $\varphi = \varphi(x,y)$.  In the
$x-y$ plane, the string becomes a monopole or point-charge, with
$\varphi(x,y)$ varying by $\pm 2\pi$ as one goes around the charge,
depending on whether the string has positive or negative orientation.
The two orientations of strings act like two signs of charges.

The analogy to electric charges turns out to be complete
\cite{Kalb:1974yc, Lund:1976ze, Vilenkin:1986ku,%
Davis:1988rw,Hecht:1990mv}. Outside of the string cores, we can
write
$\varphi = f_a e^{i\theta_a}$, and the equation of motion becomes
\be
\label{EOM_thetaa}
\partial_t^2 \theta_a = \nabla_i^2 \theta_a \qquad \mbox{or} \qquad
\partial_\mu \partial^\mu \theta_a = 0 \,.
\ee
Except in string cores, we can define dual electric and magnetic
variables
\be
\label{EMvar}
F^{\mu\nu} = f_a \epsilon^{\mu\nu\alpha} \partial_\alpha \theta_a \,,
\ee
obeying
\be
\label{Maxwell}
\partial_\mu F^{\mu\nu} = f_a \epsilon^{\mu\nu\alpha}
\partial_\mu \partial_\alpha \theta_a = 0 \,,
\qquad
\epsilon^{\mu\nu\alpha} \partial_{\alpha} F_{\mu\nu}
= -2f_a \partial^\alpha \partial_\alpha \theta_a = 0 \,,
\ee
which are the free-space Maxwell equations in 2+1 dimensions.  A
surface in 2D is a loop, and the electric flux through the surface is
$2\pi$ times the winding number around the loop,
\be
\label{Eq:wind}
\int_{C} \hat{n}_i E_i = \int_{C} \epsilon_{ij} E_i dl_j
= f_a \int_{C} dl_j \frac{\partial \theta_a}{\partial x_j}
= 2\pi f_a n_{\mathrm{encl.}}
\ee
showing that a string is the source for a flux of $\pm 2\pi f_a$.

The electrical attraction between two strings will be
\be
\label{MaxF}
F = \frac{q_1 q_2}{2\pi r} = \pm \frac{2\pi f_a^2}{r} \,,
\ee
as expected if each has an energy $\pi f_a^2 \ln(m_s r)$ as we found
above%
\footnote{The units may look strange, because an energy in 2D is an
  energy-per-length or tension in 3D units.}.
The short-distance part of this energy,
$\pi f_a^2 \ln(m_s r_0) = \pi f_a^2 \kappa$, should be interpreted as the
mass of the charge, $m = \kappa \pi f_a^2$.  Varying $\kappa$ (the log
of the ratio of separation to core length scales) is varying the
charges' mass at fixed charge magnitude.  We want to argue that making
the charges heavier will make them more non-relativistic, so they move
and annihilate less efficiently and have a higher density.

Suppose that at $t=0$ there is a very dense starting ensemble of
positive and negative charges, which evolve under Hubble drag and
their electromagnetic interactions.  We can find how the density
of charges will evolve by making parametric scaling estimates.  The
charges find each other under the influence of Coulomb attraction and
annihilate off.  Suppose that at time $t$ the mean inter-charge
separation is $\ell$, so the density of charges is $1/\ell^2$, and let
us try to estimate $\ell$.  On dimensional grounds we expect
$\ell \sim \kappa^{-n} t$, with $n$ an exponent we want to determine.

We see that the density of charges should fall, with $\OO(1)$ of the
charges annihilating in an $\OO(t)$ amount of time.  To do so, the
charges have to move a distance $\sim \ell$ in a time $t$, implying
that $v\sim \ell/t$.  The typical force on a charge is
$F \sim 2\pi f_a^2 /\ell$, so the kinetic energy a charge obtains is
force times distance,
$\epsilon = F\times d \sim 2\pi f_a^2/\ell \times \ell \sim 2\pi f_a^2$.
Equating with $\half mv^2 = \half \kappa \pi f_a^2 v^2$, we find
$v \sim 2/\sqrt{\kappa}$.  Then $\ell \sim 2 \kappa^{-1/2} \, t$.
The density of strings is $n \sim \ell^{-2} \sim \kappa / 4t^2$.
Therefore we estimate that the number density of strings should
increase linearly with the log of the scale ratio $\kappa$, and that
the velocity should scale as $1/\sqrt{\kappa}$, becoming
nonrelativistic as the logarithm becomes large.

We therefore have a robust argument that, in 2 space dimensions, the
string density will scale with the log of the scale separation
$\kappa \equiv \ln(m_s t)$ as
$n \sim \kappa/t^2$, with small squared velocities
$\langle v^2 \rangle \sim 1/\kappa$.

\subsection{Scaling in 3D}
\label{sec:3Dscale}

The argument in 2D clearly does not translate simply into 3D, since
3D strings move under tension as well as under mutual interactions.
Nevertheless we expect strong, though not necessarily linear,
$\kappa$-dependence in the string network density in 3D as well.  The
reason is that the long-range interactions of the strings provide
rather efficient mechanisms for the strings to radiate energy and to
annihilate against each other.  This is in contrast to local (gauge)
string networks, where there are no long-range interactions and only
very inefficient radiation of gravitational waves.  As a result,
the density of global strings, $\xi < 1.2$ at achievable $\kappa$
values, is an order of magnitude smaller than the value $\xi\sim 13$
found in numerical lattice field theory
\cite{Moore:2001px,Martins:2003vd} and Nambu-Goto
\cite{Dabholkar:1989ju, Bennett:1989yp, Allen:1990tv} evolutions.

However, while the interactions between strings and the radiation of
energy from accelerating strings should scale as $f_a^2$, the tension
of strings should scale as $f_a^2 \kappa$.  In the large-$\kappa$
limit, the tension should dominate the mutual string interactions and
radiation.  Therefore, as $\kappa$ is increased global string networks
should behave more like Nambu-Goto string networks.  For very large
values of $\kappa$, the network density $\xi$ should be a factor of 10
larger than at currently achievable $\kappa$ values.

\begin{figure}[ht]
  \hfill \epsfxsize=0.4\textwidth\epsfbox{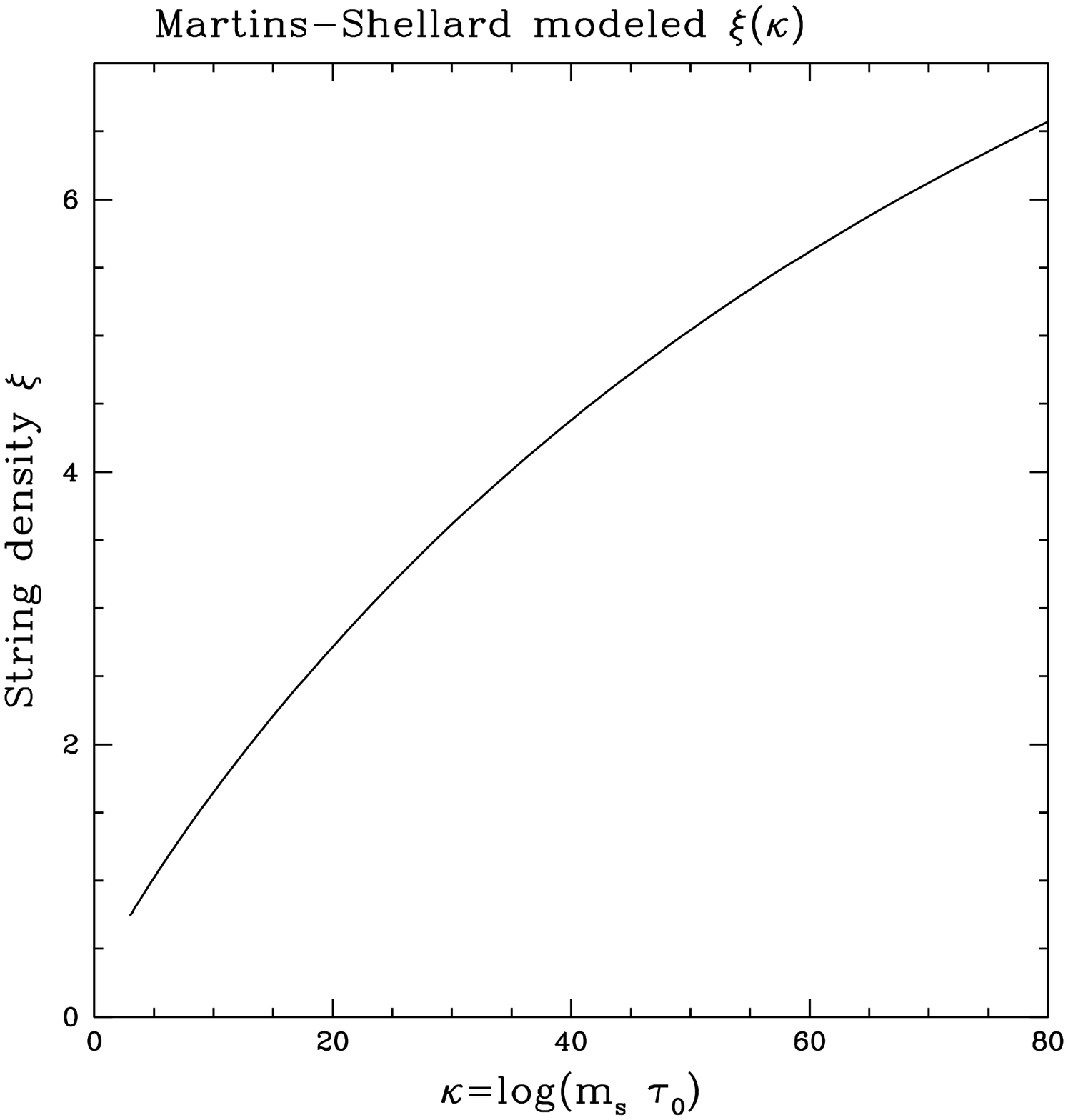}
  \hfill \epsfxsize=0.4\textwidth\epsfbox{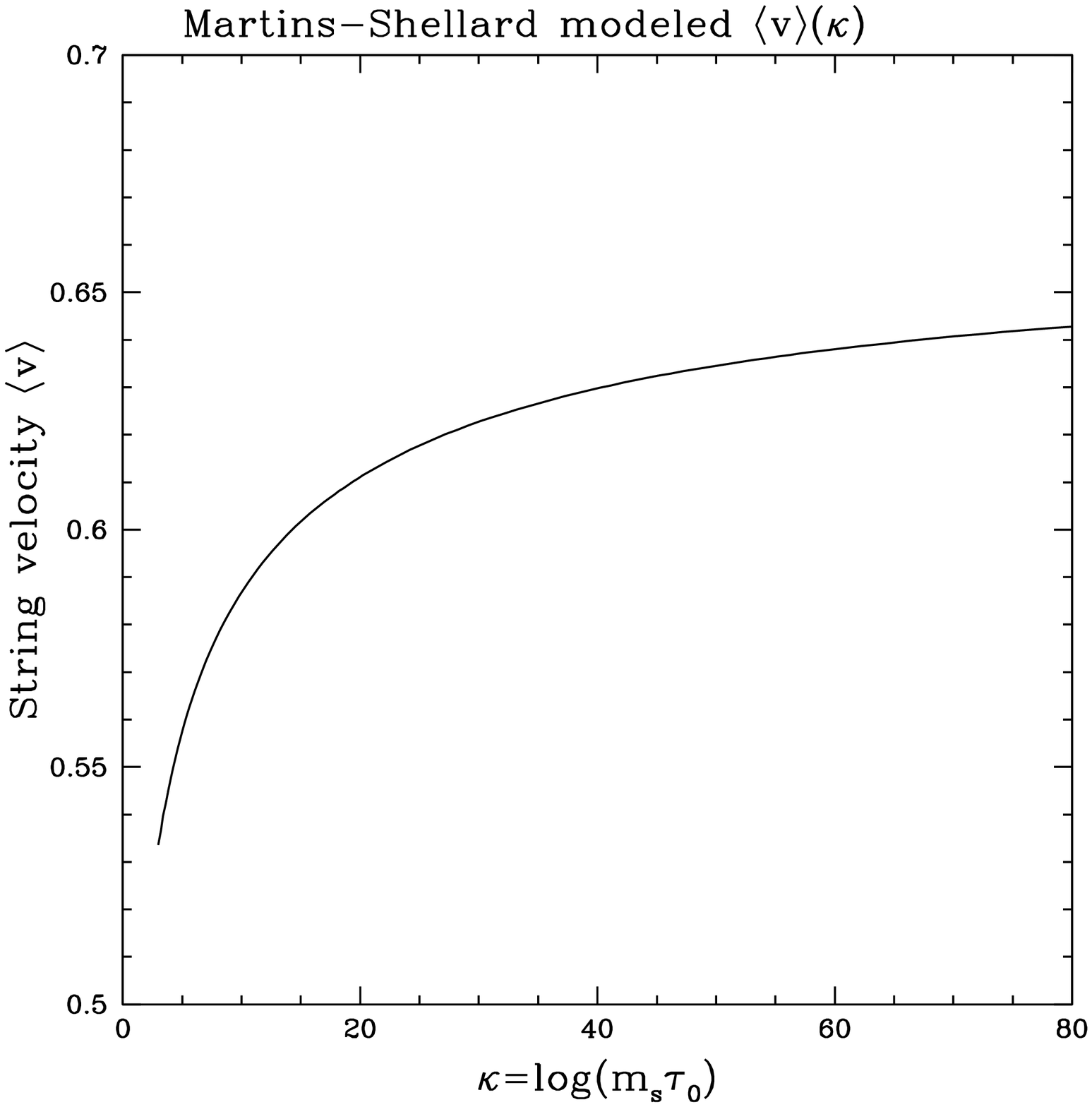}
  $\phantom{.}$ \hfill
  \caption{\label{martinsshellard}
    Estimated $\kappa$ dependence of the string density $\xi$ and
    velocity $\langle v \rangle$ according to the one-scale model of
    Martins and Shellard.}
\end{figure}

Martins and Shellard attempted to model this by amending their 1-scale
model for string networks to include the radiation effects associated
with global strings \cite{Martins:2000cs}.  Fitting the single
parameter describing radiation efficiency in their Eq.(4.2) to the
value of $\xi=1.15$ at $\kappa=6$ from Fig.~\ref{string2} below, and solving
their equations describing the string density and velocity, we get the
results plotted in Fig.~\ref{martinsshellard}.  According to this
model, the string density should vary nearly linearly with $\kappa$
for $\kappa \sim 5$--10, and should grow by roughly a factor of 5 in
going from $\kappa=6$ to $\kappa=70$.

We will compare numerical simulations to these expectations in the
next section, by considering string evolution without the ``tilt''
term, \Eq{L2}.

\section{Simulations of string networks}
\label{sec:simulation}

\subsection{Implementation}
\label{sec:implement}

To investigate these issues, we have implemented an MPI-parallelized
code to evolve the equations of motion that follow from
\Eq{Lagrangian} and \Eq{L2}.  We work in a radiation dominated FRW
background, so the Hubble parameter%
\footnote{We write the Hubble parameter as $\ah$ to avoid confusion
  with the lattice spacing $a$.}
$H = \partial_t \ah / \ah$ is
related to the time $t$ as $H=1/(2t)$.  We work in comoving
coordinates and introduce conformal time
\be
\label{conf_time}
d\tau = \frac{dt}{\ah}\,, \qquad \mbox{so that} \qquad
\tau = 2 \sqrt{\frac{t_0 t}{\aho}} \,,
\ee
with $t_0$ and $\aho$ the values of $t$ and $\ah$ at some (arbitrary)
fixed time.  Similarly, we rescale masses by $\ah$ so that
$m_{\mathrm{phys.}} dt = m_{\mathrm{conf.}} d\tau$; in what follows,
all masses are $m_{\mathrm{conf.}}$.
The temperature scales as $T \propto \tau^{-1}$, and the metric is
$g_{\mu\nu} = \frac{\aho^4}{4t_0^2} \mathrm{Diag}[-\tau^2,\tau^2,\tau^2,\tau^2]$.
In these units and writing space $(i,j)$ and time $(0)$ components
and factors of $\tau$ from the metric explicitly, the action, up to an
irrelevant multiplicative constant, is
\bea
\label{action}
S = \int d\tau \int d^3 x && \left( \vphantom{\half}
-\tau^2 \partial_\tau \varphi^* \partial_\tau \varphi
+ \tau^2 \partial_i \varphi^* \partial_i \varphi \right.
\\ \nn && \left. \phantom{\bigg(}
   {} + \frac{\lambda \tau^4}{8} \Big( 2\varphi^* \varphi - f_a^2 \Big)^2
   + \tau^4 \chi(\tau) (1 - \cos(\mathrm{arg}\: \varphi)) \right) \,.
\eea
We discretize this on a grid that is uniform in these coordinates,
with spacing $a$ and temporal spacing $a_t = a/n_t$.  We give a few
more details about our numerical implementation in
Appendix~\ref{app:A}; in particular we explain there how we determine
the density of strings and walls, and what we believe is a new
technique for establishing the string velocity directly from
information in the field variables.

Our numerical implementation is rather standard, except for two
points.  First, like many authors we replace
\be
\label{Lreplace}
\chi(T) [1 - \cos(\mathrm{arg}\: \varphi)] \; \Longrightarrow \;
\chi(T) [ 1 - f_a^{-1} \varphi_r ] \,,
\ee
with $\varphi_r = \sqrt{2} \Re\:\varphi$.  This form has the advantage
of being analytic in field variables, so no trigonometric evaluations
are required. It differs from the original form only in string cores,
where we expect this term to have little impact; the symmetry-breaking
term will always be small, so it is only important because it applies
over large regions of space where the field is near its minimum.
We assume that $\chi(\tau)$ grows as a power of $\tau$ in the relevant
time regime, so if $\chi(T) \propto T^{-n}$ then
$\tau^2 \chi(\tau) = f_a^2 (\tau/\tau_0)^{n+2}$, two powers higher
than the temperature dependence of $\chi(T)$.  We also scale out an
overall factor of $f_a^2$ from the Lagrangian and rescale
$\varphi \to \varphi/f_a$.

The other nonstandard change we make is to the potential term.
Physically, we are interested in large values of
$m_s^2 = \lambda f_a^2$.  We cannot make the value larger than
$m_s^2 a^2 \sim 1$, since otherwise we encounter numerical artifacts
(as discussed in Appendix~\ref{app:A}); but physically we want it to
be as large as possible, since we expect the physical value of $m_s$
to be very large.  Therefore we remove two powers of $\tau$ from this
term so that its $\tau$ scaling is the same as the gradient terms, so
$m_s a$ remains fixed in our simulations.  In other words, we keep the
string core size fixed in lattice units, rather than allowing it to
grow smaller in comoving coordinates due to Hubble expansion.  This
change enlarges the dynamic range where our simulations can see
cosmic string networks, making it easier for us to achieve ``scaling.''

\subsection{Results for string-only networks}
\label{sec:resultma0}

As a first application, we investigate string networks without the
symmetry-breaking term, that is, setting $\chi(\tau)=0$ or $m_a^2=0$.
In this case, no domain walls form, and the string network evolves until
we terminate the simulation at $\tau \leq L/2$.

\begin{figure}[pht]
  \epsfxsize=0.48\textwidth\epsfbox{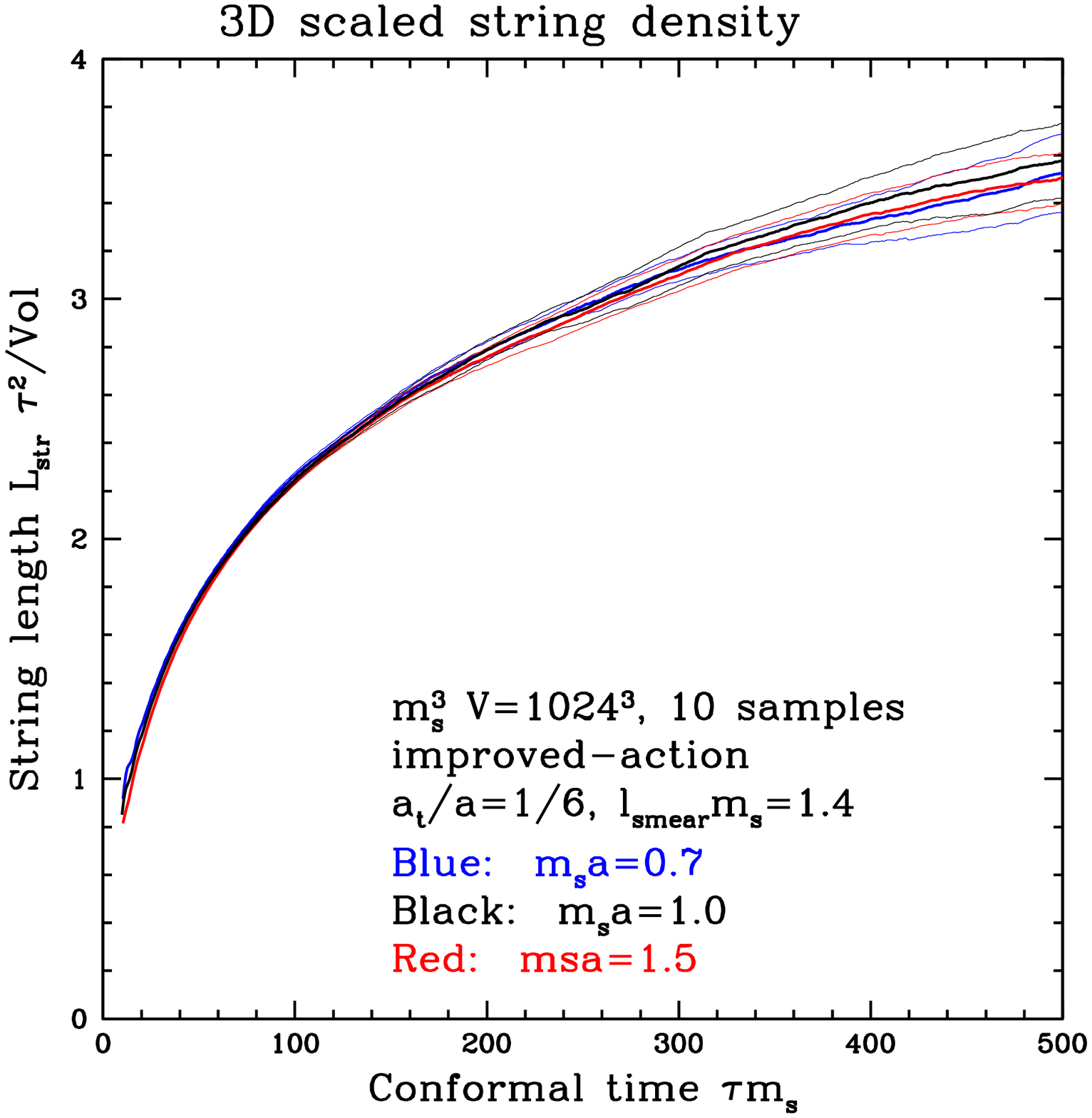}
  \hfill
  \epsfxsize=0.48\textwidth\epsfbox{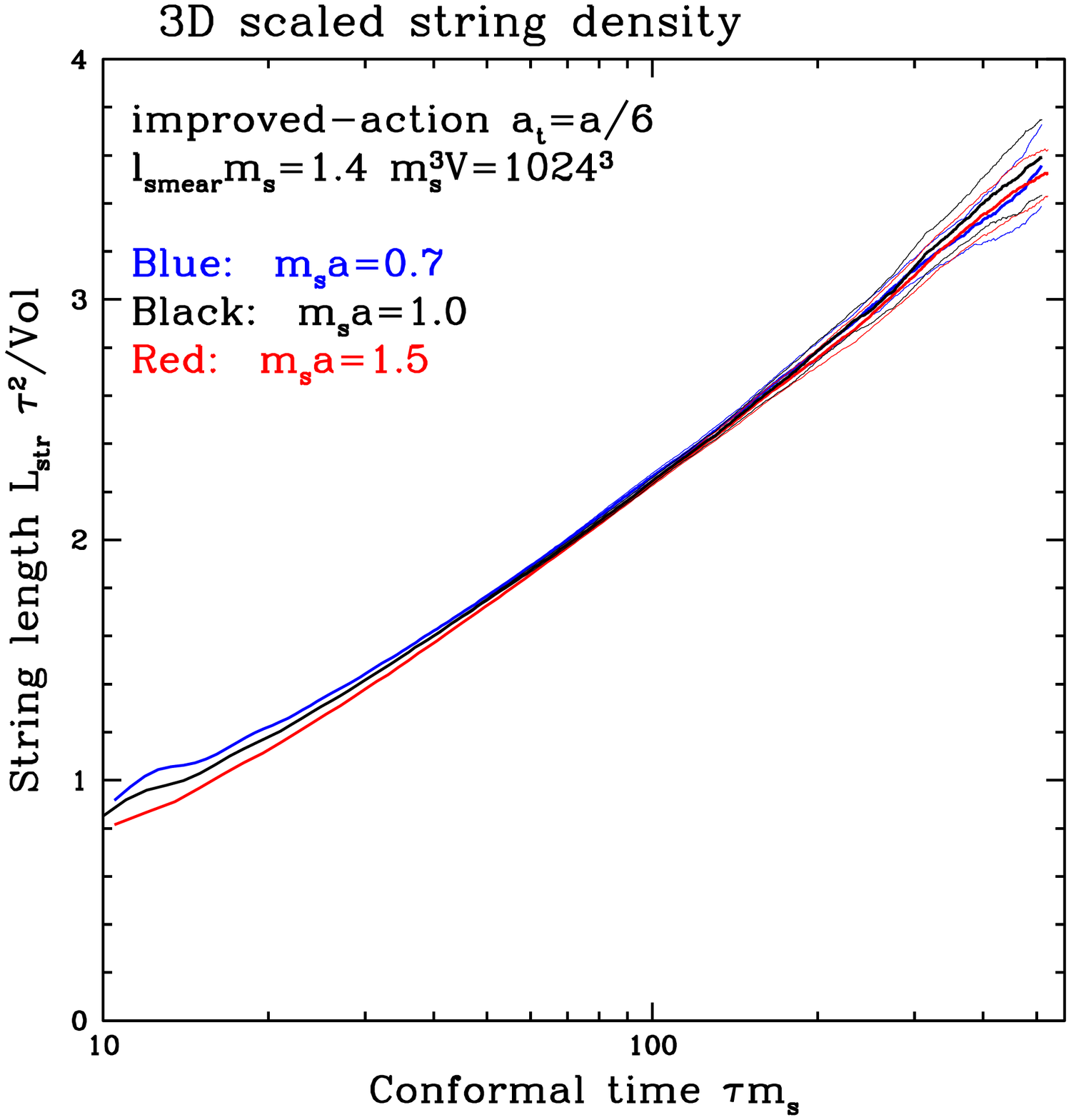}
  
  \epsfxsize=0.48\textwidth\epsfbox{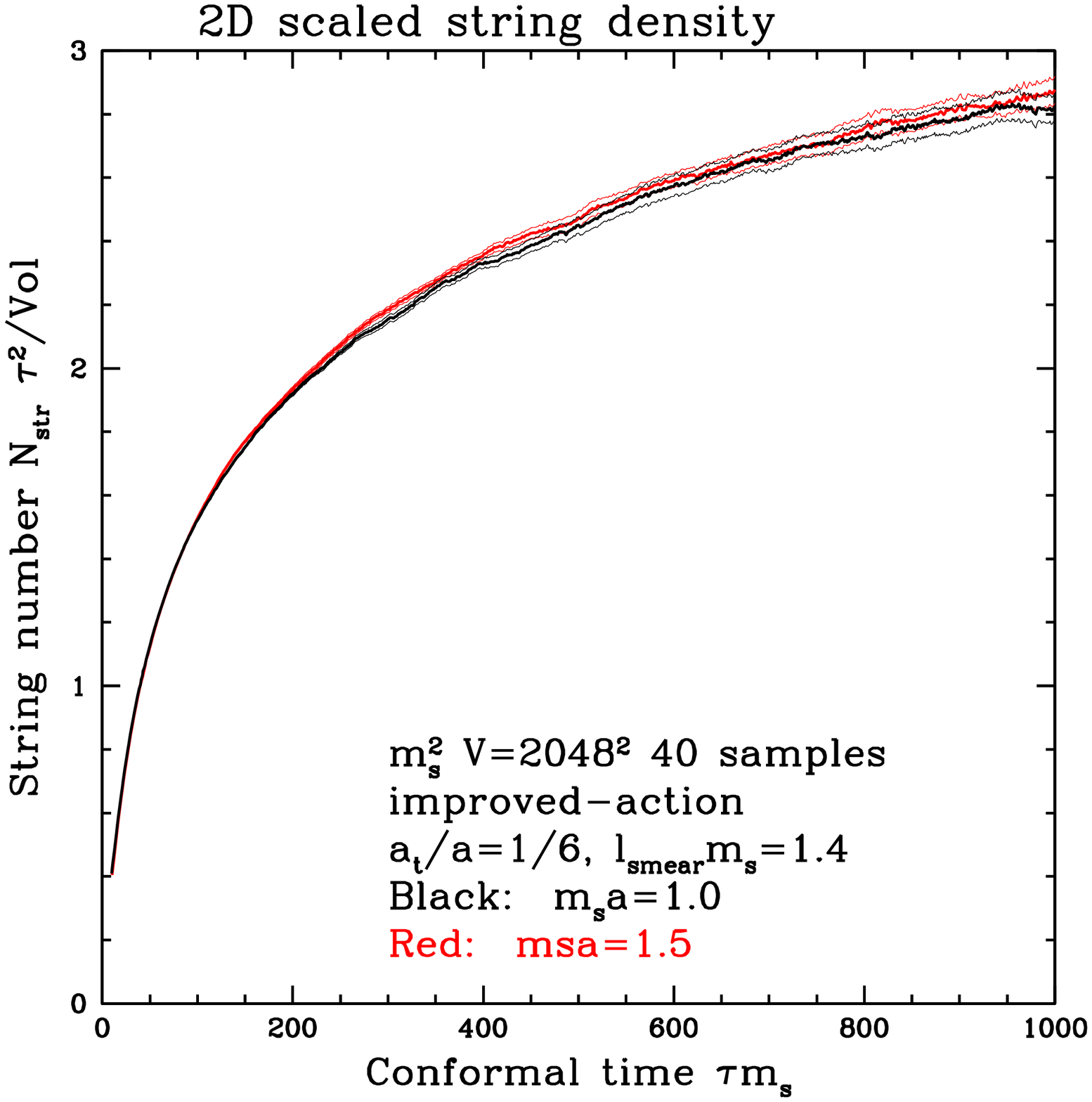}
  \hfill
  \epsfxsize=0.48\textwidth\epsfbox{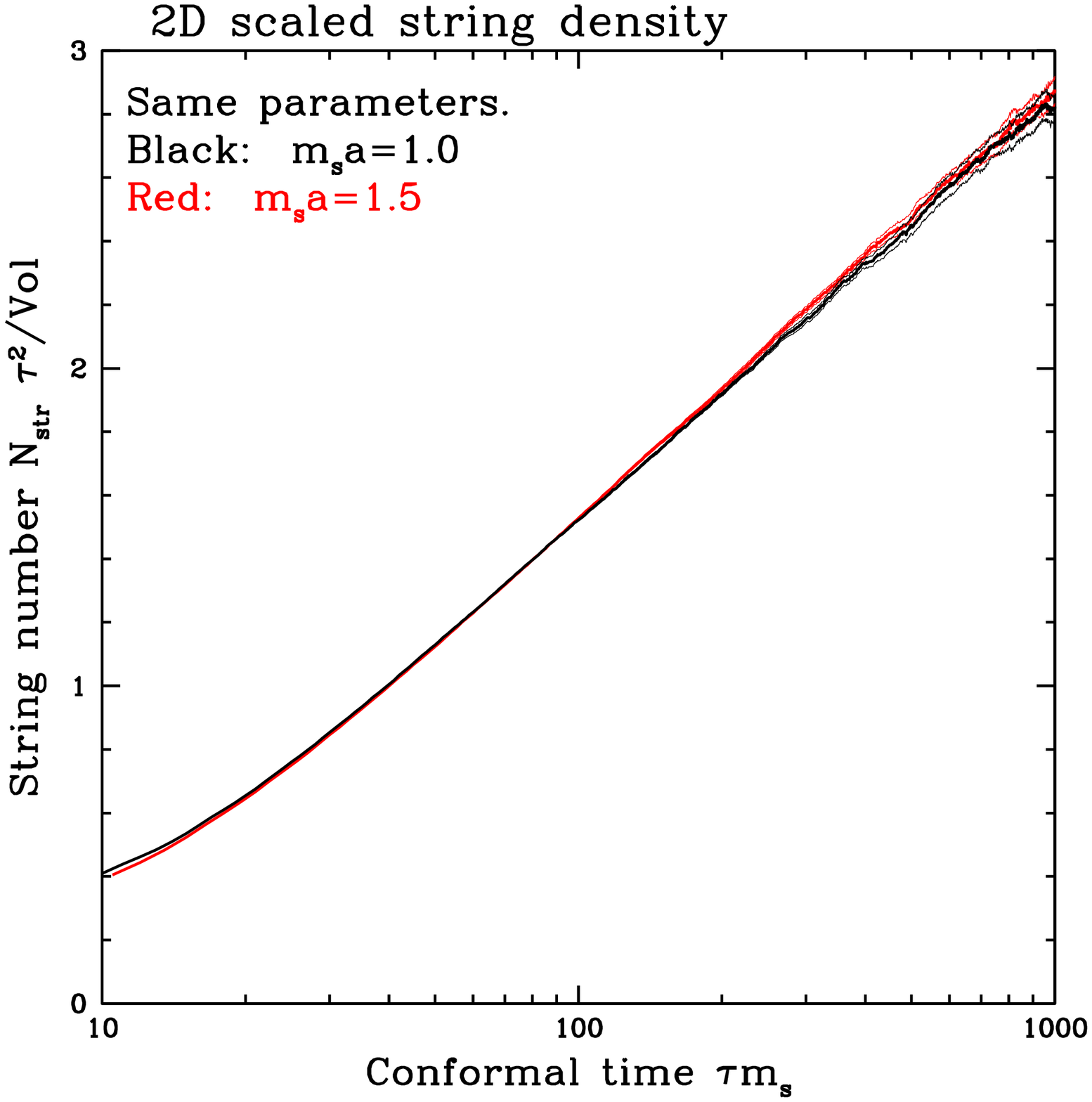}
  \caption{\label{string1}
    String length per volume, scaled by $\tau^2$, for 3D (upper) and
    2D (lower) simulations, plotted against conformal time (left) and
    its log (right).  Upper/lower curves are $\pm 1\sigma$ statistical
    errors.}
\end{figure}

We have plotted the length (number) of strings, scaled by $\tau^2$ to
account for system expansion, for 3D and 2D simulations in
Fig.~\ref{string1}.  The left plots are in terms of conformal time
measured in units of the saxion mass, $\tau m_s$; they show a rise, at
first rapid and then more gradual, in the string density when
normalized by a $\tau^2$ factor.  To test the hypothesis that this
rise represents logarithmic scaling in the separation-to-core ratio
$\kappa \simeq \ln(\tau m_s)$, we present a semilog plot on the
right-hand side of the figure.  Indeed, the nearly straight-line
behavior is striking in both 2D and 3D.  In each figure we have shown
two or more choices of lattice spacing $m_s a$, and we indicate the
$1\sigma$ statistical errors with upper and lower thin curves
accompanying each best-value thick curve.
The figure is also a check that our rather coarse lattice, $am_s=1.5$,
is sufficient to see continuum behavior; we plot curves for
$am_s=1.5$, for $am_s=1.0$, and (in 3D) for $am_s=0.7$, which fall on
top of each other to within the small statistical errors.

It is common in the cosmic string community to consider also the mean
velocity, $v^2$, and Lorentz $\gamma$-factor of the strings.
It is more customary to normalize the network density in
terms of the time rather than the conformal time; according to
\Eq{conf_time} this divides%
\footnote{The factor is still $t^2$, rather than $t$, because one must
  also work in terms of physical rather than comoving lengths and
  volumes.}
the string densities shown in Fig.~\ref{string1} by 4.  In addition,
it is customary to track not the length of string network, but the
energy in the network divided by the string tension.  This is the same
as integrating $\int \gamma dl$ rather than $\int dl$ in determining
the effective length of string in the network.  In computing the mean
velocity, $v^2$, and $\gamma$-factor one also uses this normalization,
so
\be
\label{def<v>}
\Big( \langle v \rangle \:,\: \langle v^2 \rangle \:,\:
\langle \gamma \rangle \Big)
= \frac{\int \gamma \: dl \times \Big(
  v \:,\: v^2 \:,\: \gamma \Big)}{\int \gamma \: dl} \,.
\ee

\begin{figure}[pht]
  \epsfxsize=0.48\textwidth \epsfbox{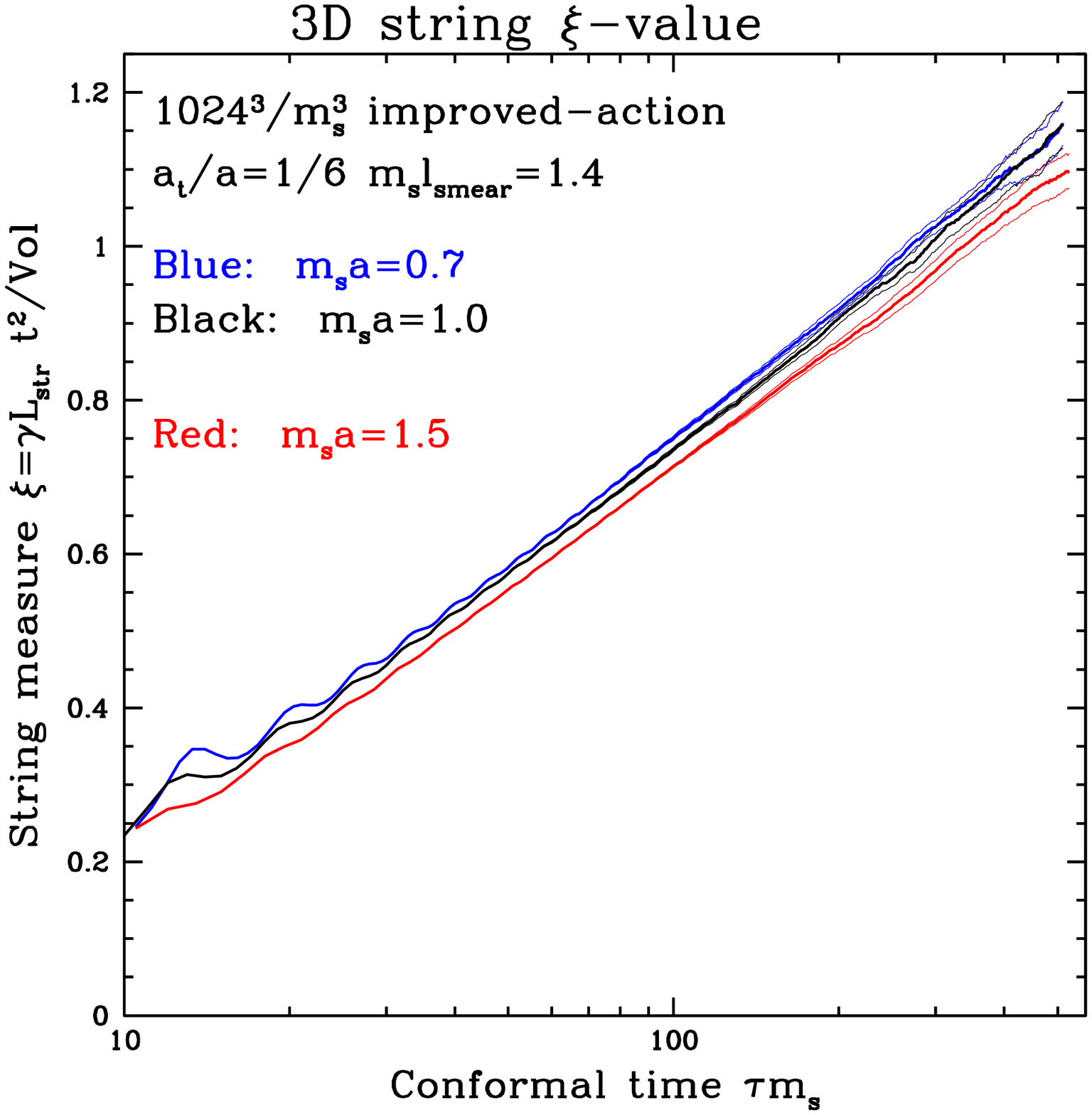}
  \hfill
  \epsfxsize=0.48\textwidth \epsfbox{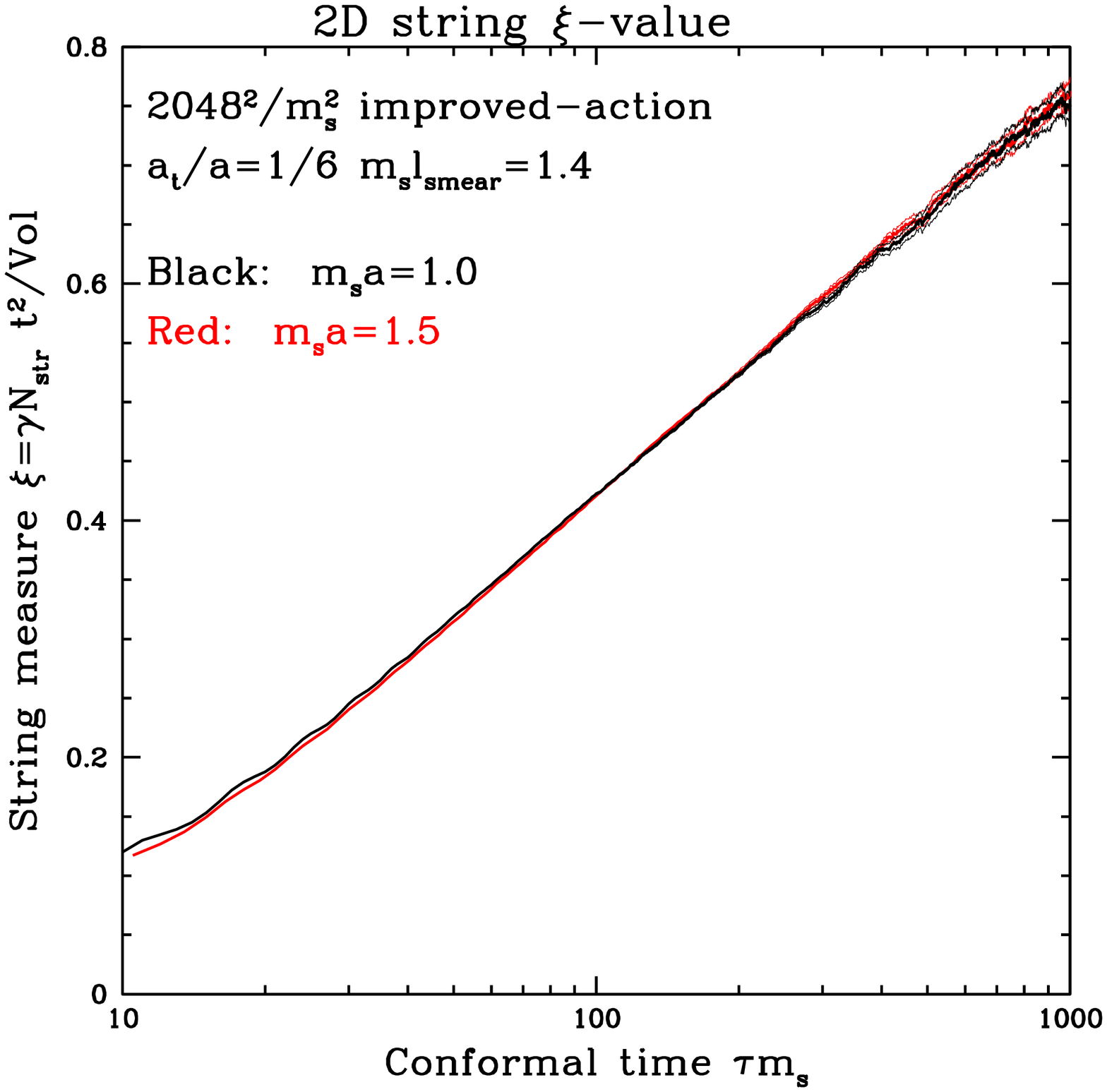}
  \caption{\label{string2}
    String network density in 3D (left) and 2D (right), varying with
    the log of conformal time.  Different curves refer to different
    lattice spacings $m_s a$. }
\end{figure}

We plot the normalized network density in 3D and 2D, using this
normalization, in Fig.~\ref{string2}.  The figures indicate that the
$\gamma$-weighted string density is also a logarithmic function of
$\tau m_s$, and that the string density in 3D is over an order of
magnitude smaller than the value $\simeq 13$ for local string networks
\cite{Bennett:1989yp, Allen:1990tv}.

\begin{figure}[pht]
  \epsfxsize=0.48\textwidth \epsfbox{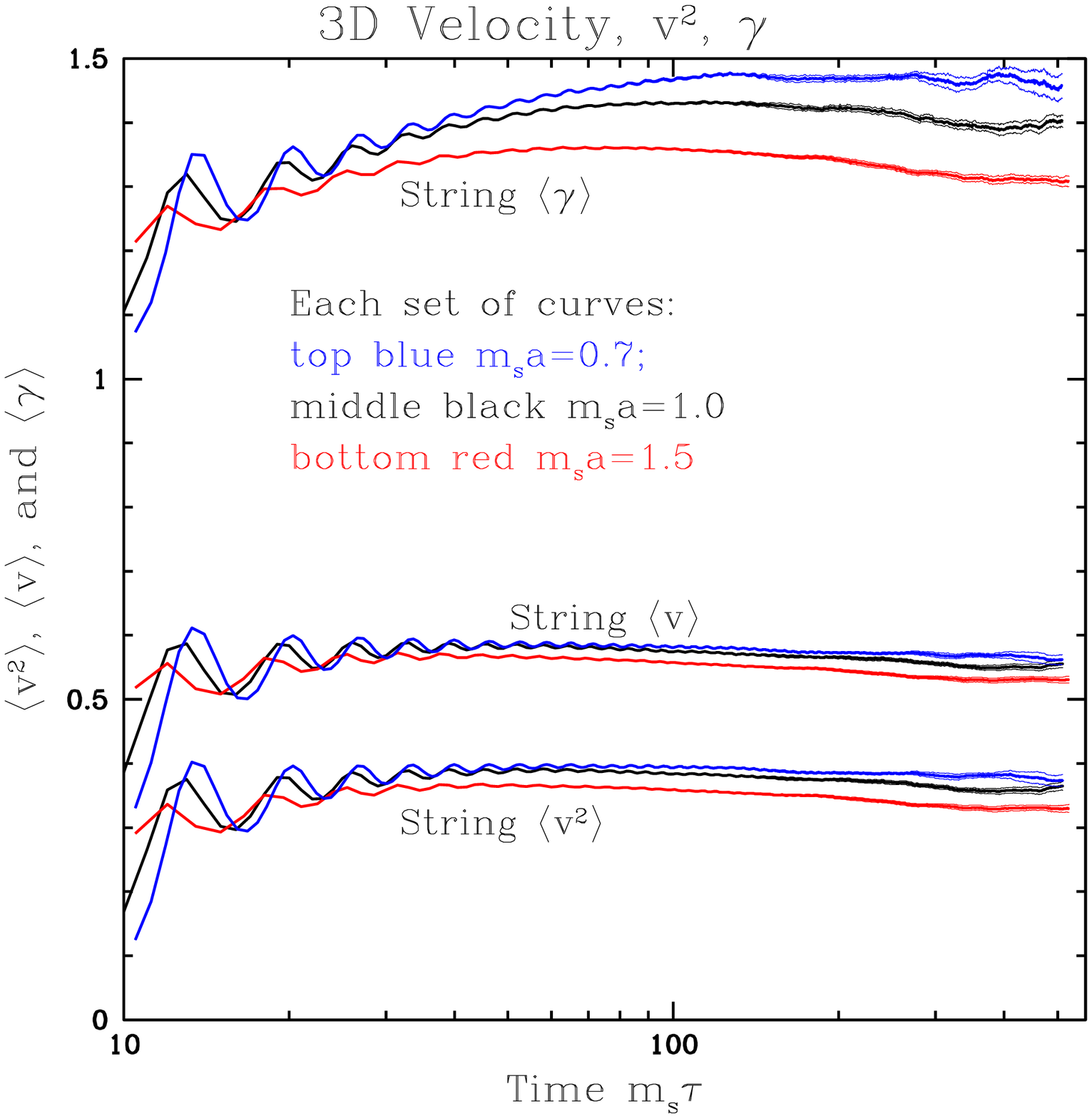}
  \hfill
  \epsfxsize=0.48\textwidth \epsfbox{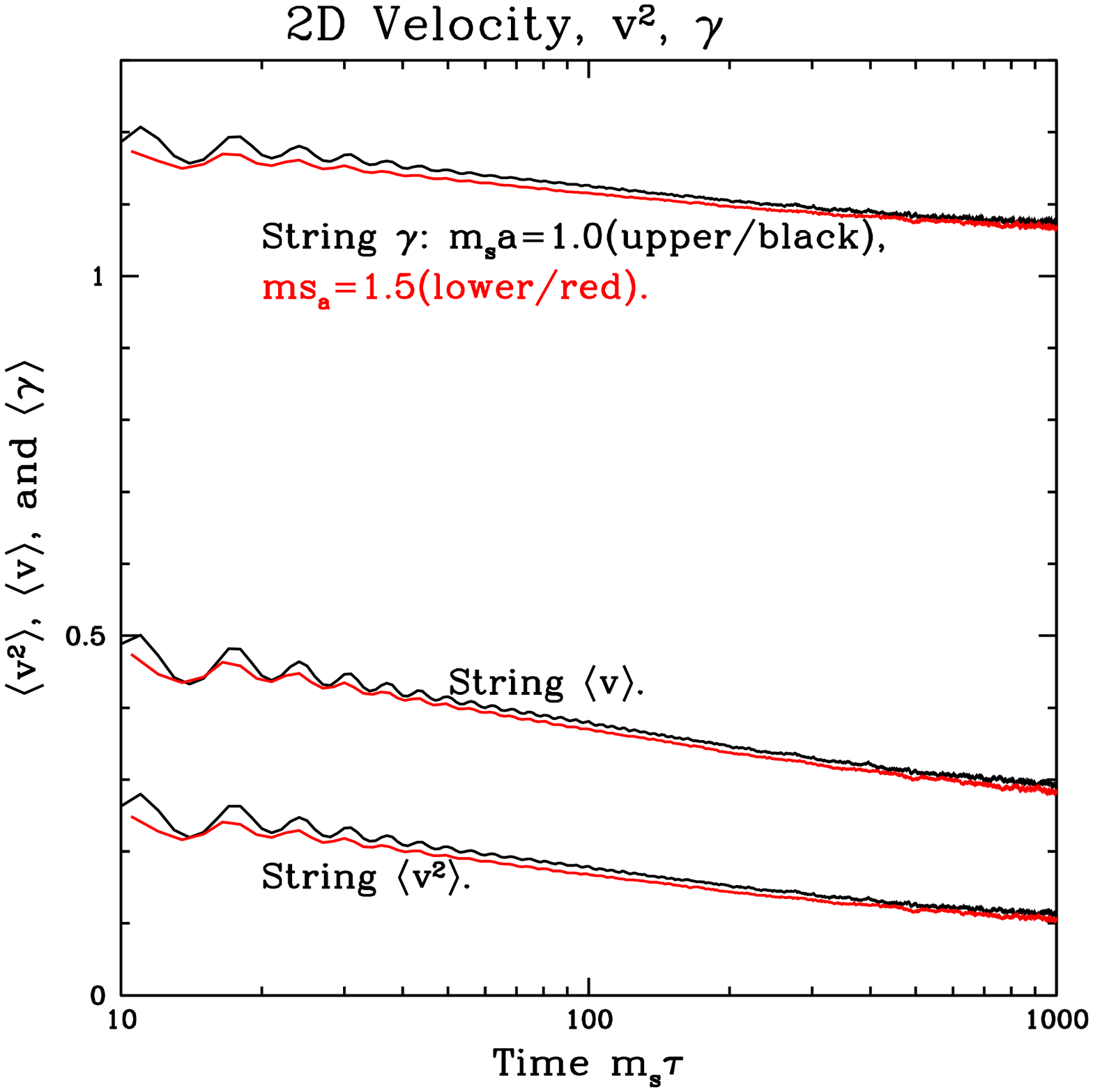}
  \caption{\label{string3}
    String mean velocity, $v^2$, and $\gamma$-factor in 3D (left) and
    2D (right), varying with the log of conformal time.}
\end{figure}

We also plot the values of
$\langle v\rangle$, $\langle v^2\rangle$, and $\langle \gamma \rangle$
in Fig.~\ref{string3}.  The string velocity falls off at large
$\tau m_s$ in 2D as we expected, but in 3D it roughly approaches a
constant, which is quite close to the value from
Fig.~\ref{martinsshellard}.  The figure shows that, while $m_s a=1.5$ was
a sufficient value to obtain continuum-limit string densities (Figure
\ref{string1} and \ref{fig:ms}), our algorithm for finding the string
velocity (see Subsection \ref{app:stringfind}) is more sensitive to
the size of the string core, and a high-quality determination of the
string velocity and especially of its $\gamma$-factor by this method
requires $m_s a \leq 1$ in 3D.  We also see oscillations in string
velocity at small $m_s \tau$; we believe that these are string-core
``breathing'' modes induced by our initial conditions, which may fool
our string-velocity method, since it assumes that the string core
takes its unperturbed form.

This section's results show that the string density in 2D evolves
exactly as we would expect:  the string density grows with
$\kappa = \ln(\tau m_s)$, while the string $\langle v^2 \rangle$ falls
as $\kappa^{-1}$.  In 3D the string network density grows in a way
which, for $\kappa$ values we have available, is roughly consistent
with linear behavior.  The string velocity in 3D is a very weak
function of $\kappa$ and is near the value predicted by the one-scale
model \cite{Martins:2000cs}.

\section{String-wall network evolution}

\subsection{Scaling expectations}

Consider the axion field in conformal coordinates.  As we discussed,
the axion mass grows strongly as temperature drops, and therefore as
conformal time progresses.  We will designate a special time $\tau_0$,
defined as when
\be
\label{def:tau0}
m_a(\tau_0) \tau_0 = 1 \qquad
\Big(\mbox{in terms of time,} \;
m_a(t_0) = H(t_0)\Big).
\ee
Roughly speaking, this is when the explicit symmetry breaking and
axion mass start becoming physically significant.  We will
parameterize the topological susceptibility near this time as
$\chi(T) \propto T^{-n}$.  In our numerical work we will take $n=7$.

When $m_a(\tau) \tau \sim \pi$ the dynamics are complex and must be
solved numerically.  But for $n=7$,
by $\tau=3\tau_0$ we have $m_a(\tau) \tau \sim 420$, which should
provide enough field oscillations to force the string-wall network
evolution to completion.  By this time $(dm_a/d\tau)/m_a^2 \ll 1$,
ensuring adiabatic evolution of the axion field.  In our
comoving and conformal coordinates, one then expects adiabatic
evolution of the form $\varepsilon \propto m_af_a^2 /\tau^2$ and%
\footnote{%
  The power $\tau^{-2}$ may look strange.  In the radiation epoch the
  conformal time is proportional to the expansion factor, so one would
  expect $\nax\propto \tau^{-3}$.  But the relation between time and
  conformal time means that a fixed particle number or energy appears
  to be growing as $\tau^2$ in conformal time. Also, we include a factor
  of the conformal-time $m_a$ in our scaling relation; and because of
  the time-scaling, a fixed $m_a$ in regular time is $m_a\propto \tau$
  in conformal time.  Together these effects provide
  $\tau^{-3+2-1}=\tau^{-2}$.}
$\nax \simeq \varepsilon/m_a \propto f_a^2/\tau^2$.
On dimensional grounds, in the radiation era and before the QCD
transition when the number of radiation degrees of freedom changes,
the axion number should be
\be
\label{nax-scaling}
\nax =
\frac{N_{\mathrm{ax}}}{V} = \frac{K \tau_0 f_a^2}{\tau^2} \,,
\ee
with $K$ a constant.  This constant determines the produced density of
axions; since little entropy is produced between GeV temperatures and
today, the axion-to-entropy ratio in the modern Universe is
\be
\label{naxK}
\frac{\nax}{s} = \frac{\nax(T=T_0)}{s(T_0)} = \frac{K H(T_0) f_a^2}
  {\frac{2\pi^2 g_*}{45} T_0^3},
\ee
where $T_0$ is the temperature when $\tau=\tau_0$ and $g_*$ is the
effective number of light degrees of freedom at $T=T_0$.  Therefore,
determining $K$ is determining the key input for the axion abundance
in the modern universe.

Our goal is to determine this constant by
evolving the axionic field, starting with a random space-varying phase
at very early time, until the string network is gone and the axion
fluctuations are small, and then evaluating
\be
\label{defK}
K = \frac{\tau^2}{\tau_0} \int \frac{d^3 k}{(2\pi)^3}
\frac{\varepsilon_k}{E_k} =
\int \frac{d^3 k}{(2\pi)^3} \left(
\frac{\sqrt{k^2+m_a^2}}{2} \langle \theta^2_a(k)\rangle
+ \frac{1}{2\sqrt{k^2+m_a^2}} \langle (\partial_t \theta_a)^2 \rangle
\right) \,,
\ee
which should be independent of the final time $\tau$ when we evaluate
it, if that time comes after adiabatic behavior sets in.
\Eq{defK} can be evaluated by fast Fourier transform (FFT) methods, or by the method of Appendix
\ref{app:B} when the FFT is not available.

As a baseline for the expected value of $K$, we can consider the
angle-average of the misalignment mechanism; we evolve a spatially
uniform initial condition for $\theta_a$, leading to
\be
\label{EOM}
\partial_\tau^2 \theta_a = -\frac{2}{\tau} \partial_\tau \theta_a
 - m_a^2 \sin(\theta_a)
\ee
with $m_a^2 = \tau^{n+2}/\tau_0^{n+4}$, and then take the average of the
resulting axion number over values of the starting angle $\theta_a$.
For our choice $n=7$ we find $K = 16.0255$.

\subsection{String-wall network evolution in 2D}
\label{sec:evol}

We have evolved the lattice axion equations of motion for both 2D and
3D systems, using the coarsest lattice that gives continuum-like
string behavior, $m_s a=1.5$ (see App.~\ref{app:test}).  We consider a
number of values for the dimensionless ratio $m_s \tau_0$.  Depending
on one's perspective, a large value for this ratio is either a large value of
$m_s$ or a large value for the time $\tau_0$; we prefer the former,
and will express all other time and frequency scales in terms of
$\tau_0$.  We make $m_a^2 \propto \tau^{n+2}$ throughout the
simulation,  that is, we assume that the QCD scale (where $m_a^2$
stops varying) is reached after the simulation ends.  We terminate
each simulation when $m_a^2 a^2 = 0.1$, so the axion mass is coming on
order the lattice spacing. At this time, we evaluate the axion-number content
of the field and extract $K$.

\begin{figure}
  \epsfxsize=0.49\textwidth\epsfbox{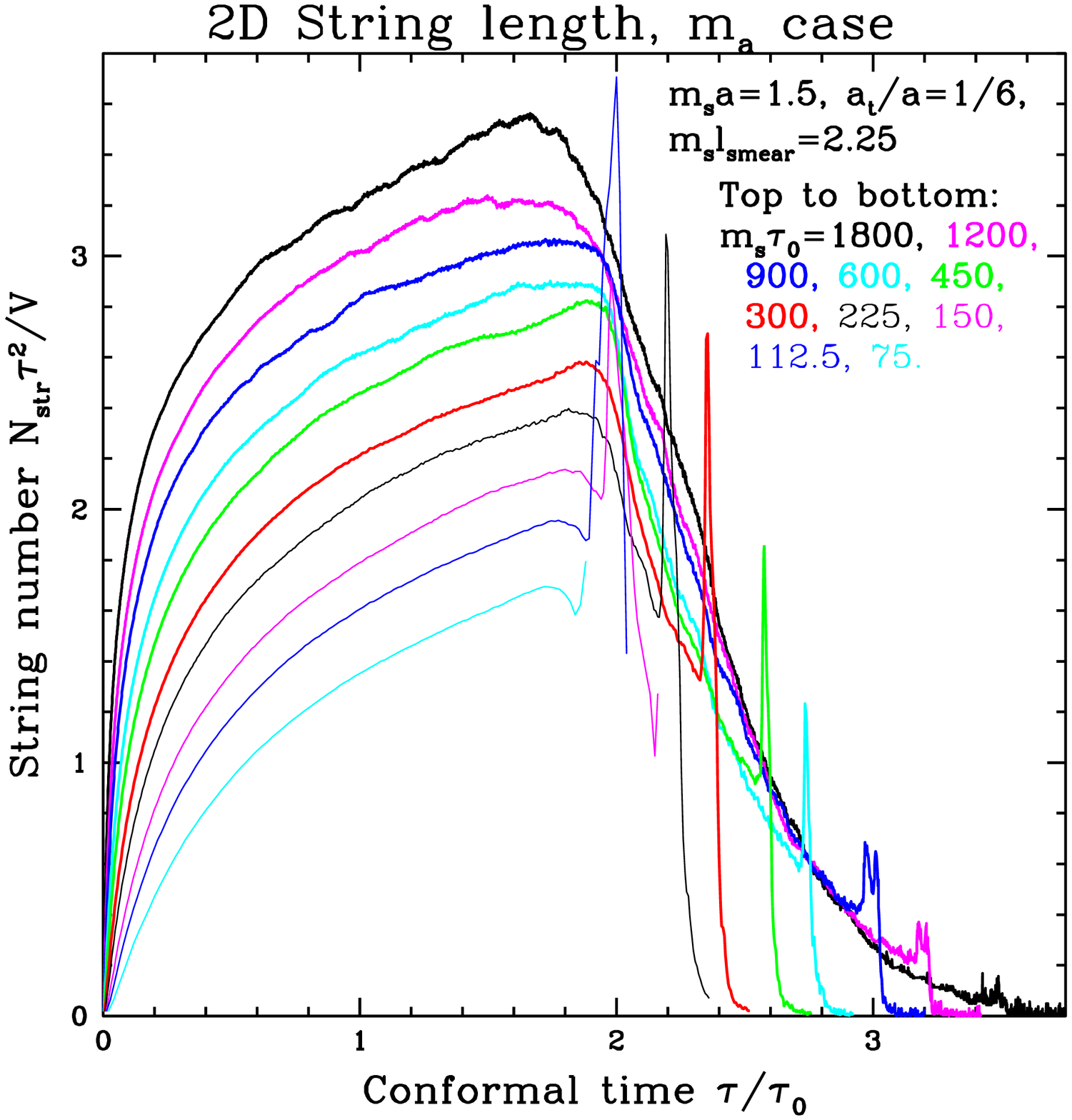}
  \hfill
  \epsfxsize=0.49\textwidth\epsfbox{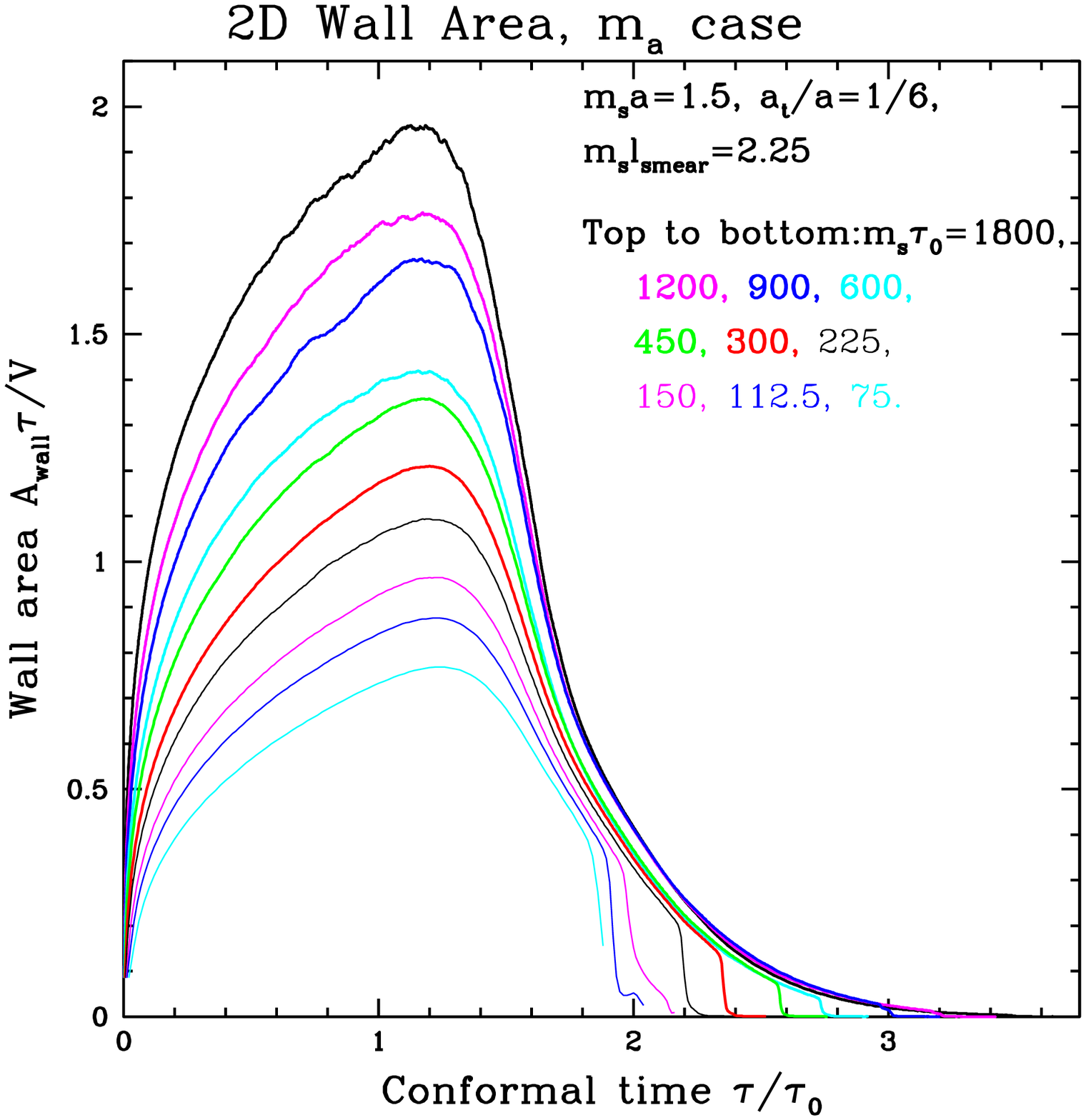}
  \epsfxsize=0.49\textwidth\epsfbox{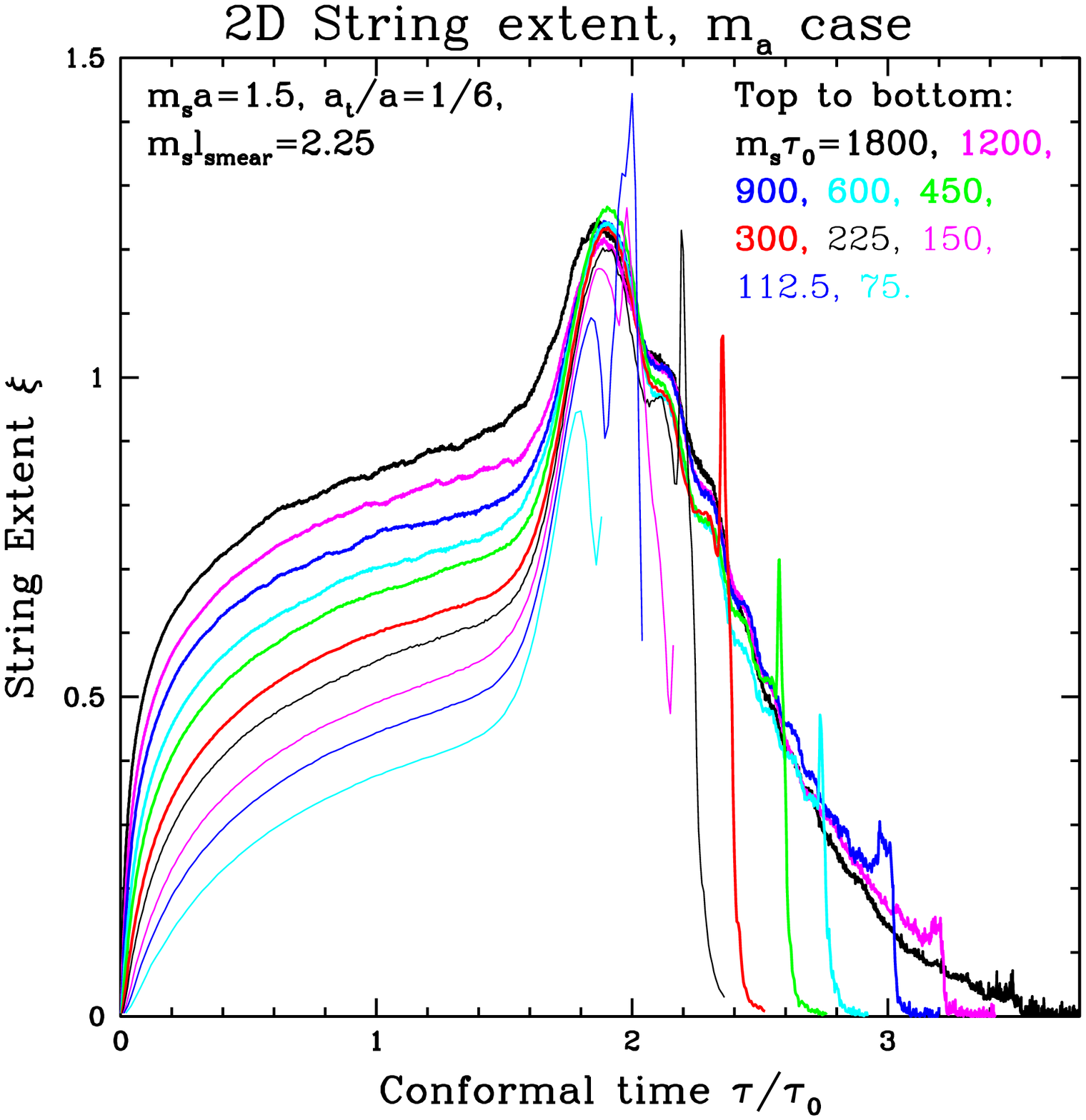}
  \hfill
  \epsfxsize=0.49\textwidth\epsfbox{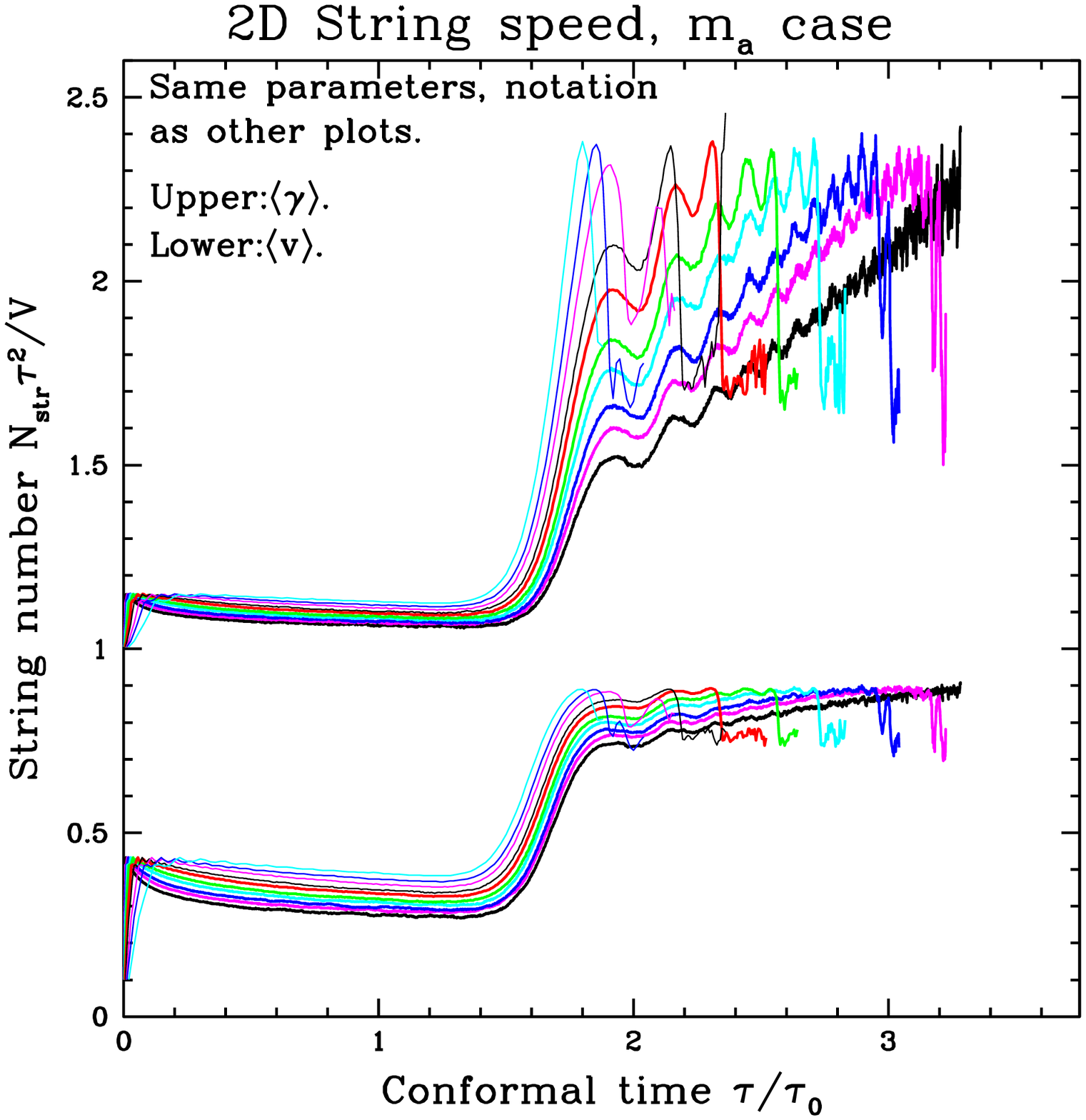}
  \caption{\label{fig:string2D1}
    String and wall density in 2D simulations, for a number of choices
    of $m_s \tau_0$.  Top left:  string number.  Top right:  wall
    length.  Bottom left:  string $\xi$-factor.  Bottom right:
    velocity, $v^2$, and $\gamma$ factor of strings.}
  \end{figure}

We begin with 2D simulations because they are numerically cheap, so we
can achieve quite large values of $m_s \tau_0$.
We look first at how the string and wall densities vary with time.
Our results, scaling out the expected $N_{\textrm{str}}\propto \tau^{-2}$,
$A_{\textrm{wall}} \propto \tau^{-1}$ behavior, appear in
Fig.~\ref{fig:string2D1}.  Much of the figure is as expected.  The
scaled string and wall densities rise with $\tau$ below about
$\tau=1.4\tau_0$, due to logarithmic scaling corrections we have
already discussed.  They are also larger for larger values of
$m_s \tau_0$, also as a result of logarithmic corrections to scaling.
The wall length starts to fall at about $\tau=1.2\tau_0$, and
the strings begin to move faster around $\tau=1.5 \tau_0$, leading to
a brief peak in $\xi$ as the strings' $\gamma$-factor rises and then
a fall in the string length and extent starting around
$\tau=1.9\tau_0$. Eventually, the extent of both strings and walls
falls near zero.  For larger values of $m_s\tau_0$, the strings are
heavier by a factor of $\ln(m_s\tau_0)$, and so they respond to the
walls with more inertia.  This delays slightly the growth in string
velocity and extent, and slightly delays the fall of the wall and
string extent.  The $\gamma$-factors we find at late times, while
large, are actually underestimates, as the $\gamma$-factor is
especially sensitive to the lattice spacing, as we explain in Appendix
\ref{app:stringfind}.

But the figure shows something that might at first be unexpected.
For each value of $m_s \tau_0$, there is a point where the wall area
abruptly crashes.  At the same value, the string number briefly
spikes, and then crashes as well.  The larger the value of
$m_s\tau_0$, the larger the $\tau_0$ value where this occurs.  In
every case it occurs when $m_a^2/m_s^2 = 1/(43\pm 3)$.

This collapse of the wall network is a numerical artifact which we
will explain in Subsection \ref{domaininstab}.  It occurs because the
ratio $m_a^2/m_s^2$ is finite in these simulations, and when it
becomes large enough, about $m_a^2/m_s^2=1/39$, the walls cease to be
even metastable and become absolutely unstable.  This leads to an
abrupt unraveling or collapse of the network.  But this collapse has
nothing to do with the physics that would occur in any circumstance where
$m_a^2/m_s^2$ remains large.  We should question the physical
relevance of any simulation where this occurs before the string-wall
network has largely broken up via the expected physics.
We see that this constrains us to consider
quite large values of $m_s \tau_0$.

\begin{figure}[htb]
  \centerline{\epsfxsize=0.6\textwidth\epsfbox{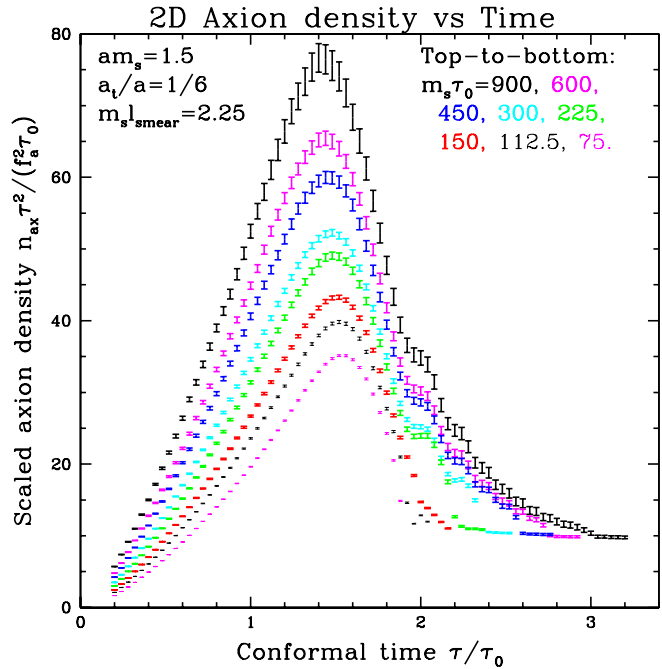}}
  \caption{\label{axtime}
    Axion number from \Eq{defK} versus time for simulations with
    $m_s\tau_0$ between 75 and 900.}
\end{figure}

To see this a little better, we plot the axion number determined via
\Eq{defK} (full details in Appendix \ref{app:B}) for all but the two
largest $m_s \tau_0$ values in Fig.~\ref{axtime}.
The figure clearly shows that larger $m_s\tau_0$ values, meaning
denser string networks, lead to larger axion number at intermediate
times.  As the string network breaks up, the axion number falls.  When
the network abruptly unravels, the axion number drops sharply, and then
goes flat after the network is gone.  Except for the smallest
$m_s\tau_0$, the final value is almost unchanged.  Still, the figure
leaves us distrustful of overinterpreting simulations where the string
network crashes rather than breaking up via natural string-wall
network dynamics.  Roughly, the $m_s \tau_0=900,$ 600, and 450
simulations look reasonable, but the abrupt drop is rather large for
the smaller-$m_s\tau_0$ simulations.

\begin{figure}
  \epsfxsize=0.48\textwidth\epsfbox{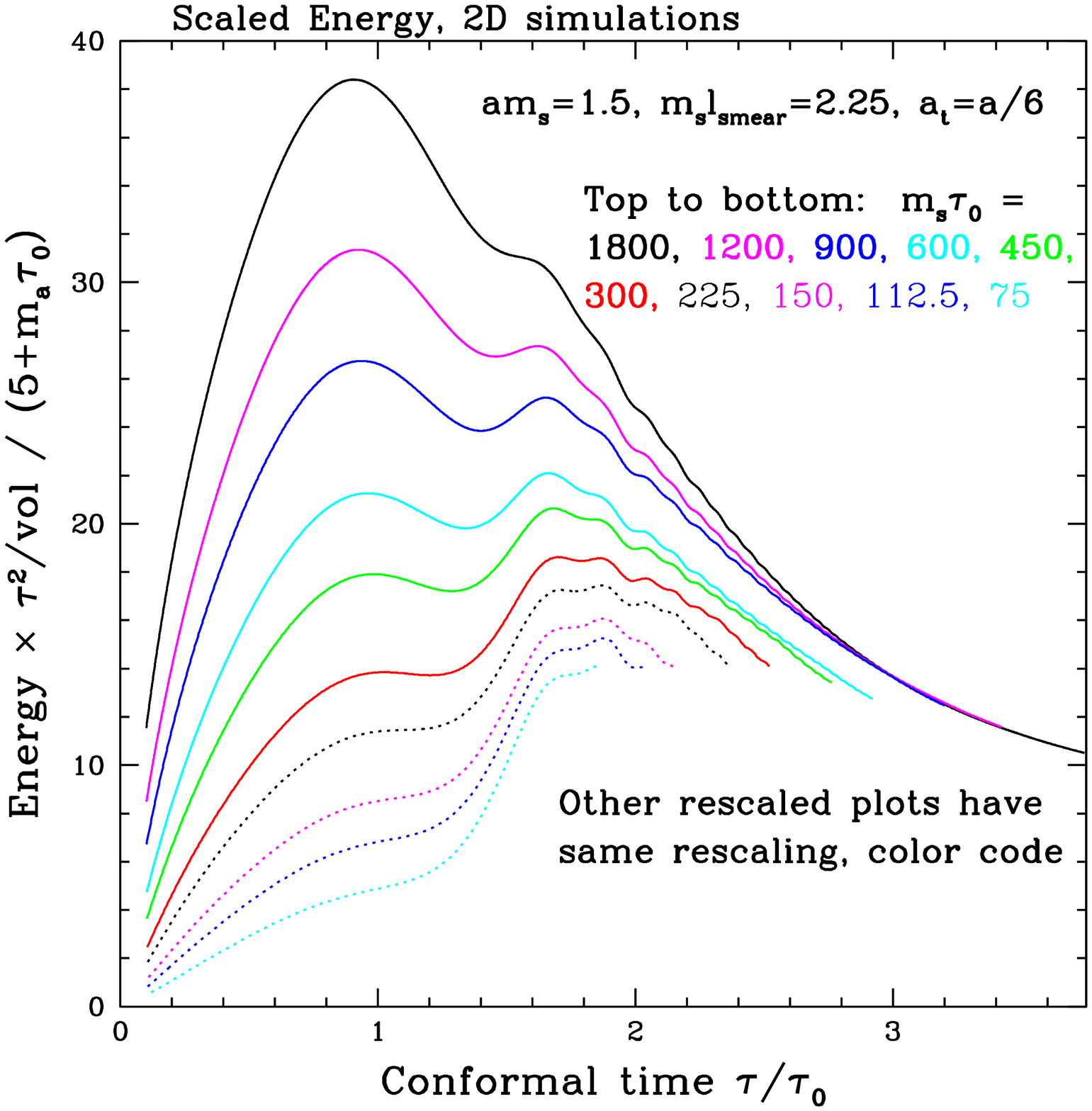}
  \hfill
  \epsfxsize=0.48\textwidth\epsfbox{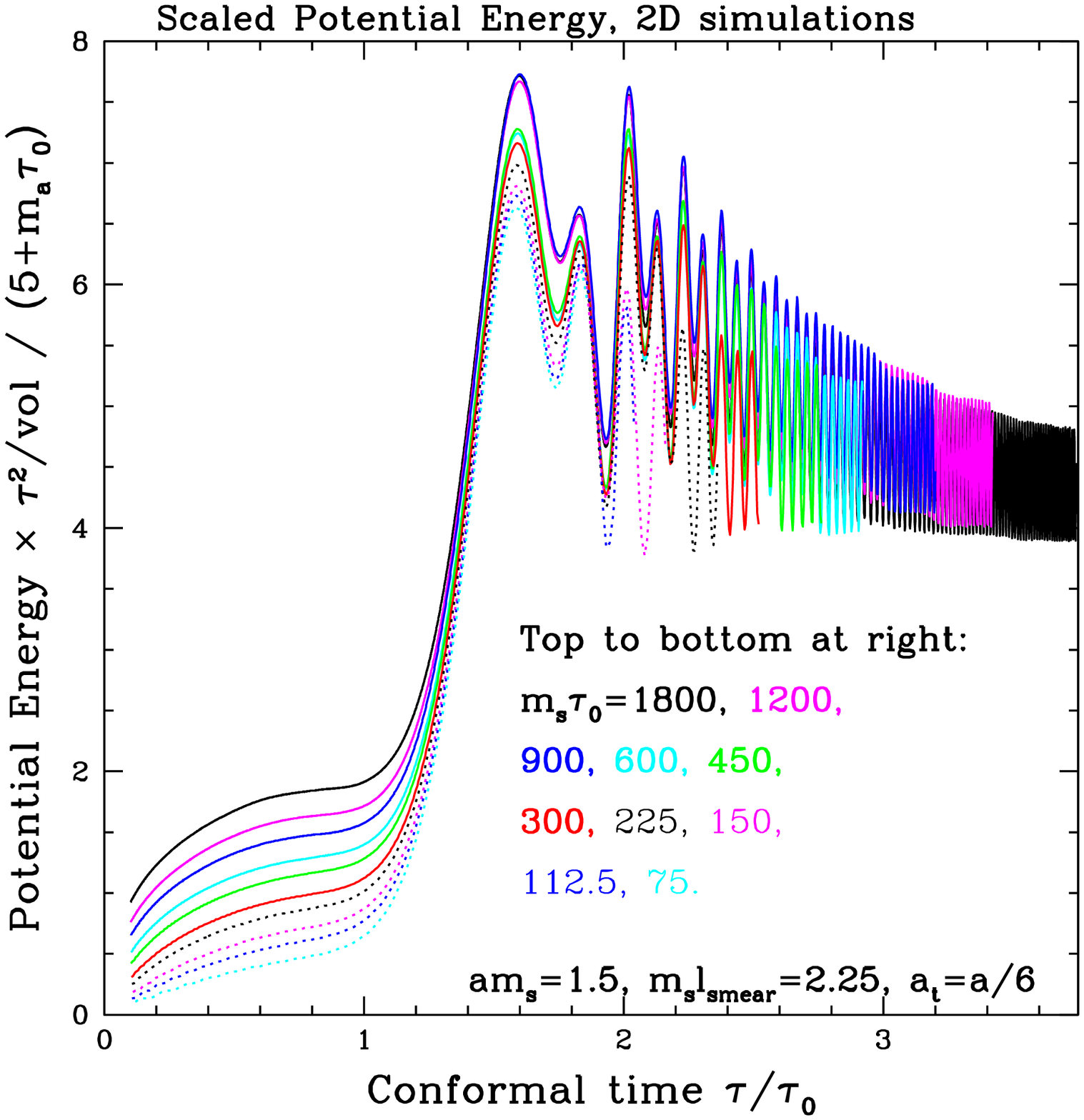}
  \epsfxsize=0.48\textwidth\epsfbox{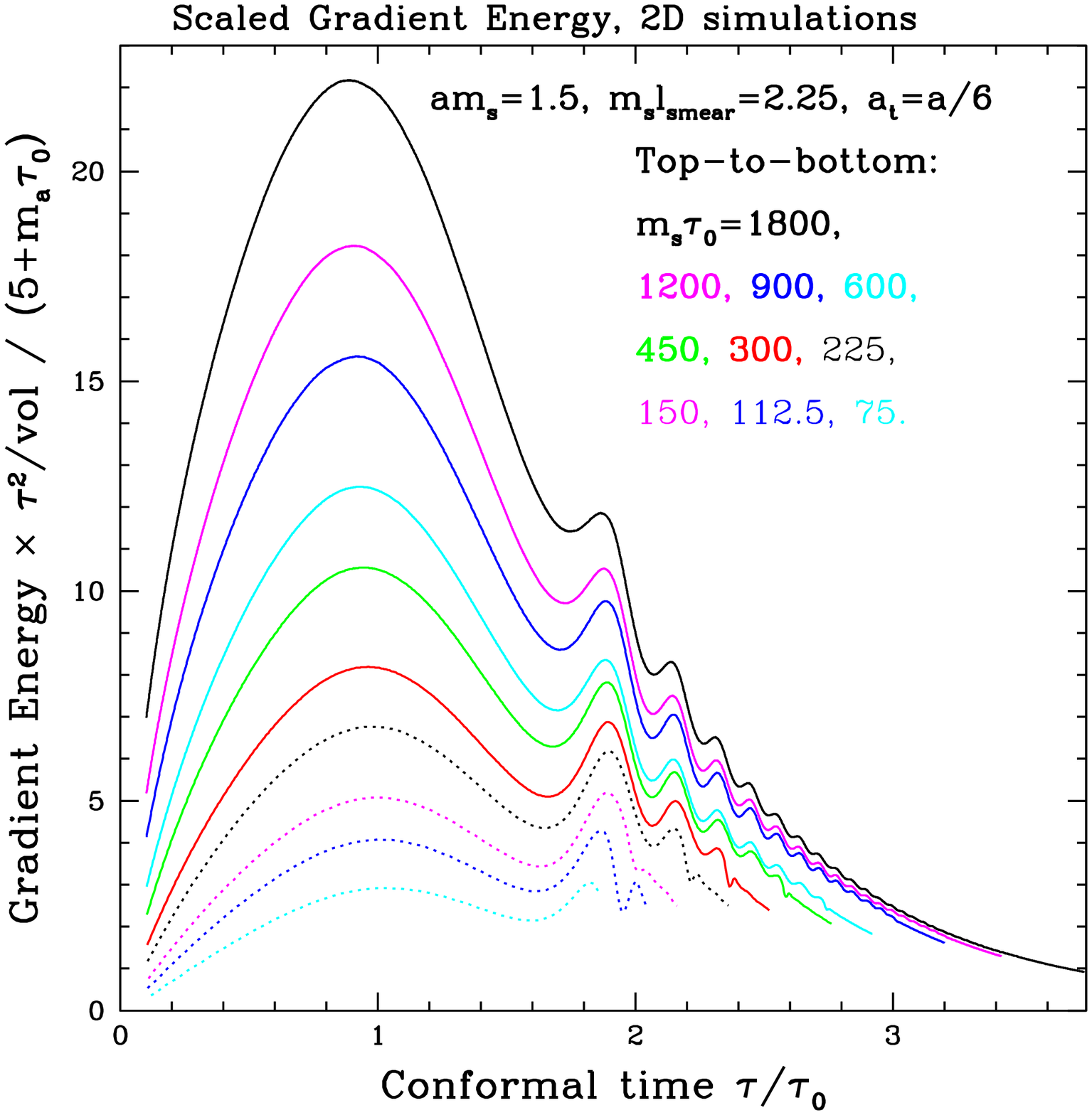}
  \hfill
  \epsfxsize=0.48\textwidth\epsfbox{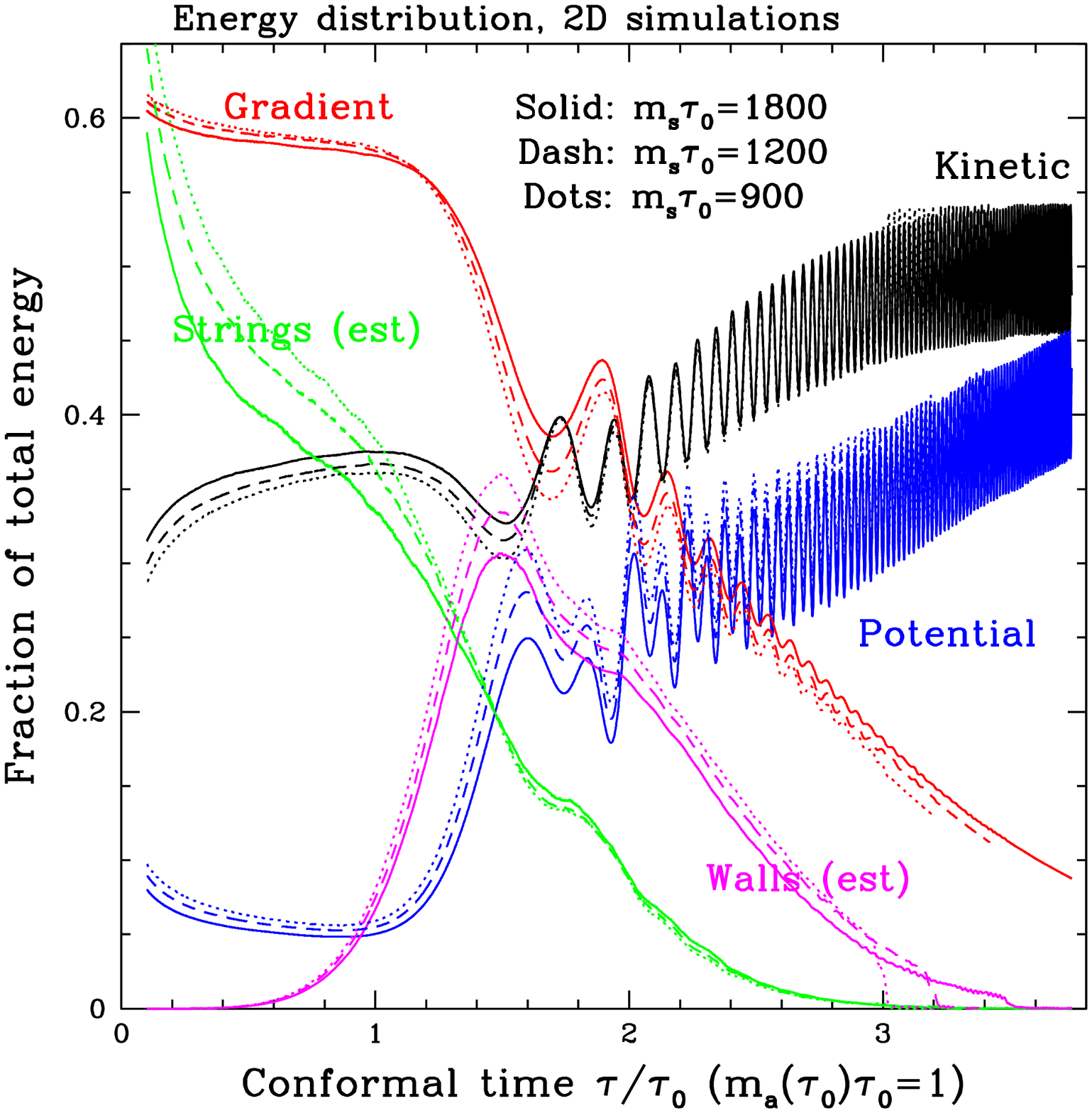}
  \caption{\label{fig:energy2D}
    Energy density, scaled by $\tau^2/(f_a^2(5+m_a\tau))$, for 2D
    simulations.  Top left:  total energy.  Top right:  potential.
    Bottom left:  gradient energy.   Bottom right:  energy fractions
    and estimated energy in strings and walls.}
\end{figure}

As a final way of looking at the network, we plot the energy content
as a function of time in Fig.~\ref{fig:energy2D}.  The energy content
naturally scales as $f_a^2/\tau^2$ on dimensional grounds and due to
Hubble expansion, so we have factored this out.  Also, at late times
the rising mass causes the energy to rise, $\varepsilon \propto m_a$;
but this behavior only sets in once the network starts responding to
$m_a$, at about $\tau=1.4\tau_0$.  Therefore we have also scaled out a
factor of $(5 + m_a\tau_0)$, with the $5$ chosen so the scaling will
turn on at about $\tau=1.4\tau_0$.  To avoid crowding the plots we
have not shown error bars. The largest $m_s\tau_0$ curves have
statistical errors around $2\%$; the smaller $m_s\tau_0$ have smaller
statistical errors.  In the energy-fractions plot we have shown only
the three largest $m_s \tau_0$ values.  The string energy is estimated
as $\gamma L_{\mathrm{str}} \pi f_a^2 \ln(m_s r)$ with
$r^{-2} = m_a^2 + \tau^{-2}$, so the IR cutoff is either the inverse string
separation or the axion mass, whichever is larger.  The domain wall
energy is estimated from domain wall area, without an attempt to
include the $\gamma$-factor for the walls' motion, thus the wall
energy is probably an underestimate. The string energy is an
underestimate after $\gamma$ gets large around $\tau=1.8\tau_0$, since
our method underestimates large string velocities when using such
large $m_s a$ values.

The figure shows that the energy starts out strongly $m_s\tau_0$
dependent, presumably because of the different density and string
tension of the string network.  By the end of the simulation this
$m_s\tau_0$ dependence has largely disappeared.  Also, the early
behavior is dominated by gradient energy, since strings at rest carry
almost all their energy in gradients (moving strings have a kinetic
energy fraction of $v^2/2$).  The late behavior is oscillating axions,
with energy equipartitioned between kinetic and (potential+gradient).
Quadratic fluctuations have $1/2$ their energy as kinetic, whereas
walls and strings have most energy in gradients and potential; so the
extent that the (phase-averaged) kinetic energy is below 1/2 is a
reasonable estimate of how much energy is still in strings an walls.
Once the energy is mostly in quadratic fluctuations (particles), there
is still a shift between gradient and potential energy, due to the
increase in $m_a$.  The oscillations at late time show that the produced
axions are not a phase-random collection, but have some phase
coherence.

We should worry about the impact of the abrupt collapse of
the string-wall network if the strings and walls still carry
significant energy fraction when it happens.  The bottom right figure
gives a nice criterion for knowing if this happens.
In each case shown in the plot, less than $5\%$ of energy in walls
when they collapse.  This is not true for smaller $m_s\tau_0$ values,
so the associated simulations cannot be completely trusted.

\subsection{Domain-wall instability}
\label{domaininstab}

The most striking feature of the string-wall network simulations we
just presented is the sudden collapse in the amount of domain wall,
accompanied by a brief spike, and then collapse, in the
amount of string.  The time value $\tau/\tau_0$ when this occurs
changes as we vary $m_s \tau_0$.  But it always occurs at the same
value of $m_a^2/m_s^2 \simeq 1/43$.  Here we show that this is not an
accident; it happens because the domain walls are only metastable, and
at $m_a^2/m_s^2=1/39$ they become completely unstable.

A domain wall is where the phase $\theta_a$ changes from 0 to $2\pi$
by going around the ``valley'' of the winebottle potential.
Since $\theta_a$ is a single real parameter, this occurs on a
codimension-1 surface.  The thickness of the region where $\theta_a$
is far from 0 or $2\pi$ is set by a competition between potential
energy, which wants to make the region thin to keep the region with
potential energy costs small, and gradient energy, which wants to make
the region thick to minimize $\int |\nabla \varphi|^2$.
The thickness of the region is $\sim 1/m_a$; when
$m_a\tau \gg 1$, this is thin compared to the horizon scale and the
walls can be approximated as planar.

In the case $m_a^2 \ll m_s^2$, we can take
$\sqrt{2}\,\varphi = f_a e^{i\theta_a}$ to good approximation.
Choosing the $\theta_a$-variation along the $z$ axis,
the energy to minimize is
\bea
\label{Ewall}
E_{\textrm{wall}} &=& \int dx\, dy \int_{-\infty}^\infty dz \left(
\frac{f_a^2}{2} (\partial_z \theta_a)^2
+ \chi ( 1 - \cos\theta_a) \right)
\\  \nn
\sigma = 
\frac{E}{A} & = & f_a^2 \int_{-\infty}^\infty dz \left(
\half (\partial_z \theta_a)^2 + m_a^2 ( 1 - \cos\theta_a ) \right) \,,
\eea
with boundary conditions $\theta_a \to_{z\to -\infty} 0$,
$\theta_a \to_{z\to \infty} 2\pi$.
Here $A$ is the area of wall we consider, and $\sigma=E/A$ is the
surface tension.  Extremization gives
\be
\label{wallEOM1}
\partial_z^2 \theta_a = m_a^2 \sin \theta_a
 \equiv \partial_{\theta_a} \bar V \,,
\ee
with $\bar V = m_a^2(1-\cos\theta_a) = V/f_a^2$.  This can be solved
by multiplying both sides by $\partial_z \theta_a$,
\bea
\label{cute}
\partial_z \theta_a \partial_z^2 \theta_a & = &
\partial_z \theta_a \partial_{\theta_a} \bar V(\theta_a) \,,
\\
\partial_z \frac{(\partial_z \theta_a)^2}{2} & = &
\frac{\partial \theta_a}{\partial z}
\frac{\partial \bar V}{\partial \theta_a}
= \frac{\partial \bar V}{\partial z} \,,
\nn
\eea
and integrating $\int dz$ to obtain a Virial-type relation
\be
\label{wallVirial}
\half (\partial_z \theta_a)^2 = \bar V(\theta_a)-\bar V(0)
= m_a^2 ( 1 - \cos\theta_a ) \,,
\ee
showing that potential and gradient terms each represent half the
wall's energy.  The surface tension is%
\footnote{Some references use the value 9.32 rather than 8.  This
  estimate originates from \cite{Huang:1985tt}, who find it for the domain
  wall at $T=0$, essentially by incorporating corrections to the strict
  $(1-\cos\theta_a)$ form of the potential.  Such corrections arise at
  zero temperature because instantons form a correlated liquid; but at
  high temperatures the dilute instanton gas approximation should be
  good and we expect such corrections to be tiny.}
\bea
\label{sigma1}
\frac{\sigma}{f_a^2} &=& \int_{-\infty}^{\infty} dz (\partial_z\theta_a)^2
= \int dz \frac{d\theta_a}{dz} \sqrt{2\bar V-2 \bar V_0}
\nn \\
&=& \int_0^{2\pi} d\theta_a \sqrt{2\bar V(\theta_a) - 2\bar V(0)}
= m_a \int_0^{2\pi} d\theta_a \sqrt{2-2\cos \theta_a} = 8 m_a.
\eea
Numerically, we estimate the energy in domain walls as
$8m_a f_a^2$ times the area where $\theta_a=\pi$, which we determine as
described in Appendix \ref{app:A}.

Now suppose that $m_s^2/m_a^2$ is large but not enormous.  Then
$\varphi(z)$ will not strictly lie on the circle $f_a e^{i\theta_a}$,
and \Eq{Ewall} becomes (introducing $\bar\varphi = \varphi/f_a$)
\bea
\label{Ewall2}
\frac{\sigma}{f_a^2} &=& \int dz \left( \half \left(
       [\partial_z \bar\varphi_r]^2 + [\partial_z \bar\varphi_i]^2 \right)
       + \bar{V}(\bar\varphi_r,\bar\varphi_i) \right) \,,
\\
\label{Vwall}
\bar{V}(\bar\varphi_r,\bar\varphi_i) &=&
\frac{m_s^2}{8} \Big( \bar\varphi_r^2 + \bar\varphi_i^2 - 1 \Big)^2
 + m_a^2 (1 - \bar\varphi_r) \,.
\eea
The equations of motion are
\be
\label{EOMwall}
\partial_z^2 \bar\varphi_{r,i}
 = \frac{ \partial \bar{V}}{\partial \bar\varphi_{r,i}} \,.
\ee
As $z$ is varied, the field $\bar\varphi$ will follow a curve in the
complex $\bar\varphi$ plane,
\be
\label{affine}
\bar\varphi(z) = \bar\varphi(\ell(z)) \,,
\ee
where $\ell$ is the affine parameter describing the curve that
$\bar\varphi$ follows; $|d\bar\varphi/d\ell|=1$ and
$|\partial_z \bar\varphi| = d\ell/dz$.
The complex equation of motion, \Eq{EOMwall}, can be decomposed into
the $(r,i)$ component tangent to the curve and the component normal to
the curve. The tangent EOM reads
\be
\label{Eqtangent}
\partial_z^2 \ell = \partial_{\ell} \bar{V}
\quad
\Longrightarrow
\quad
(\partial_z \ell)^2 = 2 (\bar{V}(\bar\varphi(\ell)) - \bar{V}_0) \,,
\ee
integrating into a Virial relation in the same way as before.  The
surface tension is again
\be
\label{sigma2}
\frac{\sigma}{f_a^2} = \int d\ell
\sqrt{2 \bar{V}(\bar\varphi(\ell)) - 2 \bar{V}(0)} \,.
\ee

The equation of motion in the field direction normal to the curve is
what determines the curve $\bar\varphi(\ell)$; it reads
\be
\label{eqnormal}
(\partial_z \ell)^2 \frac{\partial^2 \bar\varphi(\ell)}{\partial \ell^2}
= \frac{\partial \bar{V}(\bar\varphi)}{\partial \bar\varphi_{\hat{n}}}  \quad
\Longrightarrow \quad
\frac{\partial^2 \bar\varphi(\ell)}{\partial \ell^2}
= \frac{\partial \bar{V}(\bar\varphi)/\partial \bar\varphi_{\hat n}}
  {2 \bar{V}(\bar\varphi(\ell)) - 2 \bar{V}_0} \,.
\ee
This equation is equivalent to asking:  what curve
$\bar\varphi(\ell)$ will produce the lowest surface tension in
\Eq{sigma2}?  By choosing a path with a small value of $\bar{V}$, we
keep the integrand small; but by choosing a short path, we keep the
integration range short.  The LHS of \Eq{eqnormal},
$\partial^2 \bar\varphi(\ell)/\partial \ell^2$, is the extrinsic
curvature of the path.  The larger its value, the more we can shorten
the curve by rounding it off at this point.  This must be balanced
against how fast $\bar{V}$ will rise when rounding off the curve.
The equation tells the exact
balance between the gain of shortening the curve and the cost of
increasing $\bar{V}$ along the curve.  Now $\bar{V}-\bar{V}_0$ in the
denominator is mostly provided by $m_a^2$, while the normal derivative
in the numerator is almost purely provided by $m_s^2$.  The larger we
make $m_a^2/m_s^2$, the less there is to stop the domain wall from
rounding off its path and exploring
$\bar\varphi_r^2 + \bar\varphi_i^2 < 1$.

\begin{figure}[ht]
  {\epsfxsize=0.45\textwidth \epsfbox{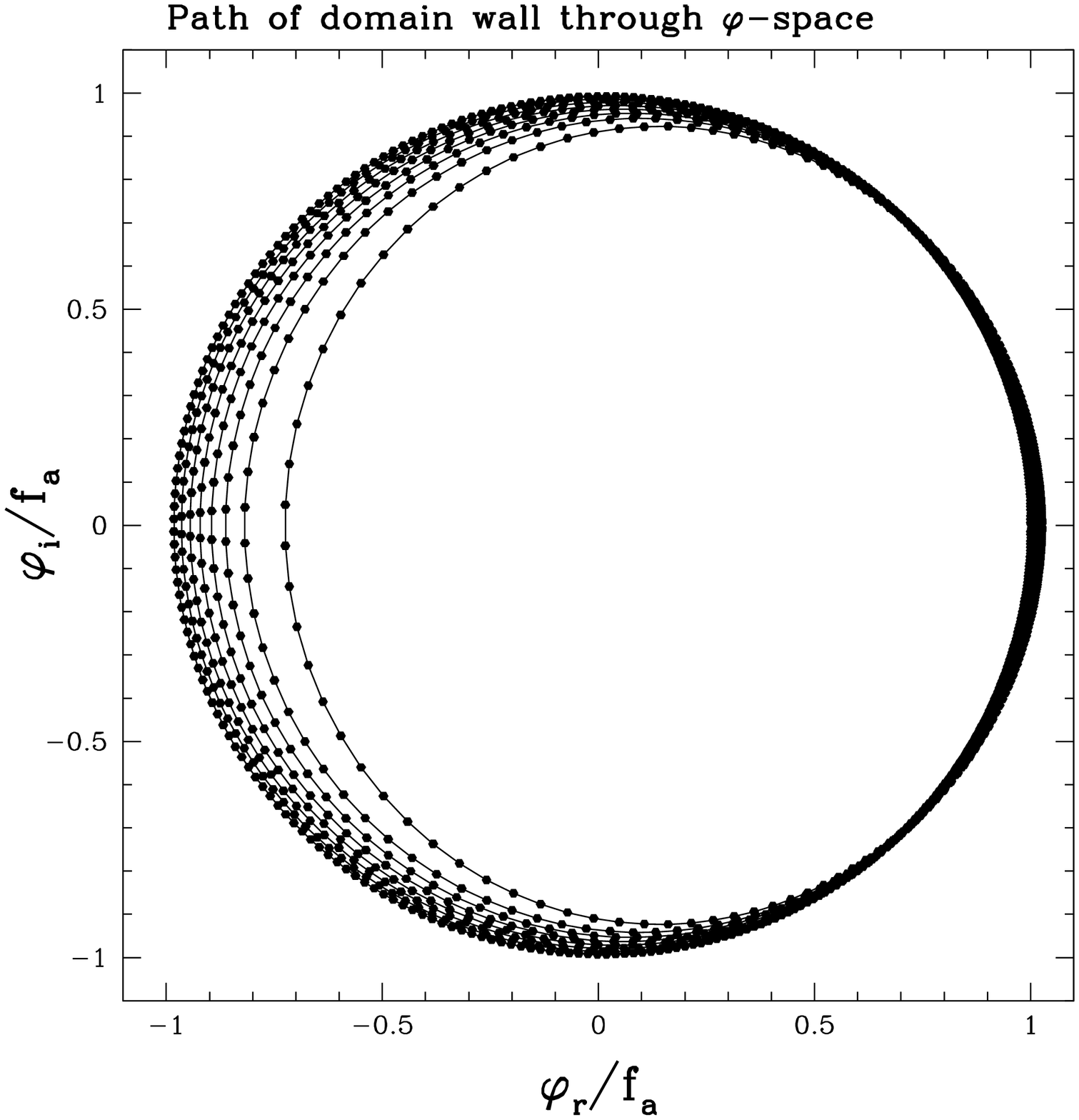}}
  \hfill
  {\epsfxsize=0.4\textwidth \epsfbox{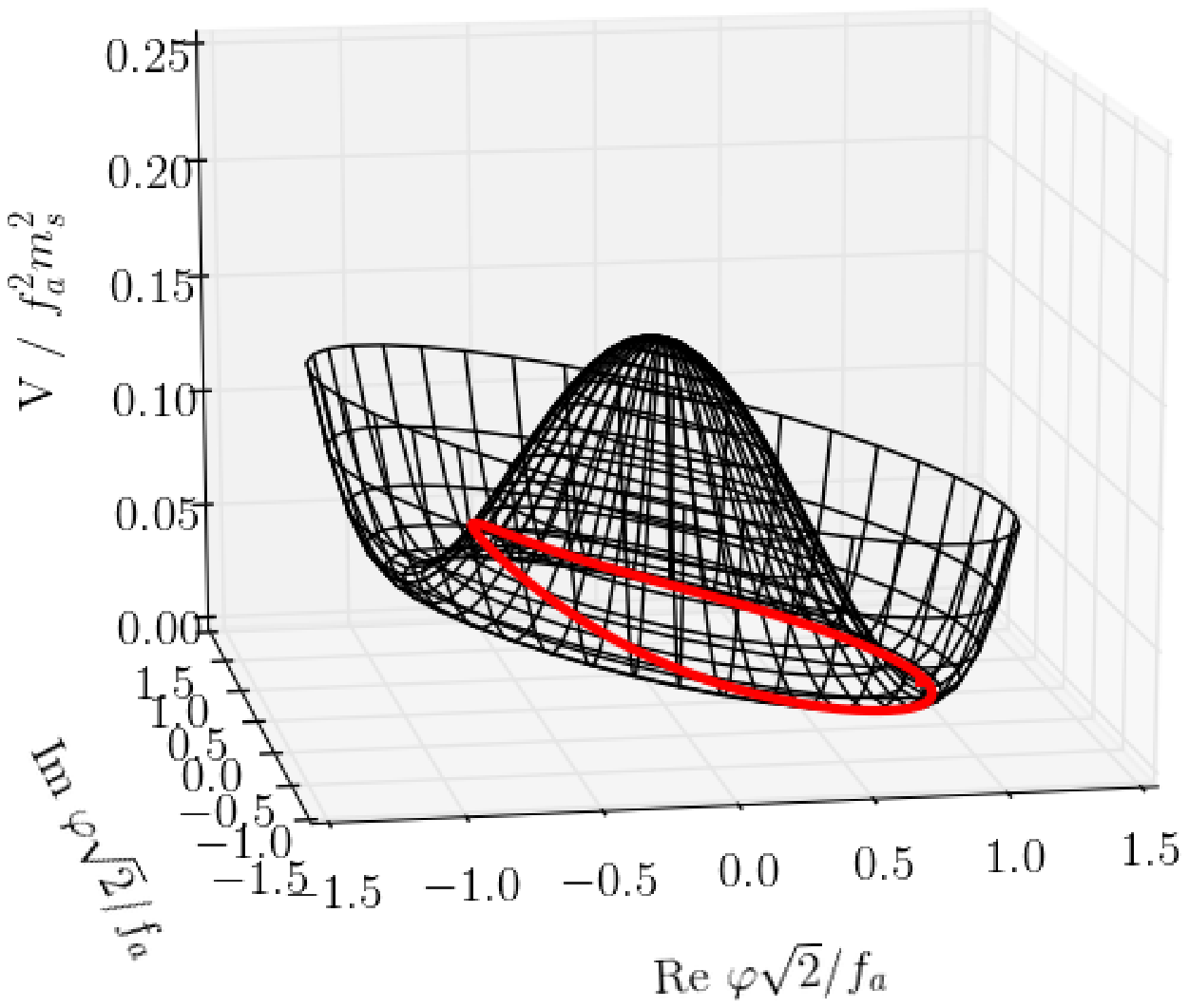}}
  \caption{Left:  domain wall solution, as seen in
    $\bar\varphi$-space, for several values of $m_a^2/m_s^2$.
    Each curve is the path $\bar\varphi$
    follows through field-space for a given $m_a^2/m_s^2$ value.
    Neighboring dots are points that are separated by $1/(4m_s)$ in
    coordinate space.  Right:  the largest-$m_a^2/m_s^2$ path as it
    appears on the $(\varphi_r,\varphi_i)$ potential, illustrating how
    the curve departs from the ``valley'' of lowest potential.
    \label{fig:wall}}
\end{figure}

We have solved explicitly for the domain wall's shape and surface
tension, using $V$ from \Eq{Vwall} with different values of
$m_a^2/m_s^2$.  We illustrate the results in Fig.~\ref{fig:wall}.  The
left plot in the figure shows several $\varphi(\ell)$ curves in the
$\varphi_r,\varphi_i$ plane; the curves shown have
$m_a^2/m_s^2$ values spaced in intervals of 0.0032.  The dots are
the values of $\bar\varphi(\ell(z))$ at a series of $z$-coordinates,
with neighboring dots on a curve separated by $\Delta z = 1/(4m_s)$.
The wall becomes thinner as the dots become more spread out.  The last
curve is the last metastable domain wall; a tiny further increase in
$m_a^2$ will cause the curve to pull over the top of the potential
peak, and the domain wall spontaneously collapses.  On the right in
the figure, we have shown the path of this domain wall in terms of the
potential $V(\varphi_r,\varphi_i)$, indicating how the curve ``pulls
up'' partway onto the bump in the potential rather than staying in the
``valley'' of the potential.
    
Our study finds that the domain walls lose their metastability
when $m_a^2/m_s^2$ reaches about $1/39$.  For larger $m_a^2/m_s^2$
values, there is no longer any domain wall solution.  In simulations we
observe the collapse of the wall area to start when the ratio is $\simeq 1/43$.
We believe that this is because fluctuations in the field,
hitting the wall, can induce a collapse when $m_a^2$ has not quite
reached the value where instability occurs.  This will not happen
everywhere at once, but only locally where larger fluctuations impact
the wall.  At these spots the wall will ``break''; a loop of string
forms the boundary of this break, temporarily raising the total
length of string in the simulation.  The hole in the wall then rapidly
grows and is joined by new breaks, leading to the collapse in the wall
area; as the holes in the wall grow and percolate, the amount of
string at first rises but then falls essentially to zero as the wall
network disappears.  This is a good description of both the timing and
behavior of the wall collapse we observe.

Physically, small values of $m_s$ are experimentally excluded.  And it
is most natural to expect $\lambda \sim 1$ so $m_s \sim f_a$, which
should be orders of magnitude larger than $m_a$.  So while this
physics is the correct dynamics of a string-wall network with
$m_a^2 / m_s^2 \sim 1/40$, it does not describe the evolution of
physical interest.  Therefore we cannot rely on the results of any
simulation in which a significant amount of energy still resides in
the string-wall network when this collapse occurs.  This criterion
places a limit on what values of $m_s \tau_0$ give reliable answers,
pushing us towards fine lattice spacings and towards the largest
values of $m_s a$ that still give continuum behavior.

\subsection{3D simulations}

Our experience with 2D simulations tells us that we must consider the
largest possible values of $m_s \tau_0$ to avoid the unphysical
collapse of the string network.  This means we need to choose the
largest $m_s a$ we can, subject to the constraint $m_s a \leq 1.5$,
found in Sec.~\ref{app:test} to ensure continuum behavior.
We should also choose the largest $a \tau_0$ value we can, subject to the
constraints that $\tau/\tau_0$ must get large enough for the network
evolution to complete, with $L \geq 2\tau$ to ensure no finite-volume errors.
Unfortunately, our limited numerical resources limit us to boxes of
$1600^3$ or smaller, constraining us to consider $a\tau_0=300$ or
$m_s\tau_0=450$ but not larger.  We have not considered $m_s\tau_0$
smaller than 225, since the 2D simulations showed that the wall
network breaks up too early in such simulations to learn anything of
value.  So we have less dynamic range than in the 2D simulations.

\begin{figure}
  \epsfxsize=0.48\textwidth\epsfbox{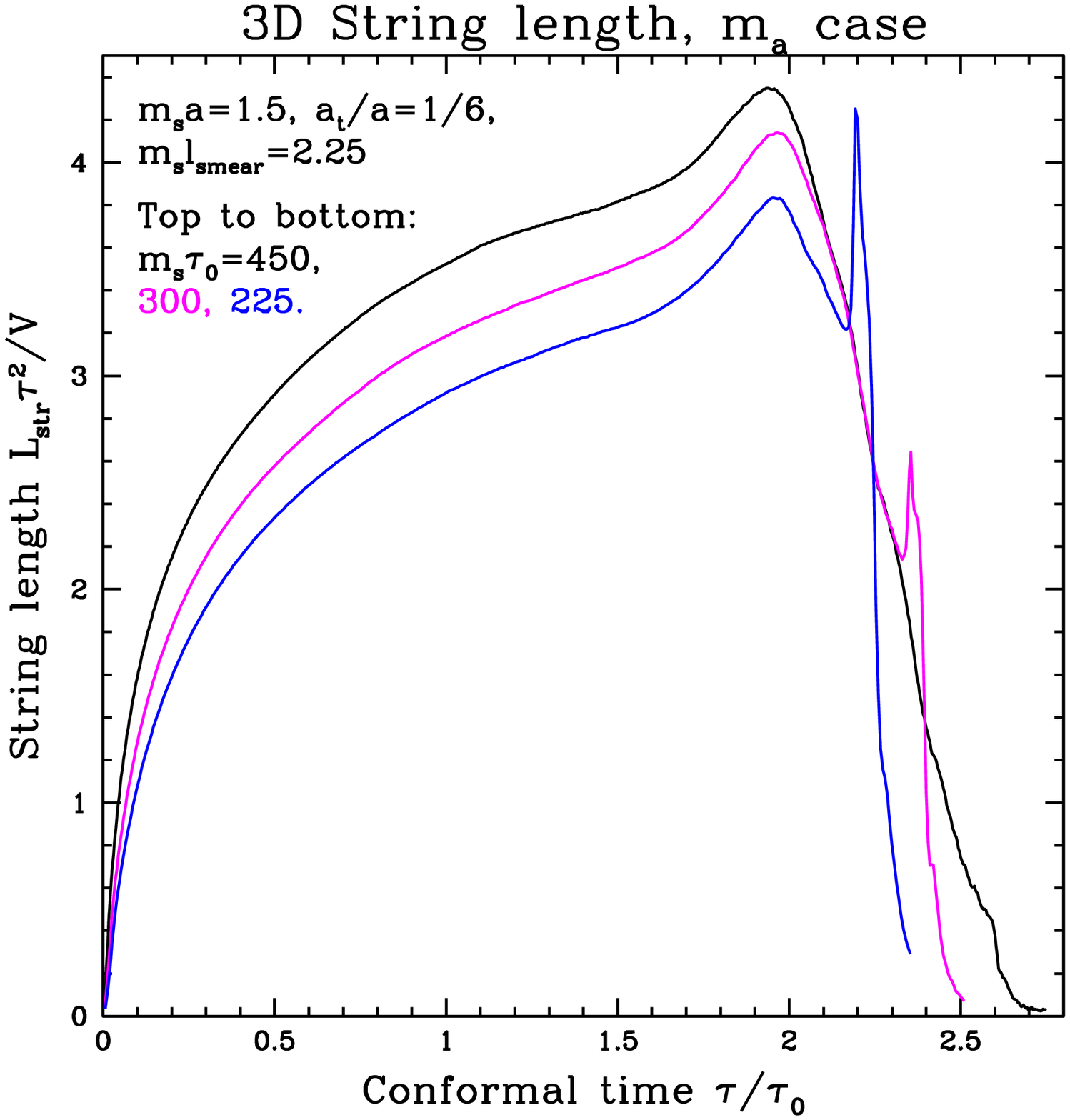}
  \hfill
  \epsfxsize=0.48\textwidth\epsfbox{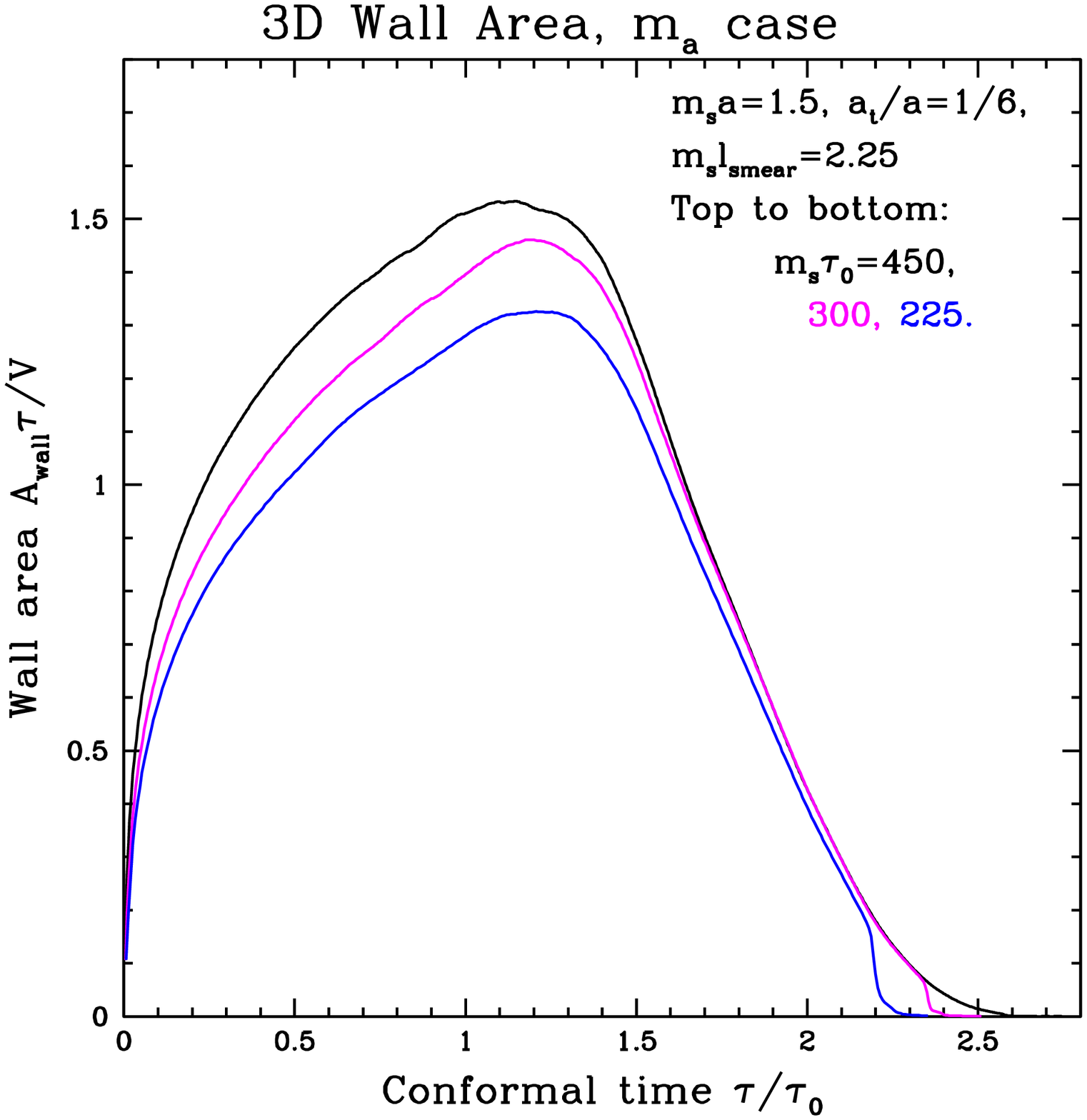}
  \epsfxsize=0.48\textwidth\epsfbox{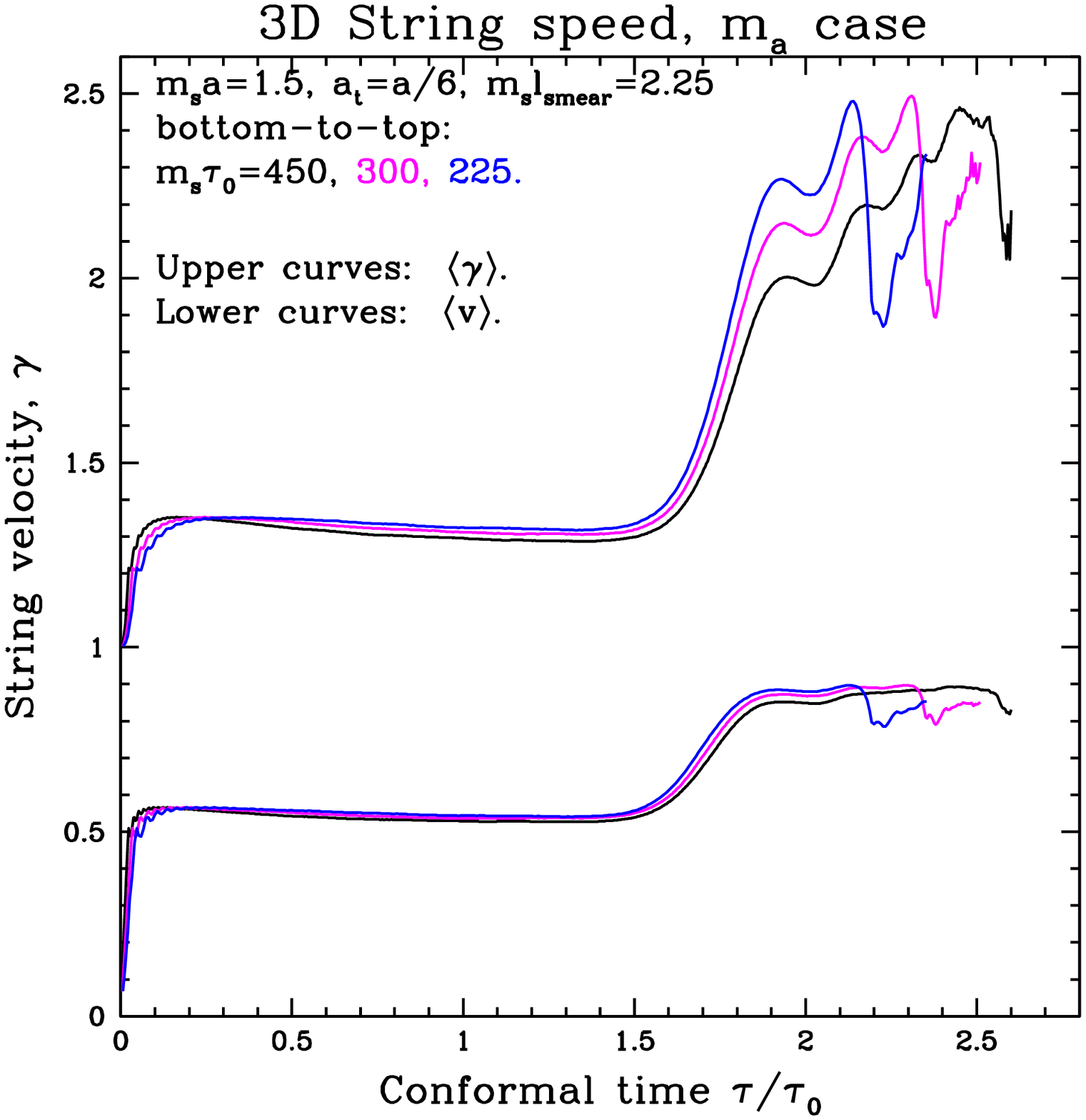}
  \hfill
  \epsfxsize=0.48\textwidth\epsfbox{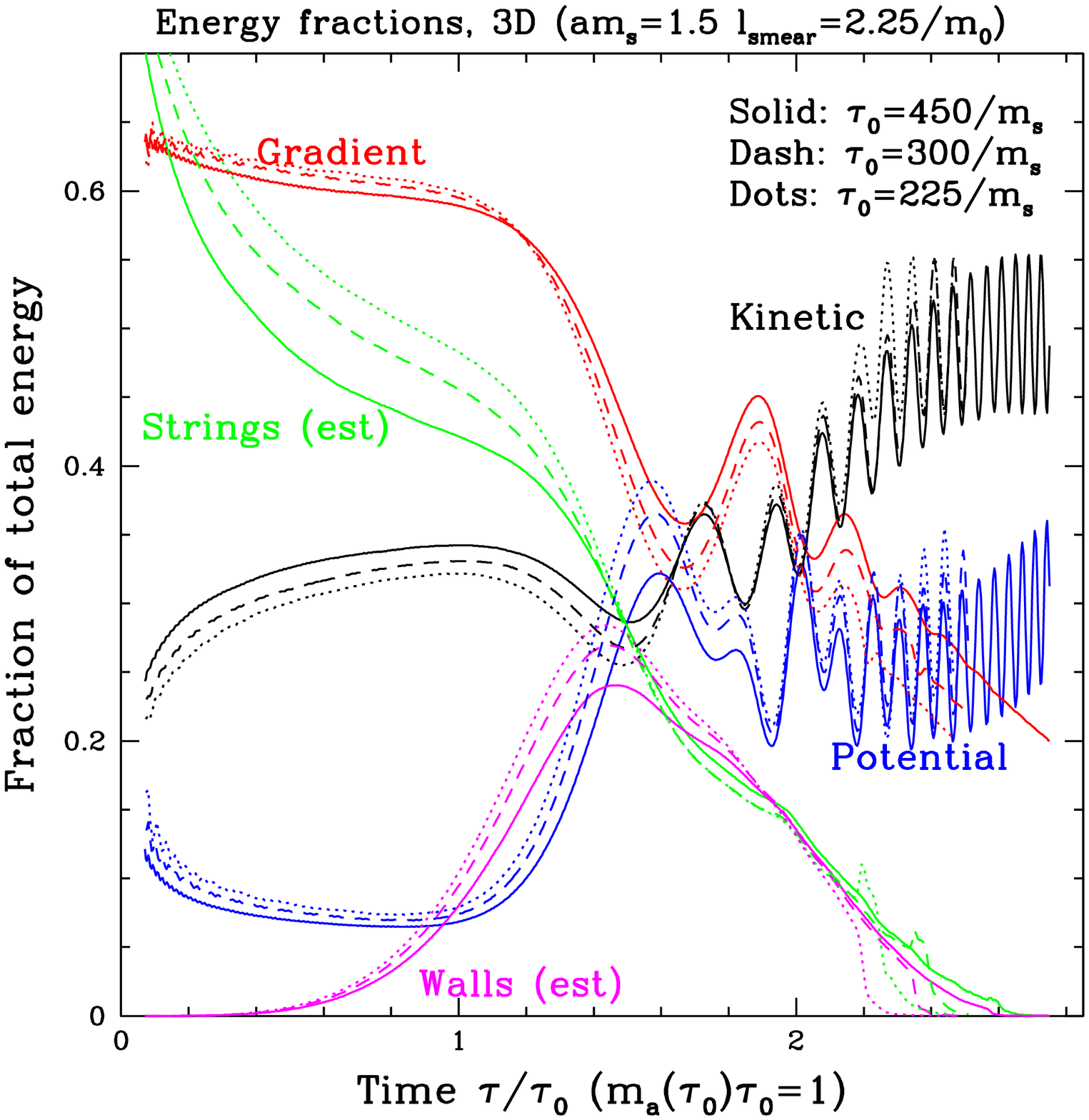}
  \caption{\label{fig3D}
    String length, wall area, string velocity, and energy ratios as a
    function of time in 3 dimensions for 3 values of $m_s \tau_0$.}
\end{figure}
    
The 3D simulations show surprisingly similar behavior to those in 2D.
Rather than repeating all plots we presented in 2D, we plot just
the string length and speed, the wall area, and the energy ratios in
Fig.~\ref{fig3D}.  The most significant difference from 2D is that,
while the strings and walls start to decay at about the same time, the
network decays more quickly, with very little string left by
$\tau=2.57\tau_0$, when the network collapse occurs for
$m_s\tau_0=450$.  In particular, the collapse of the walls, in the
plot of wall area and of energy ratios, is almost invisible for this
largest $m_s \tau_0$ value.  Therefore, while this final simulation
does not provide the right string tension due to the missing core
tension, it at least presents a case where the
network breaks up via a physical mechanism rather than the loss of
wall stability.

\begin{figure}
  \epsfxsize=0.48\textwidth\epsfbox{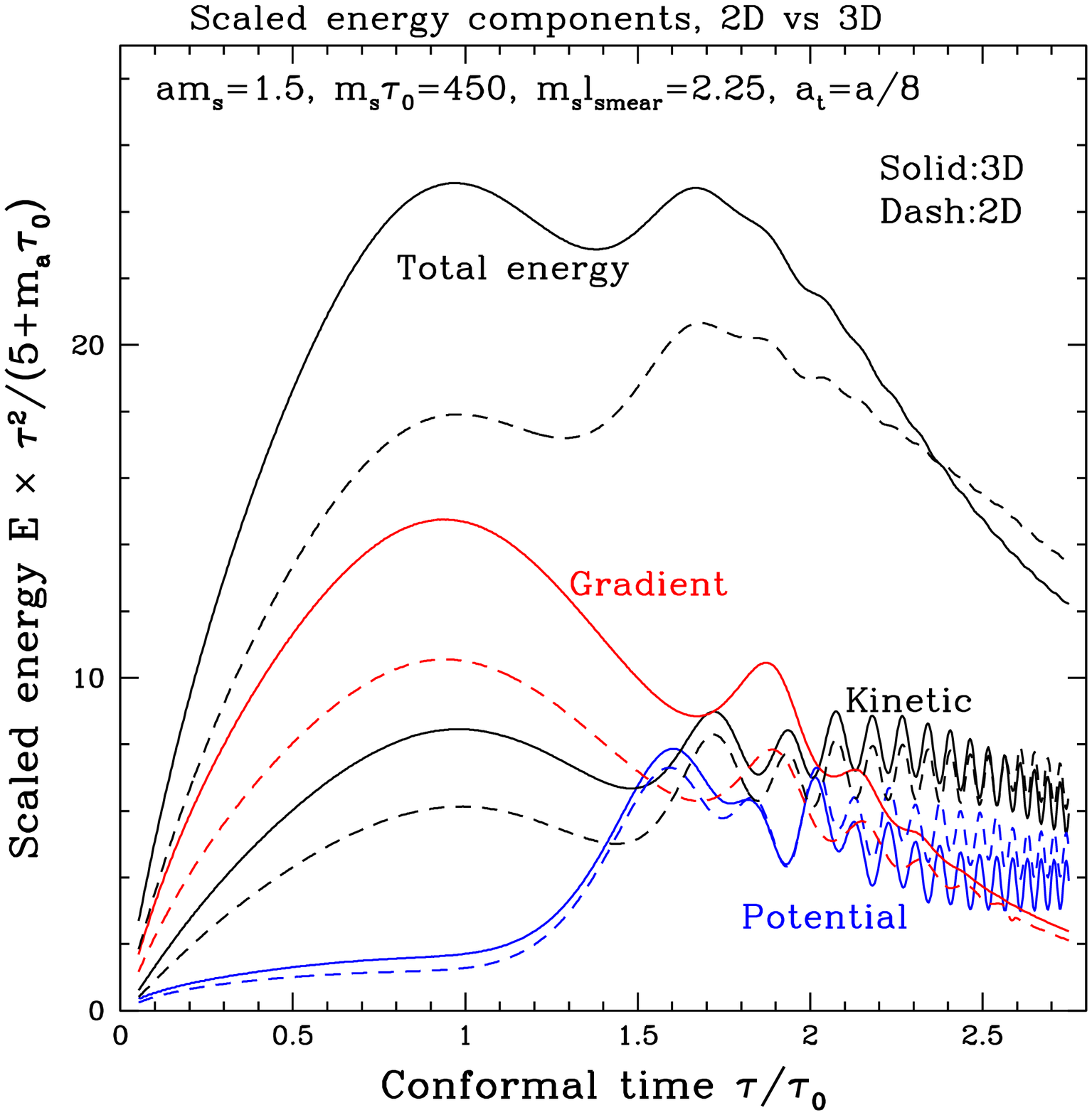}
  \hfill
  \epsfxsize=0.48\textwidth\epsfbox{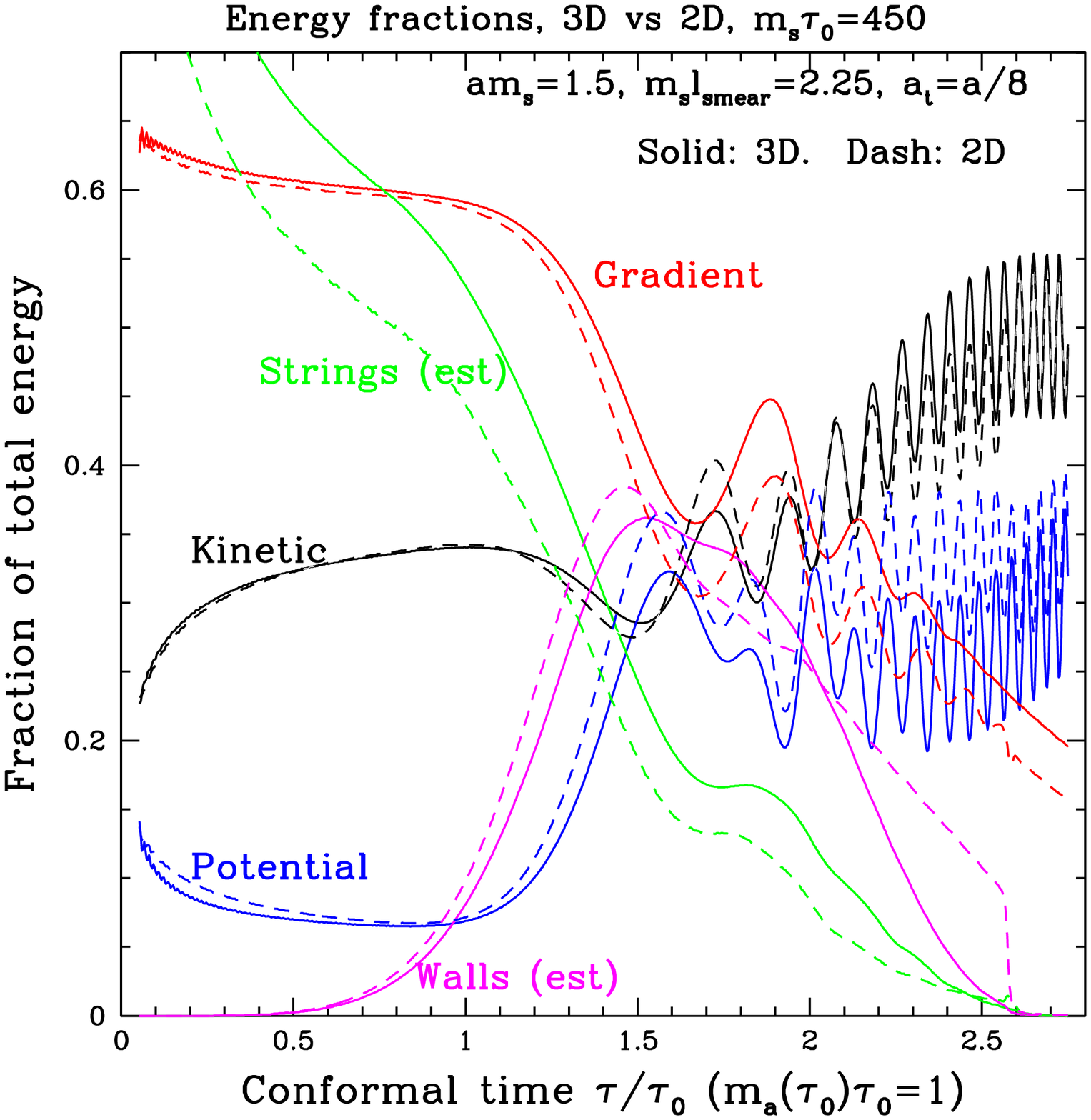}
  \caption{\label{300compare}
    Comparing 2D and 3D simulations at the same $m_s\tau_0$ value:
    Scaled energy (left) and energy fractions (right).}
\end{figure}

For further comparison between the 2D and 3D cases, we have plotted
the energy and the energy fractions for each dimensionality at a common value of
$m_s \tau_0 = 450$, in Fig.~\ref{300compare}.  The energy density
starts out somewhat higher in the 3D case, corresponding to the larger
string density obtained in 3D; but the later stages of the
evolution are strikingly similar, except that the 3D string/wall
network decays before the unphysical collapse, while the 2D network
still carries some energy when the collapse occurs.

\begin{figure}
  \centerline{\epsfxsize=0.6\textwidth\epsfbox{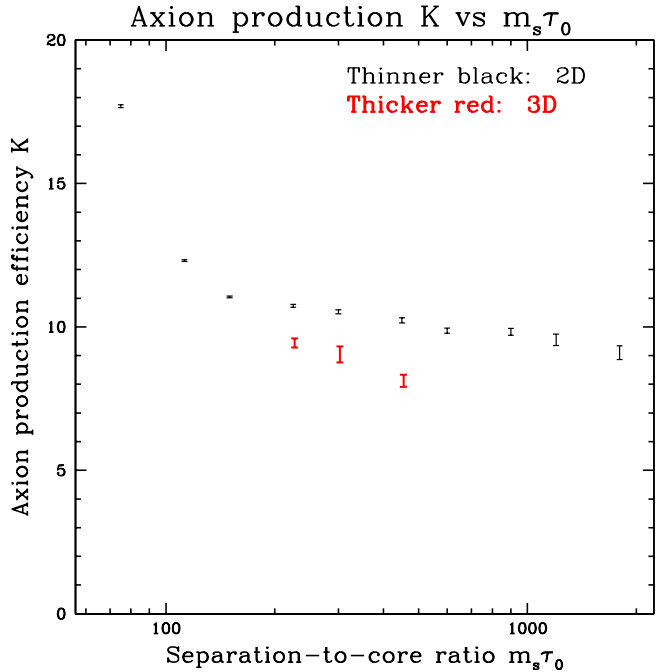}}
  \caption{\label{figfinal}
    Axion production efficiency $K$ as a function of $m_s \tau_0$, in
    2D and in 3D}
\end{figure}

We also plot the final axion-production efficiency $K$ as a function
of $m_s \tau_0$, for both 2D and 3D, in Fig.~\ref{figfinal}.  Besides
the very smallest values of $m_s \tau_0$, it is a very weak function.
Surprisingly, $K$ is about half of the angle-averaged
misalignment-mechanism estimate, and it appears to be a weakly
declining function of $m_s \tau_0$.  Note however that it is very
dangerous to extrapolate based on this result, since the physical
value of $m_s \tau_0 \sim 10^{30}$ is about 27 orders of magnitude
larger.

\section{Discussion and Conclusions}

The axion abundance is determined by dynamics occurring around the
conformal time $\tau_0$ where $\tau m(\tau)=1$; we have seen that the
axion number is essentially fixed by $\tau=3\tau_0$.  Entering this
time period, there is a network of axionic cosmic strings, which
evolves and breaks up due to the appearance of the axionic mass.

We have shown that \textsl{in both 2D and in 3D}, the initial density
of the cosmic string network is a strong, roughly linear, function of
the log of the horizon-to-core ratio $\ln(m_s \tau_0)$.  In
simulations we can make $\ln(m_s \tau_0)$ as large as $\ln(450)=6$ in
3D and $\ln(1800)=7.5$ in 2D; but in nature we expect it to be
$\ln(\sim 10^{30}) \sim 69$.  Therefore we can
only study axion production (so far) for cases with a much more dilute
starting string network than we expect in the cosmological axion
context.

We also showed that, for too small a value of $m_s \tau_0$, the
network dynamics suffer an additional numerical artifact; the axionic
walls and strings become absolutely unstable and the network suddenly
collapses.  This occurs whenever the ratio of angular-to-radial masses
exceeds $m_a^2 / m_s^2 = 1/39$.  This effect compels us to work at
large (numerically expensive) values of $m_s \tau_0$; but we can
achieve values large enough that the instability only occurs after
almost all strings and walls have disappeared.

We do not believe that this problem influenced
the results of \cite{Hiramatsu:2012gg} because their $m_a^2/m_s^2$
ratio stopped increasing partway through their simulations and
then remained fixed, preventing it from growing too large.  On the
other hand, for the axion to play a part in dark matter we expect
$\tau_0$ to occur at temperature $T_0 \sim 1.5$ GeV, so this
flattening-off behavior is unphysical, raising some questions about
their results.

Perhaps the most significant, and to us surprising, result of our
simulations is that, even over a range of $\kappa=\ln(m_s \tau_0)$
where the starting string density varies by a factor of 2, the final
axion number is unchanged at the $20\%$ level, and in fact trends
slightly down with increasing string network density.  So at least
within the range of string network densities we have been able to
study, the string density appears to play little role in establishing
the final axion number density.

Suppose this is the case, so the produced
axion number is well described by \Eq{naxK} with $K=8$.  What do we
learn about the axion decay-constant $f_a$ and mass $m_a$?  Combining
\Eq{naxK}, the Friedmann equation for a hot gas,
\be
H^2 = \frac{8\pi}{3 m_{\mathrm{pl}}^2} \frac{\pi^2 g_* T^4}{30},
\ee
an estimate of the hot topological susceptibility from Wantz and
Shellard \cite{Wantz:2009it},
\be
\label{Wantz}
\chi(T\gg T_c) \simeq \alpha_{\sss{WS}} \Lambda^4 (\Lambda/T)^n \,,
\ee
with $\Lambda \equiv 400$ MeV, $n\simeq 6.68$, and
$\alpha_{\sss{WS}}=1.68\times 10^{-7}$, and the relevant cosmological
information from Planck \cite{Ade:2015xua}
\bea
\label{Planck}
\frac{n_b}{s} &\simeq& 8.59\times 10^{-11} \,, \nn \\
\frac{\rho_{\sss{DM}}}{s} &=& \frac{\Omega_{\sss{DM}} h^2}{\Omega_b h^2}
\frac{m_p n_b}{s} \simeq \frac{0.1194}{0.0221}(938\:\mathrm{MeV})
(8.59\times 10^{-11})
\simeq 0.39 \:\mathrm{eV},
\eea
along with an estimate $g_*=64$ based on a plasma of
($e\mu\tau,\nu_{123}$,$\gamma$,g,$udsc$) with the QCD degrees of freedom
contributing 80\% of the free-particle value to account for strong
interaction corrections at this temperature \cite{Borsanyi:2013bia},
we find
%
%Details:  Define $\kappa = \sqrt{8\pi^3 g_*/90}$,
% $\kappa_s = 2\pi^2 g_*/45$, and $K=8$ (presumably).  Write the
% vacuum topological susceptibility as $\chi(T=0) = M^4$ with $M=76$
% MeV.  Friedmann says:
% $H_0^2 = \kappa^2 T_0^4 / m_{pl}^2$.
% Definition of $T_0$ is $m(T_0)=H_0$, with
% $m(T_0) = \alpha_{ws} \Lambda^{4+n}/T_0^{n}$
% Combining, we find
% $T_0 = \Lambda \left( \frac{\alpha_{ws} m_{pl}^2}{\kappa^2 f_a^2}
%  \right)^{1/(4+n)}$
% The Planck result is that
% $ \frac{0.39eV}{m_a} = \frac{K H_0 f_a^2}{\kappa_s T_0^3}$
% Using $m_a = M^2/f_a$ and the previous expression for $T_0$, we find
% $ f_a = \frac{m_{pl}}{\kappa} \alpha^{1/(n+6)} \times
% \left(\frac{\kappa_s \Lambda (.39eV)}{K M^2} \right)^{(n+4)/(n+6)}$
% which is how we got both numbers and scaling dependence.
%
%
\be
\label{fa_final}
f_a = 3.3\times 10^{11}\:\mbox{GeV}\,
\times \left( \frac{K}{8} \right)^{-0.84}
\left( \frac{g_*}{64} \right)^{0.34}
\left( \frac{\alpha_{\sss{WS}}}{1.68\times 10^{-7}} \right)^{0.079}\,.
\ee
This value gives $m_a = (76\:\mathrm{MeV})^2/f_a = 18\,\mu$eV.
The transition temperature is $T_0\sim 1.5$ GeV.  In our simulations,
most of the nontrivial 3D network evolution took place between
$\tau=1.4\tau_0$ and $\tau=2.6\tau_0$, corresponding to
$T=1100$--$580$ MeV.
This is the range where we need to understand the
topological susceptibility better -- though $f_a$ is only sensitive to
a rescaling of the topological susceptibility through
the $0.079$ power, so even a factor of 10 error in the
estimate of \Eq{Wantz} (as suggested by recent work
\cite{Borsanyi:2015cka}) makes a modest 20\% shift in $f_a$ and $m_a$.
A significant change in our estimate for $K$ would have a larger
effect (though a very small effect on the relevant $T_0$).

Over the range we studied, there was very little change to $K$ in
going from 2D to 3D and in varying the string network density by about
a factor of 2.  We have made no attempt to separate which axions arise
from misalignment, which from strings, and which from walls; indeed it
is not clear to us that doing so is either well defined or terribly
useful.  But it appears that there is not a large component strictly
proportional to the density of strings.  Nevertheless, we find it
rather brave to extrapolate that the same value $K=8$ should apply
when the string network is 5 to 10 times denser than in the
simulations we considered.

We believe that it is well motivated to look for a way of simulating
axion production from global networks with much larger core tension.
It is clear that the enormous factor we need cannot be achieved simply
by shrinking the lattice spacing.  Rather, a new strategy is needed.
We see the need for a method to excise and treat explicitly the physics
of the string core, by adding degrees of freedom to describe its
tension and inertia.  We are close to presenting such a technique for
2 dimensional networks, taking advantage of the dual electromagnetic
description.  Details will appear elsewhere.

\section*{Acknowledgments}

We would like to thank Leslie Rosenberg for useful
conversations, and Zoltan Fodor and collaborators for useful
conversations and for communicating the results of their study of
topology on the lattice \cite{Borsanyi:2015cka} before its
publication.  This work was supported by the Natural Science and
Engineering Research Council (NSERC) of Canada.

\appendix

\section{Numerical implementation}
\label{app:A}

\subsection{Action approach}
\label{app:action}

The starting point of our implementation is the action, \Eq{action},
with $\tau^4 \to \tau^2$ in the first potential term.  We will
directly discretize the action; the extremization with respect to each
field value then gives us an update rule which will be a leapfrog
update.  It is also straightforward in this approach to make the
temporal spacing vary, for instance having finer time step at very
early times when the $\tau^2$ behavior (which gives Hubble drag) is
rapidly changing from timestep to timestep.  Specifically, if we
discretize on a cubic lattice with spacing $a$ and a set of conformal
times $\tau_n$ with spacing $a_{\tau,n} \equiv \tau_{n}-\tau_{n-1}$,
then the action is
\bea
\label{S_discrete}
S_{\mathrm{latt}} &=&   - \sum_{n} \sum_{x} a^3 a_{\tau,n}
\left( \frac{\tau_n \tau_{n-1}}{2} \left[
  \frac{(\varphi_r(x,\tau_n)-\varphi_r(x,\tau_{n-1}))^2
%    + (\varphi_i(x,\tau_n) - \varphi_i(x,\tau_{n-1}))^2
  + (r\to i)
  }{a_{\tau,n}^2} \right]
\right) \\
&& + \sum_{n} \sum_{x} a^3 \tau_n^2 \frac{a_{\tau,n} + a_{\tau,n+1}}{4}
\left( \sum_{i=1,2,3} \left[
  \frac{(\varphi_r(x+a\hat{i},\tau_n) - \varphi_r(x,\tau_n))^2
%    + (\varphi_i(x+a\hat{i},\tau_n) - \varphi_i(x,\tau_n))^2
    + (r\to i)
  }{a^2} \right] \right) \nn \\
&& + \sum_{n} \sum_{x} a^3 \tau_n^2 \frac{a_{\tau,n} + a_{\tau,n+1}}{2}
\left( \frac{m_s^2}{8}
\left[ \varphi_r^2(x,\tau_n) + \varphi_i^2(x,\tau_n) - 1 \right]^2
+ m_a^2(\tau) ( 1 - \varphi_r ) \right) .
\nn
\eea
Here the time derivative term stretches between times $\tau_{n-1}$ and
$\tau_n$, so we replace $\tau^2$ in its coefficient with the product
of the starting and ending time.  The coefficient $a_{\tau,n}$ on this
term accounts for the $\int d\tau$ running from $\tau_{n-1}$ to
$\tau_n$.  For the space terms, which occur at time $\tau_n$, we
replace $\tau^2$ with $\tau_n^2$, and we treat the amount of
$\int d\tau$ contributing to the term to be half the interval before
the term appears plus half the interval after the term appears, hence
the factor $(a_{\tau,n} + a_{\tau,n+1})/2$.  We have also implemented
a next-neighbor improved gradient term, but most of our results are
based on the above gradient term.

For uniform temporal spacing, variation with respect to $\varphi_i$
gives
\bea
\label{update_uniform}
\varphi_i(x,\tau_{n+1}) - \varphi_i(x,\tau_n) &=&
 \frac{\tau_n \tau_{n-1}}{\tau_n \tau_{n+1}} \left(
  \varphi_i(x,\tau_n) - \varphi_i(x,\tau_{n-1}) \right)
 \nn \\
 && + \frac{\tau_n^2}{\tau_n \tau_{n+1}} \left(
  \frac{\sum_i \Big[\varphi_i(x+a\hat{i},\tau_n)
     -2 \varphi_i(x,\tau_n) + \varphi_i(x-a\hat{i},\tau_n)\Big]}
   {a^2} \right. \nn \\
 && \phantom{\frac{\tau_n^2}{\tau_n \tau_{n+1} \bigg(}} \left.
     \vphantom{\frac{\Big[}{1}} {} + \frac{m_s^2}{2} \varphi_i
     (1-\varphi_r^2 - \varphi_i^2) \right) \,.
\eea
One often writes this in terms of the conjugate momentum
$\varphi_i(x,\tau_{n+1}) - \varphi_i(x,\tau_n) \equiv \pi_i(x,\tau_n)$.  The factor $\tau_{n-1}/\tau_{n+1}$, which
arises from the overall $\tau^2$ factor in the action, accounts
for Hubble drag.  The middle term is the gradient term, the final term
is the radial potential term, and the $\varphi_r$ equation is the same
but with the addition of a linear $m_a^2$ term.

In practice we use a very fine temporal spacing at small $\tau$,
$a_\tau = a/40$ for $\tau<a$ and $a_\tau < \tau/20$ thereafter, to
avoid errors when the Hubble drag is large.  This is probably
unnecessary.  At larger times we go over to a fixed value for
$a_\tau/a$.

As initial values, we first set the field to be of unit magnitude and
independent random phase ($\varphi_r(x) = \cos \theta_x$ and
$\varphi_i(x) = \sin \theta_x$, with each $\theta_x$ chosen uniformly
from $[0,2\pi)$) and then apply smearing to a coherence length
$\ell_{\mathrm{smear}}$.  We will check for dependence on parameters
such as $m_s a$, $\ell_{\mathrm{smear}}$, and $a_t/a$ momentarily.

\subsection{Strings and string velocities}
\label{app:stringfind}

We want to know the density of the string network, which requires
identifying where the strings are.  To do this we define the
plaquettes that are pierced by a string, and we count these plaquettes,
applying a statistical correction which we now explain.

\begin{figure}[ht]
\centerline{\epsfxsize=0.6\textwidth \epsfbox{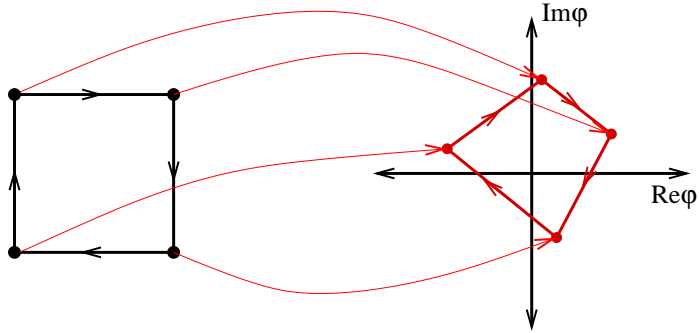}}  
  \caption{\label{corners} Four points around a plaquette map to four
    points on the complex $\varphi$ plane.  If the resulting
    tetragon encloses the origin, we identify the plaquette as being
    pierced by a string.}
\end{figure}

Each plaquette has four corners, and we say that a string goes through
a plaquette if the tetragon in $\varphi$ space, whose corners are the
$\varphi$ values at the corners of the plaquette, encloses the point
$\varphi=0$, see Figure \ref{corners}.  The direction or sense of the
string is determined by whether the origin is enclosed in a clockwise
or a counterclockwise sense.

Our algorithm for determining if a string pierces a plaquette is as
follows.  For the tetragon in Fig.~\ref{corners} to contain a string,
the real axis must be crossed twice with the same handedness
(clockwise or counterclockwise); that is, the number of strings
piercing the plaquette is half the signed sum of real axis crossings,
where the sign is determined by whether the crossing is clockwise or
counterclockwise.  The real axis is crossed between points
$x_1$ and $x_2$ with values $\varphi_1=\varphi(x_1)$ and
$\varphi_2=\varphi(x_2)$ if $\Im(\varphi_1) \Im(\varphi_2) < 0$.
The axis crossing is clockwise if $\Im(\varphi_1 \varphi_2^*) > 0$ and
is counterclockwise if $\Im(\varphi_1 \varphi_2^*) < 0$.  The winding
number is the sum of $+ \frac{1}{2}$ ($- \frac{1}{2}$) for each
clockwise (counterclockwise) axis crossing, as we consider each pair of
corners going clockwise around the plaquette.

We also identify which
links have a domain wall go through them by finding links where
$\Im(\varphi_1) \Im(\varphi_2^*) < 0$ (the real axis is crossed) with
$\Im(\varphi_1 \varphi_2^*) \Im(\varphi_1 - \varphi_2) < 0$ (it is the
negative half-axis which is crossed).
This occurs an odd number of times for each plaquette containing a
string, and an even
number of times for any other plaquette.  Each algorithm (string and
wall) requires only multiplication and comparison; no divisions or
trigonometric functions are needed, providing good numerical
efficiency.  We can sample the full lattice for strings every lattice
unit of time ($\Delta \tau = a$) while taking less numerical effort
than the field updates.
We believe that our plaquette identification is equivalent to that of
Hiramatsu \textsl{et al} \cite{Hiramatsu:2010yu}, but by avoiding explicit
reference to angles we avoid the trigonometric evaluations needed
there.

We make no attempt to connect plaquettes pierced by strings to identify
the strings and their lengths.  Instead we rely on counting plaquettes
with string and links with wall.  In 2 dimensions this works for
strings, but in 3d (and in both 2d and 3d for walls) there are
normalization issues.  If the unit normal along a string is
$(l_x,l_y,l_z)$, then a length $L$ of string will pierce $L|l_x|$
$yz$-plaquettes, $L|l_y|$ $xz$-plaquettes, and $L|l_z|$
$xy$-plaquettes; for different directions the total number of
plaquettes is between $L$ and $L\sqrt{3}$. We are only trying to
determine the string density statistically, and all directions are
statistically equally likely, so we can find the right average string
density statistically by counting string-pierced plaquettes and
dividing by the angle-averaged number of plaquettes per unit length of
string, which is
\be
\label{count_factor}
\mbox{count-factor} = \int_0^{2\pi} \frac{d\phi}{2\pi} \int_{0}^\pi
\frac{\sin \theta d\theta}{2} ( |\Omega_x| + |\Omega_y| + |\Omega_z| )
= 3 \int_0^{1} d\cos(\theta) \times \cos(\theta) = \frac{3}{2} \,,
\ee
where $\Omega_{x,y,z}$ are the $x,y,z$-components of the unit vector
in the $\phi,\theta$ direction.
The same overcounting factor applies for domain walls in 3D, since the
number of links piercing the wall depends on the wall-normal in the
same way as plaquettes depend on the string-normal.  In 2D the domain
walls are overcounted by
$\int_0^{2\pi} \frac{d\phi}{2\pi} (|\Omega_x| + |\Omega_y|) = 4/\pi$.

To determine the average string velocity $v$, $v^2$, and
$\gamma=(1-v^2)^{-1/2}$, we use the time derivative of the field near
the string core.  Consider a string core stretched along the $z$ axis
and moving in the $x$ direction.  For the string at rest, the minimum
of the string's energy, \Eq{Estring}, occurs for
\be
\label{core}
\varphi(x,y,z) = e^{-i\phi_0} f_a \frac{x+iy}{r} f(r) \,,
\ee
with $f(r)$ solving the equation of motion
\be
\label{f_eom}
f'' + \frac{f'}{r} + \frac{f}{r^2} + \frac{m_s^2}{2} f(1-f^2) = 0
\,, \qquad
f(0)=0, \quad f(r\to \infty) \to 1 .
\ee
The equation of motion near 0 enforces that
\be
\label{small_f}
f(r) = c m_s r - \frac{c}{16} (m_sr)^3 + \frac{c+16c^3}{768} (m_sr)^5 +
\ldots
\ee
with $c$ a constant; solving \Eq{f_eom} via
overshoot-undershoot determines $c=0.41238$. Therefore, near the
string core, the field is
\be
\label{small_phi}
\varphi(x,y) = f_a c m_s (x+iy) e^{-i\phi_0}
 \left( 1 - m_s^2 \frac{x^2+y^2}{16} + \ldots \right) \,.
\ee
The moving string solution at $t=0$ is found by replacing $x$ with
$\gamma x$; and the time derivative for the moving string is
$\partial_t \varphi = -v \partial_x \varphi$.  Therefore
\be
\label{order2}
-e^{i\phi_0} \partial_t \varphi(x,y;v) = f_a c m_s \gamma v
- f_a c m_s^3 \gamma v \left( \frac{3\gamma^2 x^2 + 2i\gamma xy
  + y^2}{16} \right) + \ldots.
\ee
At lowest order in $m^2r^2$ we find
\be
\label{phit_LO}
\partial_t \varphi^* \partial_t \varphi = f_a^2 m_s^2 c^2 \gamma^2 v^2
\quad \to \quad
\gamma^2 v^2 = \frac{\partial_t \varphi^* \partial_t \varphi}
  {f_a^2 m_s^2 c^2} \,.
\ee
The next-order term in Eq.~(\ref{order2}) can be used to engineer a
correction for fields close to but not at the center of the string:
\be
\label{eq:vsq}
\gamma^2 v^2 \simeq 
  \frac{\partial_t \varphi^* \partial_t \varphi}
       {m_s^2 c^2 f_a^2}
  \left( 1 + \frac{\varphi^* \varphi}{8c^2 f_a^2} \right)
  + \frac{(\varphi^* \partial_t \varphi + \varphi \partial_t
    \varphi^*)^2}{16 m_s^2 c^4 f_a^4} \,,
\ee
which we found by establishing the leading-order small-distance
behavior of $\varphi^* \varphi$ and $|\varphi^* \partial_t \varphi|$
and finding a combination that would cancel the subleading
contributions in $\partial_t \varphi^* \partial_t \varphi$.

We apply this approach by sampling over points that are near the
string's center, so that the second-order, $\sim m_s^2 r^2$ terms are
small, and higher (uncomputed) corrections should be negligible.  We
choose for our sample of points near the string core the set of points
on corners of plaquettes that are pierced by a string; points on
more than one string-pierced plaquette are counted once per plaquette.
For each plaquette we find $v^2/(1-v^2)$ explicitly by averaging
Eq.~(\ref{eq:vsq}) over the four plaquette corners; we then compute the local
value of $\langle v \rangle$, $\langle v^2 \rangle$, and
$\langle \gamma \rangle$.  We average $(1,v,v^2,\gamma)$ over all
pierced plaquettes, weighting with weight $\gamma$ in order to follow
the literature convention that the string network should be weighted
by $\int \gamma dl$ (energy content), not $\int dl$ (length).
The numerical overhead is small because the computation need only be
performed on sites that are identified as corners of a string-pierced
plaquette.

The method assumes that points with distance $\sim a$ away from the
string core are still ``relatively near'' the string's core, which is
an expansion in $(c m_s a)^2/16 < 1$.  Note that, in 3D, the average
distance-squared of a point on a pierced-plaquette to the nearest
point on the piercing string is $\langle x^2 \rangle = a^2/2$, so the
sampled points are surprisingly close to the strings, and the
higher-order corrections are expected to be small.

Another weakness of the approach is that it is based on the structure
of a straight string; it will commit errors for bent, accelerating
strings.  However, in this case the string's length and velocity are
anyways not uniquely defined; it is not clear to us that the method
works any worse than other approaches.  Similarly, it could be
confused by any radial fluctuations (breathing modes) in $f(r)$; but
these modes are heavy and we do not expect them to occur with large
amplitude except perhaps at first due to initial conditions.
Also note that, since the algorithm
determines $\gamma^2 v^2$ in terms of a manifestly positive
expression, it never concludes that a string is moving faster than
light speed, something which could in principle occur for some
string-velocity approaches.

We mention one final challenge for our method.  We have assumed that
the string's shape correctly reflects the Lorentz contraction of the
string's structure.  On a sufficiently coarse lattice, Lorentz
invariance is not respected in the string core and the Lorentz
contraction will not be as strong as it should be in the continuum.  This
means that the method will under-estimate the $\gamma$ factor of very
fast strings on coarse lattices.  If the main goal is a high-precision
determination of string velocities and $\gamma$-factors, one should
use a smaller value of $m_s a$ than we have used in most of our
studies.

\begin{figure}[ht]
\centerline{\epsfxsize=0.6\textwidth \epsfbox{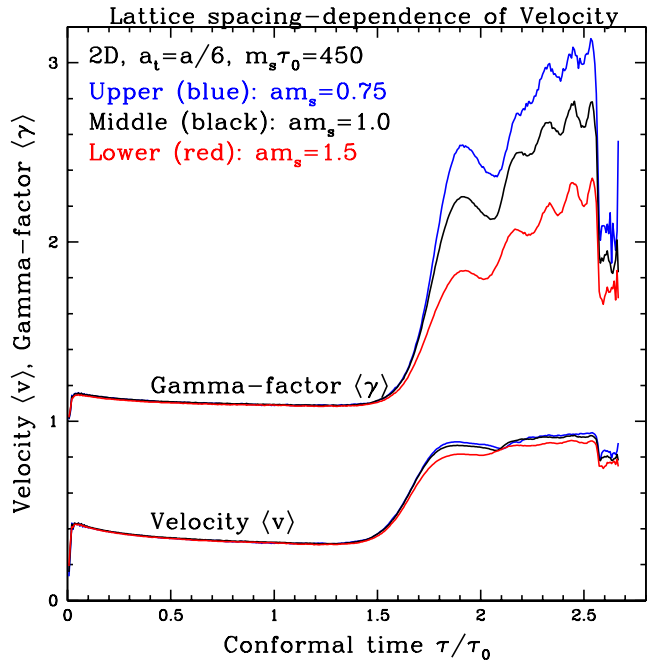}}
  \caption{\label{fig:appgamma}
    String velocity for 2D, $m_s\tau_0=450$ simulations with three
    values of lattice spacing: $m_s a=0.75,$ 1.0, 1.5.  Our string velocity
    method finds a higher $\gamma$-factor on the finer lattice.
  (The crash in the $\gamma$-factor at $\tau=2.57\tau_0$ coincides
    with the collapse of the network by direct wall instabilities and
    should not be taken seriously.)}
\end{figure}

To test this last point, we have repeated the simulation of a
$m_s\tau_0=450$ 2D string network using $m_s a=0.75$ and $m_s a=1.0$
lattices, to
compare with our results in the main text using $m_s a=1.5$.  Most
properties (energy, string length, wall extent, final axion number)
agree very well; the final axion number is the same within 1\%
statistical errors.  But once the strings start to move very fast late in
the simulation, the finer lattice observes a significantly larger
$\gamma$-factor for the strings, as shown in Fig.~\ref{fig:appgamma}.

We think a similar velocity-finding approach could be applied to
field-theory simulations of local strings (Abelian Higgs model
simulations), but since the string's structure then
depends on two parameters (the Higgs mass and the gauge boson mass),
the application is more complicated, with the constant $c$ above
replaced by some $m_{\sss H}/m_{\sss A}$ dependent value; finding the
NLO corrections would also be significantly more complicated.

\subsection{Numerical tests}
\label{app:test}

We want to find the largest values of $m_s a$ and $a_t/a$ that are
compatible with a continuum interpretation, and we want to check for
sensitivity to initial conditions such as $\ell_{\textrm{smear}}$.
We will test these parameters using $m_a=0$ or string-only
simulations, because there are then fewer scales to consider.

We start with $m_s a$.  If we choose this too large, the string core
is $\lesssim 1$ lattice site across, which is too small for the lattice
to resolve properly.  In this case, the UV edge of the integral in
\Eq{tension} becomes sensitive to the exact location of the string
relative to the lattice, leading to a string energy varying
periodically with period $a$, rather than being translation
independent.  This makes it possible for a string to ``stick'' in the
most energetically favorable lattice location, which will interfere
with string evolution and impede the annihilation of the network.
To test for this problem, we made a series of evolutions with
different values of $m_s a$, but identical other properties as
measured in terms of $m_s$.  These can be interpreted as varying the
lattice spacing.

\begin{figure}[ht]
{\epsfxsize=0.48\textwidth \epsfbox{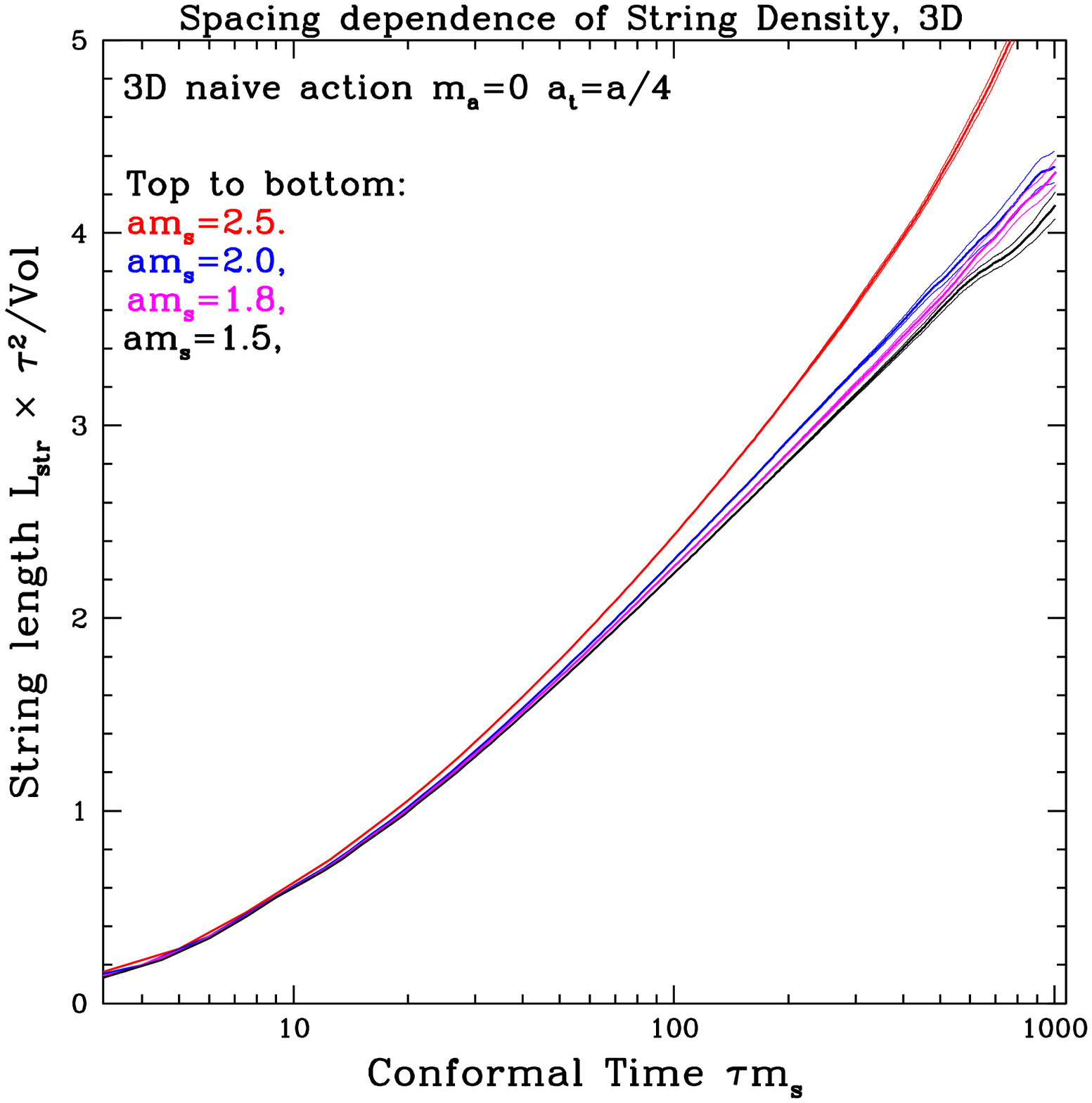}}
     \hfill
{\epsfxsize=0.48\textwidth \epsfbox{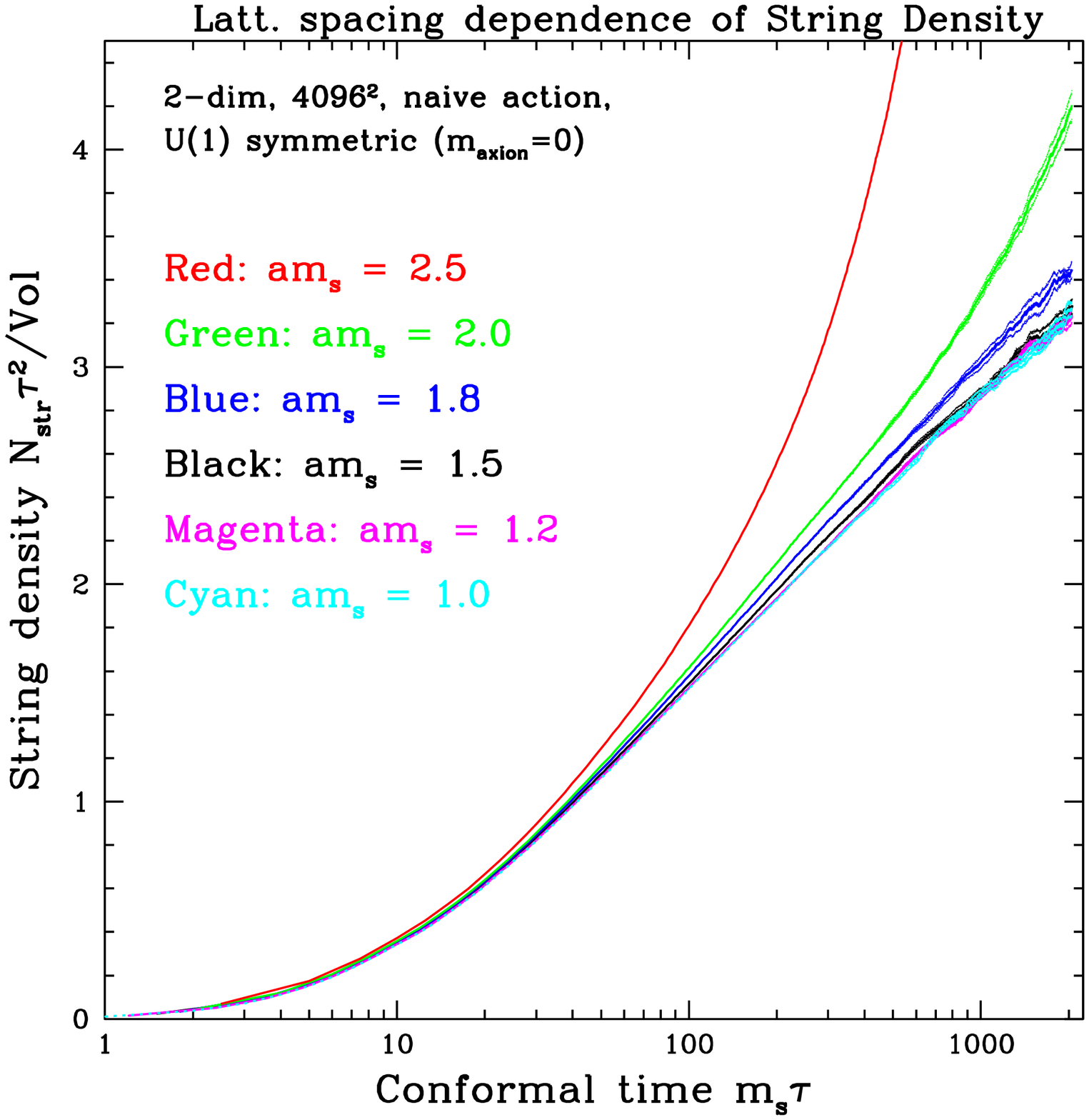}}
\caption{\label{fig:ms}
  Dependence of string density on the lattice spacing (in units of the
  saxion mass, $am_s$).  Left:  string length in 3 dimensions.
  Right:  number of strings in 2 dimensions.  Lines represent
  $1\sigma$ statistical range based on averaging several samples.}
\end{figure}

We show how the string length (or, in 2D, the number
of strings) vary with time for several values of $m_s a$, in both 2
and 3 dimensions, in Fig.~\ref{fig:ms}.  In both cases,
the results are consistent within 2\% for all $m_s a \leq 1.5$,
but deviate for larger $m_s a$, especially at later times.  We did a
similar study with the next-neighbor improved action (not shown),
which showed continuum behavior slightly sooner, at $m_s a = 1.8$.
The reduced number of lattice points this allows, $(1.8/1.5)^{d+1}$ in
$d$ space dimensions, does not quite make up for the factor of 2 in
numerical cost for that algorithm, so we have generally stuck with
nearest-neighbor interactions in the remainder of our work.

Note that the result in Fig.~\ref{fig:ms} clearly shows that the
string length rises as $\ln(m_s t)$, rather than approaching a flat
value, both in 2D and in 3D.  Therefore we already see the logarithmic
corrections to scaling in this figure.  If we re-plot the figure in
terms of $\tau/a$, rather than $m_s \tau$, the lines do \textsl{not}
fall on top of each other; the relevant physical length scale for
comparison is $m_s$, which sets the string core size, not the lattice
spacing. 

\begin{figure}[ht]
  \centerline{\epsfxsize=0.6\textwidth \epsfbox{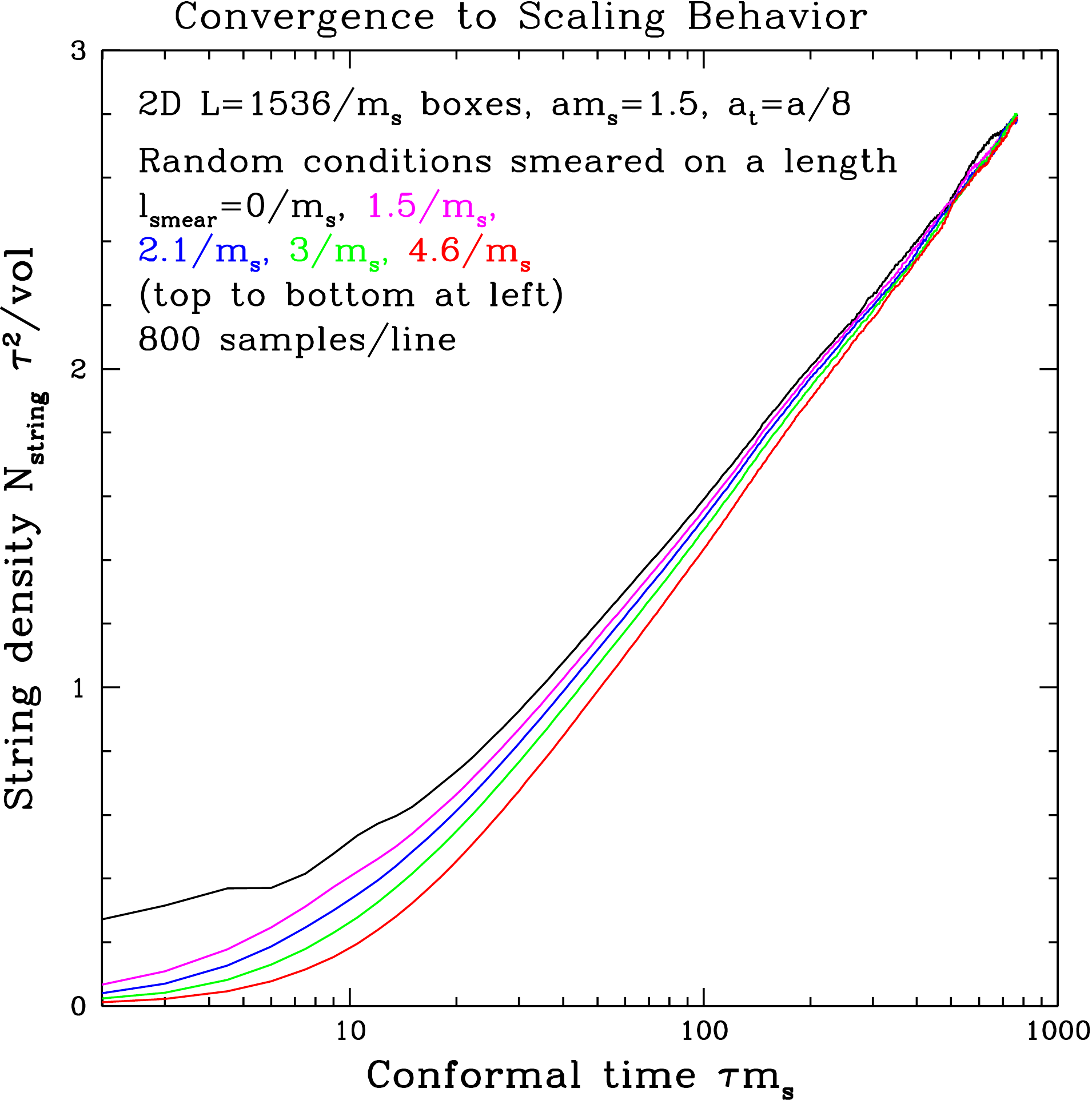}}
  \caption{\label{fig:smear}
    Dependence of string density, in 2D, on the initial correlation
    length of the random initial conditions.}
\end{figure}

Next we check for independence on initial conditions, by varying the
length scale $\ell_{\mathrm{smear}}$ over which the initial conditions
are smeared.  The result, shown in Fig.~\ref{fig:smear}, indicates
that different initial conditions rather quickly converge to the same
string network density, which again scales logarithmically with the
system age as measured in $m_s$ units.  This means the choice for this
smearing length is not important.  We typically choose $2.1/m_s$ in
this work.

\begin{figure}[ht]
\epsfxsize=0.48\textwidth \epsfbox{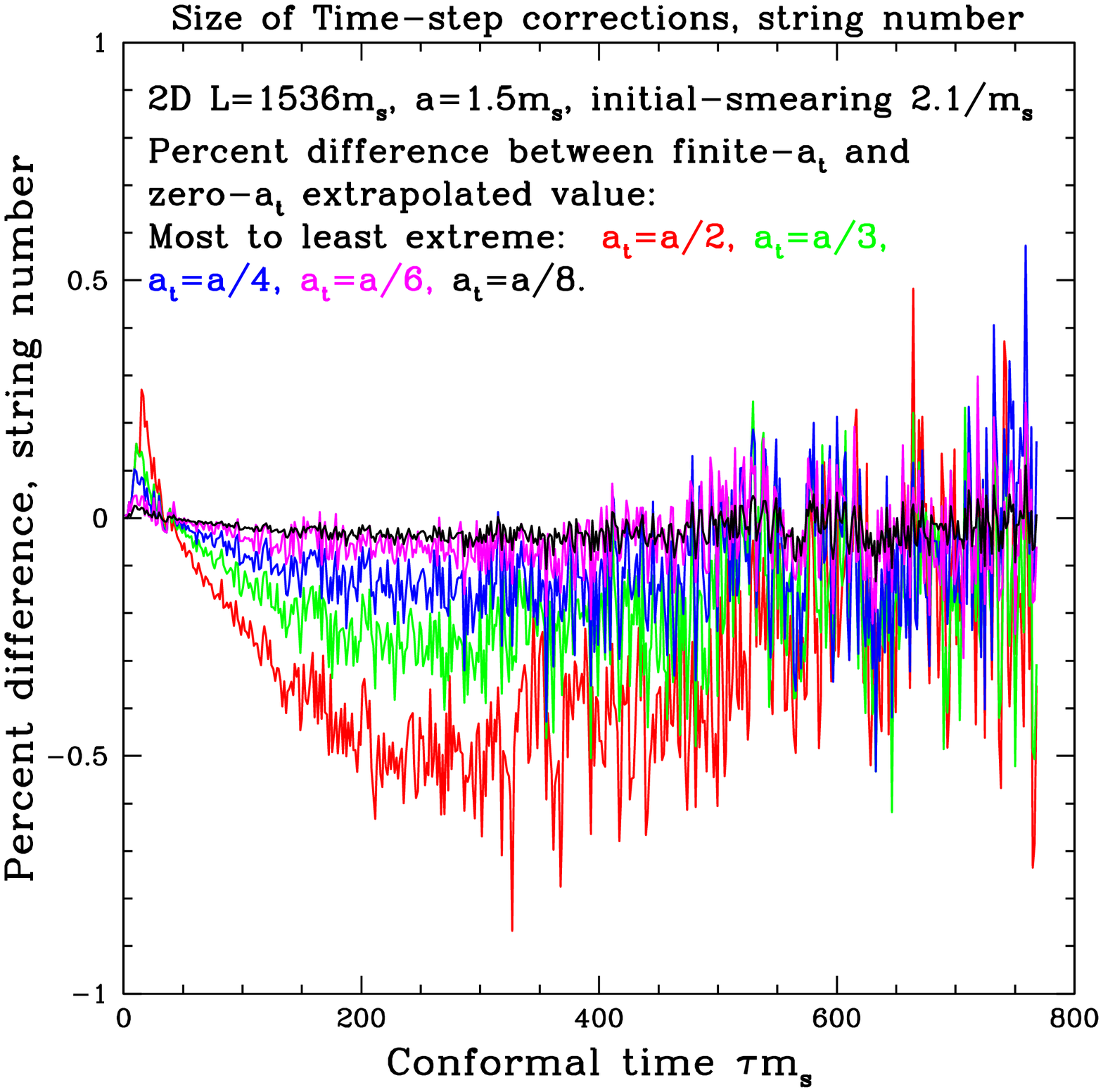}
\hfill
\epsfxsize=0.48\textwidth \epsfbox{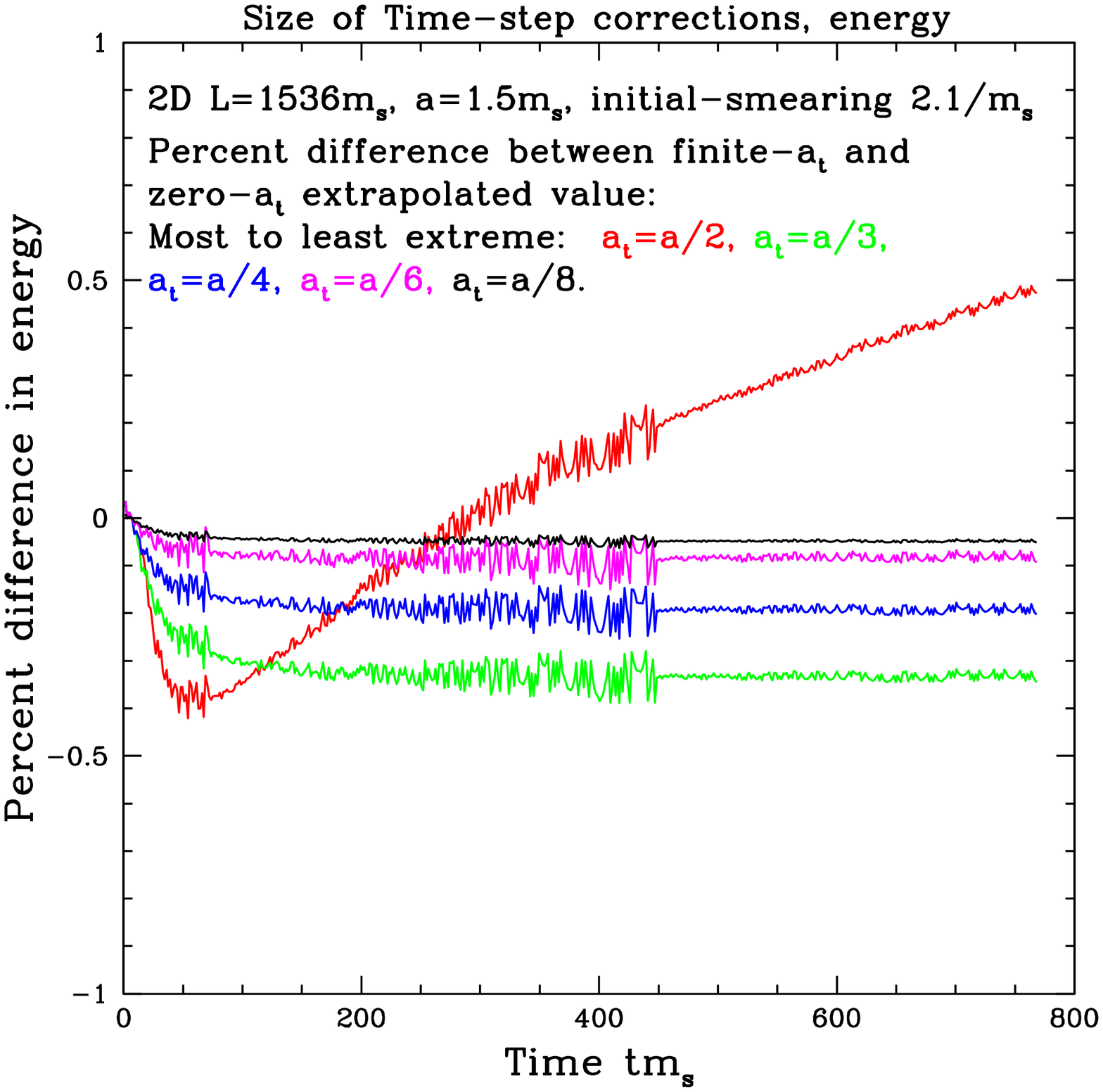}
\caption{\label{t-space}
  Temporal spacing dependence of string density (left) and energy
  (right).  In each case we compare with an extrapolation to zero
  spacing, based on $a_t/a=1/6$ and $1/8$ and assuming $(a_t/a)^2$
  scaling of errors, which is the expected scaling form.}
\end{figure}

We should also check what temporal spacing is sufficient to
approximate continuous time.  There is a Courant condition beyond
which the update algorithm becomes unstable:
$a_t/a = 2/\sqrt{4d+m_s^2}$ with $d=2,3$ the number of space
dimensions.  We choose $a/a_t$ integer, so this limits the available
values to $a/a_t = 2,3,\ldots$.  The axion is very light, and at late
(interesting) times we expect that the physics is primarily in long
wave lengths except near string cores; so one might expect less severe
dependence on this ratio than for some lattice problems.  This turns
out to be the case; as we show%
\footnote{We suppressed statistical fluctuations by using identical
  initial conditions for each $a_t$ value we considered.
  The noise in the lines on the right in the figure arise
  because we write out too few digits before processing the data; they
  drop in size when the energy becomes one digit shorter so one more
  digit is written out.}
in Fig.~\ref{t-space}, a temporal
spacing of $a_t/a = 1/2$ actually only leads to percent differences in
string length and energy, compared to small temporal spacing.
Nevertheless, out of paranoia we have generally used $a_t/a=1/6$ or
$1/8$ in other measurements, which should keep $a_t$ errors below the
per-mille level.

\begin{figure}[ht]
  \centerline{  \epsfxsize=0.6\textwidth \epsfbox{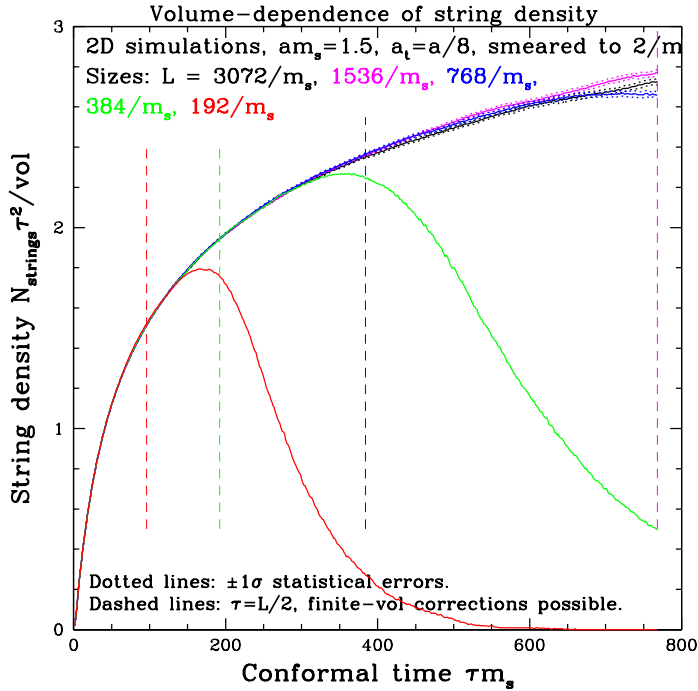}}
  \caption{\label{vary_vol}
    Volume dependence of string length, in 2 dimensions, indicating
    $1\sigma$ error ranges; based on 800, 3000, 12800, 12800, 12800
    samples.}
\end{figure}

Finally, although there are strong theoretical arguments that the
volume cannot affect the network's statistical properties so long as
$\tau<L/2$ (in conformal coordinates), we check this nevertheless in
Fig.~\ref{vary_vol}.  The figure shows that $\tau$ must grow to nearly
twice $L/2$ before significant finite-volume corrections occur; but
they then rapidly become severe.  All results in the rest of the paper
consistently use volumes large enough that evolutions stop at
$\tau < L/2$.

We have generally re-checked the 2D-only results of this section with
3D simulations, which give consistent results but with poorer
statistics because the added numerical costs of treating the 3D
system make it harder to achieve high statistics.

\section{Counting axions}
\label{app:B}

At the end of a simulation we have $\varphi(x)$, from which we want to
extract an axion number.  To do so we first convert to the axion field
amplitude $\theta_a$, and then extract the axion number stored in the
field.  The axion field amplitude $\theta_a$ is determined as
\be
\label{thetaa}
\theta_a = \mathrm{arg} \varphi \,, \quad
\partial_\tau \theta_a = \frac{\Im \varphi^* \partial_\tau \varphi}
        {\sqrt{\varphi^* \varphi}} \,.
\ee
Since the linear term, \Eq{Lreplace}, shifts the absolute minimum of
the potential slightly away from $\varphi_r=f_a$, we make an addative
shift to $\varphi_r$ so that its minimum is at $f_a$ before applying
these rules.

We then assume that this axion field obeys the quadratic Hamiltonian
(writing $\dot\theta = \partial_\tau \theta$)
\be
\label{Hthetaa}
H(\theta_a) = f_a^2 \int d^3 x \half \left(
\theta_a (m_a^2 - \nabla^2) \theta_a + \dot\theta_a^2 \right).
\ee
The Hamiltonian is diagonal in Fourier space ($V$ is the volume of
space)
\be
\label{Hkspace}
H = f_a^2 V \int \frac{d^3 k}{(2\pi)^3} \half \left(
(k^2+m_a^2) \theta_a^2(k) + \dot\theta^2_a(k) \right).
\ee
The particle number associated with a mode with oscillation frequency
$\omega = \sqrt{k^2 + m_a^2}$ is the energy over the frequency, so the
particle number density is
\be
\label{nkspace}
\nax = \int \frac{d^3 k}{(2\pi)^3} \half \left(
\sqrt{k^2+m_a^2} \theta_a^2(k) + \frac{1}{\sqrt{k^2+m_a^2}}
\dot\theta^2_a(k) \right).
\ee

Evaluating this in Fourier space is straightforward and is efficient
when the FFT is available.  If the box size is not a power of two or if the
data is divided over processors in an inconvenient way, we can use
Laplace methods instead.  It is easy to write an algorithm to
evolve $\theta,\dot\theta$ in dissipative ``time'' $\tilde\tau$,
with boundary conditions that $\theta(\tilde\tau=0)$ is the original
value of $\theta_a$:
\be
\label{tildetau}
\partial_{\tilde\tau} \theta_a(x,\tilde\tau)
= (\nabla^2 - m_a^2) \theta_a(x,\tilde\tau) \,, \qquad
\theta_a(x,\tilde\tau=0) = \theta_a(x) \,,
\ee
\textsl{and the same evolution for} $\dot\theta_a$ (in terms of
$\tilde\tau$ evolution, $\dot\theta_a$ is considered an independent
field).  The reason to do so is that we can find
$\theta_a(x,\tilde\tau)$ by position-space methods, but we know that
the evolution in Fourier space will be
\be
\label{tildek}
\theta_a(k,\tilde\tau) = \theta_a(k)\: \exp(-(k^2+m_a^2)\tilde\tau)
\,,
\ee
so large-$k$ modes are more quickly suppressed than small-$k$ modes.
Note that
\be
\label{secret}
\sqrt\frac{2}{\pi} \int_0^\infty \frac{d\tilde\tau}{\sqrt{\tilde\tau}}
\: e^{-2(k^2+m^2) \tilde\tau}
= \frac{1}{\sqrt{k^2+m_a^2}} \,,
\ee
so therefore
\bea
\label{tildemagic}
&& \sqrt\frac{2}{\pi} \int_0^\infty \frac{d\tilde\tau}{\sqrt{\tilde\tau}}
\int d^3 x \left[ \theta_a(x,\tilde\tau) (-\nabla^2 + m_a^2)
  \theta_a(x,\tilde\tau)   + \dot\theta_a^2(x,\tilde\tau) \right]
\nn \\
& = & V \int \frac{d^3 k}{(2\pi)^3} \left[
  \sqrt{k^2+m_a^2} \theta_a^2(k) + \frac{1}{\sqrt{k^2+m_a^2}}
  \dot\theta_a^2(k) \right] \,.
\eea
By evolving in $\tilde\tau$ and numerically implementing the first
integral, which can all be done in coordinate space, we get the
desired second integral,
which correctly counts axion number.  We have compared this method to
the FFT method where both are applicable and have confirmed that they
give the same answer up to controllable numerical
($\tilde\tau$-spacing and large-$\tilde\tau$ extrapolation) issues,
which are easily held below 1\% with smaller numerical cost than the
time evolution needed to find the final $\theta_a$ configuration.
However, this method is not practical for making frequent
determinations of $\nax$ over the course of a simulation.

\bibliographystyle{unsrt}
\bibliography{paper_refs}

%dualcharge = Kalb:1974yc, Lund:1976ze, Vilenkin:1986ku, Davis:1988rw
%*Note: These are all for the 3d case. It is straightforward to see that in 2d
%the dual tensor of the Kalb-Ramond action reduces to the electromagnetic field strength
%tensor.
%
%nambu_string = Dabholkar:1989ju, Bennett:1989yp, Allen:1990tv

\end{document}